%% file: deepthought_dpr1_combined.tex
\documentclass{natureprintstyle}
\usepackage{natbib}
\usepackage[hyphens]{url}
\input{glossary.tex}
\newglossaryentry{qam}{name=QAM, description=first=quartile agreement matrix, first=quartile agreement matrix (QAM)}
\usepackage{comment}
\usepackage{float}
\usepackage{textcmds}
\usepackage{etoolbox}
\BeforeBeginEnvironment{framed}{\begin{minipage}{\linewidth}}
\AfterEndEnvironment{framed}{\end{minipage}\par}
\usepackage{caption}
\usepackage{aas_macros}
\usepackage{framed}
\usepackage{amsmath}
\usepackage{hyperref}
\usepackage{graphicx}
\usepackage{siunitx}
\usepackage{txfonts}
\usepackage{comment}
\usepackage{xr}
\newcommand{\dt}{\textsc{DeepThought}\xspace}

\newcommand{\dataurl}{\url{https://zenodo.org/record/2634598}}
\title{Distributed peer review enhanced with natural language processing and machine learning}

\input{author_natastro.tex}

\begin{document} 


\maketitle
\input{affil_natastro.tex}

\begin{abstract}
While ancient scientists often had patrons to fund their work, peer review of proposals for the allocation of resources is a foundation of modern science. A very common method is that proposals are evaluated by a small panel of experts (due to logistics and funding limitations) nominated by the grant-giving institutions. The expert panel process introduces several issues - most notably: 1) biases introduced in the selection of the panel. 2) experts have to read a very large number of proposals. Distributed Peer Review promises to alleviate several of the described problems by distributing the task of reviewing among the proposers. Each proposer is given a limited number of proposals to review and rank. 

We present the result of an experiment running a machine-learning enhanced distributed peer review process for allocation of telescope time at the European Southern Observatory.

In this work, we show that the distributed peer review is statistically the same as a `traditional' panel, that our machine learning algorithm can predict expertise of reviewers with a high success rate, and we find that seniority and reviewer expertise have an influence on review quality. The general experience has been overwhelmingly praised from the participating community (using an anonymous feedback mechanism).    
\end{abstract}
   
%

\input{deepthought_dpr1_main.tex}
\input{statements.tex}
\bibliographystyle{apj.bst}
\bibliography{wekerzendorf.bib}
\clearpage
\begin{center}
    \thispagestyle{empty}
    \vspace*{\fill}
    \Huge
    Supplementary Information
    \vspace*{\fill}
\end{center}
\clearpage
\input{deepthought_dpr1_si_main.tex}

\input{deepthought_dpr1_si_appendix.tex}

\end{document}

%% file: glossary.tex
\PassOptionsToPackage{draft}{hyperref}
\usepackage{xspace}

\usepackage[xindy, toc, hyperfirst=false, nolist, nostyles, sanitize={name=false,description=false,symbol=false}]{glossaries}
\glsdisablehyper
\usepackage[hyperref,x11names, table]{xcolor}

\newglossaryentry{nlp}{name=NLP, description={Natural Language Processing}, first={Natural Language Processing (NLP)}}

\newglossaryentry{vrad}{name={radial velocity~}, text={radial velocity}, symbol={\ensuremath{v_\textrm{rad}}}, description={radial velocity}, sort=vrad}
\newglossaryentry{vrot}{name={stellar rotation~}, name={stellar rotation}, symbol={\ensuremath{v_\textrm{rot}}}, description={radial velocity}, sort=vrot}

\newcommand{\xray}{X-ray}

\makeatletter \newcommand{\ion}[2]{#1 \textsc{\@roman{#2}}} \makeatother

\newcommand{\sn}[2]{SN~#1#2\xspace}

\newglossaryentry{angstrom}{name=\AA, description={unit of length $10^{-10}$\,m}, sort=angstrom}
\newglossaryentry{nir}{name=NIR,description={near infrared},first = {near infrared (NIR)}}
\newglossaryentry{psf}{name=PSF,description={point-spread function},first = {point-spread function (PSF)}}
\newglossaryentry{fwhm}{name=FWHM,description={Full Width Half Maximum},first = {FWHM}}
\newglossaryentry{rms}{name=RMS,description={Root Mean Square},first = {RMS}}
\newglossaryentry{signalnoise}{name=S/N,description={signal to noise}}
\newglossaryentry{uv}{name=UV,description={ultra violet},first = {ultra violet (UV)}}
\newglossaryentry{halpha}{name=\ensuremath{\textrm{H}\alpha}, description={First line of the Balmer series at 6563\,\AA}, sort=halpha}
\newglossaryentry{mgb}{name={Mg \textsc{i} b}, description={Triplet at 5167\,\AA, 5173\,\AA and 5184\,\AA}}
\newglossaryentry{sobolevapprox}{name={Sobolev approximation}, description={Lines are approximation with an infinitley thin interaction region \citep[e.g. no broadening][]{1960mes..book.....S}}, first={Sobolev approximation }}
\newglossaryentry{radeq}{name={radiative equilibrium}, description={The net flux of energy between matter and radiation field is zero}}
\newglossaryentry{nebularapprox}{name={nebular approximation}, description={Assumes that the plasma condition are controlled by a central radiation source. The radiation field decreases with the distance to the source by geometrical dilution. See \citet{1978stat.book.....M} for details}}
\newglossaryentry{modnebularapprox}{name={modified nebular approximation}, description={In contrast to \gls{nebularapprox} where only geometrical dilution is taken into account, the modified nebular approximation also takes dilution by other radiative processes into account }, first={modified nebular approximation}, parent=nebularapprox}
\newglossaryentry{thompsonscat}{name={Thomson scattering}, description={Scattering of photons on low energy electrons}}
\newglossaryentry{lte}{name={LTE}, description={Local Thermodynamic Equilibrium}, first={local thermodynamic equilibrium (LTE)}}
\newglossaryentry{lsr}{name={LSR}, description={Local Standard of Rest}, first={\textit{local standard of rest} (LSR)}}
\newglossaryentry{mc}{name={MC}, description={Monte Carlo}, first={\textit{Monte Carlo} (MC)}}
\newglossaryentry{wcs}{name={WCS}, description={world coordinate system}, first={world coordinate system (WCS)}}
\newglossaryentry{cmf}{name=CMF, text=CMF, first=Comoving Frame (CMF henceforth), description={Comoving Frame}}

\newglossaryentry{uvoir}{name=UVOIR, text=UVOIR, first=UV/optical/Near-IR (UVOIR), description={UV/optical/Near-IR}}
\newglossaryentry{ccd}{name=CCD,description={Charged Coupled Device}, first={charged coupled device (CCD)}, firstplural={charged coupled devices (CCDs)}}

\newglossaryentry{ew}{name=Equivalent Width, text={EW}, description={width of a rectangle that has the same area as a spectral line when taken to zero flux}, first={equivalent width (EW)}, firstplural={equivalent widths (EWs)}}
\newglossaryentry{agb}{name=AGB,description={Asymptotic Giant Branch}, first={Asymptotic Giant Branch (AGB)}}
\newglossaryentry{cmb}{name=CMB,description={Cosmic Microwave Background}}
\newglossaryentry{csm}{name=CSM,description={Circumstellar Medium}, first={circumstellar medium (CSM)}}
\newglossaryentry{csi}{name=CSI,description={Circumstellar Interaction}, first={circumstellar interaction (CSI)}}
\newglossaryentry{ism}{name=ISM,description={Interstellar Medium}, first={interstellar medium (ISM)}}
\newglossaryentry{ige}{name=IGE,description={Iron Group Element}, first={iron group element (IGE)}, firstplural={iron group elements (IGEs)}}
\newglossaryentry{epm}{name=EPM,description={Expanding Photosphere Method \citep{1974ApJ...193...27K}}, first={Expanding Photosphere Method (EPM)}}
\newglossaryentry{aic}{name=AIC,description={Accretion Induced Collapse}, first={accretion induced collapse (AIC)}}
\newglossaryentry{ime}{name=IME,description={Intermediate Mass Element}, first={intermediate mass element (IME)}, firstplural={intermediate mass elements (IMEs)}}
\newglossaryentry{h0}{name=\ensuremath{H_0},description={Hubbles constant}}
\newglossaryentry{nse}{name=NSE,description={Nuclear Statistical Equilibrium}, first={nuclear statistical equilibrium (NSE)}}
\newglossaryentry{cdm}{name=CDM,description={Cold Dark Matter}}
\newglossaryentry{grb}{name=GRB,description={Gamma Ray Burst}, first={Gamma Ray Burst (GRB)}, firstplural={Gamma Ray Bursts (GRBs)}}
\newglossaryentry{xps}{name=XPS, description={X-ray point source}, first={X-ray point source (XPS)}, firstplural={X-ray point sources (XPS)}}
\newglossaryentry{donor}{name=donor,description={non-degenerate companion in the \gls{sds}}}
\newglossaryentry{mainsequence}{name=main sequence,description={main sequence star}}
\newglossaryentry{redgiant}{name=red giant,description={red giant star}}
\newglossaryentry{mlcs}{name=MLCS,description={Multicolor Light Curve Shape method \citep[MLCS;][]{1996ApJ...473...88R}}, first={Multicolor Light-Curve Shape method \citep[MLCS;][]{1996ApJ...473...88R}}}
\newglossaryentry{rsoph}{name=RS~Ophiuci ,description={white dwarf accreting from a red giant - assumed progenitor of the \gls{sds}}, sort=rsoph}
\newglossaryentry{usco}{name=U~Scorpii,description={white dwarf accreting from a main sequence star - assumed progenitor of the \gls{sds}}, sort=usco}
\newglossaryentry{rcw86}{name=RCW~86,description={supernova remnant sometimes associated with \sn{185}{}}, sort=rcw86}
\newglossaryentry{casa}{name=Cas~A,description={Cassiopeia A supernova remnant - probably a \gls{snib} event}}
\newglossaryentry{cepheid}{name=Cepheid,description={very luminous variable star with a strong luminosity period relationship}}
\newglossaryentry{urca}{name=Urca, text=\textit{Urca}, description={process predominatly contributing to cooling in stars. The \textit{Urca} process consists of alternating electron-capture and $\beta^{-}$ decay of two nuclei pairs.},sort=urca}
\newglossaryentry{alphacen}{name=Alpha Centauri,description={one of the brightest stars in the night sky and a close binary}}
\newglossaryentry{pcygni}{name={P Cygni}, text={P Cygni},description={a hypergiant luminous blue variable with strong winds. Often referred to as a description for their line profiles showing a emission peak at the rest wavelength of the line and a blue-shifted absorption trough.}}

\newglossaryentry{teff}{name={effective temperature~}, text={effective temperature}, symbol={\ensuremath{T_\textrm{eff}}}, description={Temperature of a blackbody emitting the same total energy}, sort=teff}

\newglossaryentry{logg}{name={surface gravity~}, text={surface gravity}, symbol={\ensuremath{\textrm{log}\,g}}, description={gravity at the surface of a star}, sort=logg}
\newglossaryentry{feh}{name={metallicity~}, text={metallicity}, symbol=\textrm{[Fe/H]},description={iron abundance relative to the sun}, sort=feh}

\newglossaryentry{texp}{name={time since explosion~}, text={time since explosion}, text={time since explosion}, symbol={\ensuremath{t_{\rm exp}}},description={time since explosion (measured in days)}, sort=texp, first={time since explosion (\ensuremath{t_{\rm exp}})}}

\newglossaryentry{lmc}{name=LMC,description={Large Magellanic Cloud}, first={Large Magellanic Cloud (LMC)}, sort=lmc}
\newglossaryentry{smc}{name=SMC,description={Small Magellanic Cloud}, sort=smc}
\newglossaryentry{z}{name=\ensuremath{z},description={redshift}, sort=z}

\newglossaryentry{sfit}{name=SFIT, text=\textsc{sfit}, description={spectral fitting program for hot stars \citep{2001A&A...376..497J}}, first={\textsc{sfit} \citep{2001A&A...376..497J}}}
\newglossaryentry{iraf}{name=IRAF, text=\textsc{iraf}, description={Image Reduction and Analysis Facility maintained by NOAO}, first={\textsc{iraf}\protect\footnote{IRAF: the Image Reduction and Analysis Facility is distributed by the National Optical Astronomy Observatory, which is operated by the Association of Universities for Research in Astronomy (AURA) under cooperative agreement with the National Science Foundation (NSF).}}}
\newglossaryentry{pyraf}{name=PyRAF, text=\textsc{PyRAF}, description={Python wrap of \gls{iraf} maintained by STSCI}, first=\textsc{PyRAF} \protect\footnote{PyRAF is a product of the Space Telescope Science Institute, which is operated by AURA for NASA.}}
\newglossaryentry{astropy}{name=ASTROPY, text=\textsc{astropy}, description=\textsc{astropy} framework, first = \textsc{astropy} \citep{2013A&A...558A..33A}}
\newglossaryentry{numpy}{name=NUMPY, text=\textsc{numpy}, description=\textsc{numpy} framework, first = \textsc{numpy} \citep{walt2011numpy}}
\newglossaryentry{scipy}{name=SCIPY, text=\textsc{scipy}, description=\textsc{scipy} framework, first = \textsc{scipy} \citep{Jones:2001fk}}
\newglossaryentry{matplotlib}{name=matplotlib, text=\textsc{matplotlib}, description=\textsc{matplotlib} framework, first = \textsc{matplotlib} \citep{hunter2007matplotlib}}
\newglossaryentry{pandas}{name=pandas, text=\textsc{pandas}, description=\textsc{pandas} framework, first = \textsc{pandas} \citep{mckinney2010data}}
\newglossaryentry{ipython}{name=ipython, text=\textsc{ipython}, description=\textsc{ipython} framework, first = \textsc{ipython} \citep{perez2007ipython}}
\newglossaryentry{jupyter}{name=jupyter, text=\textsc{jupyter}, description=\textsc{jupyter} framework, first = \textsc{jupyter} \citep{kluyver2016jupyter,perez2015project,ragan2014jupyter}}
\newglossaryentry{aplpy}{name=aplpy, text=\textsc{aplpy}, description=\textsc{aplpy} framework, first = \textsc{aplpy} \citep{2012ascl.soft08017R}}
\newglossaryentry{nltk}{name=nltk, text=\textsc{nltk}, description=\textsc{nltk} framework, first = Natural Language ToolKit \citep[\textsc{NLTK};][]{bird2009natural}}
\newglossaryentry{scikit-learn}{name=scikit-learn, text=\textsc{scikit-learn}, description=\textsc{scikit-learn} framework, first = \textsc{scikit-learn} \citep[][]{scikit-learn}}
\newglossaryentry{scikit-image}{name=scikit-image, text=\textsc{scikit-image}, description=\textsc{scikit-image} framework, first = \textsc{scikit-image} \citep[][]{scikit-image}}
\newglossaryentry{moog}{name=MOOG,text={\textsc{moog}}, description={spectral synthesis software \citep{1973ApJ...184..839S}}, first={\textsc{Moog} \citep{1973ApJ...184..839S}}}
\newglossaryentry{atlas9}{name=ATLAS9,description={grid of stellar atmospheres \citep{2004astro.ph..5087C}}, first={ATLAS9 \citep{2004astro.ph..5087C}}}
\newglossaryentry{vald}{name=VALD,description={Vienna Atomic Line Database \citep{2000BaltA...9..590K}}, first={Vienna Atomic Line Database \citep[VALD;][]{2000BaltA...9..590K}}}
\newglossaryentry{sextractor}{name=SExtractor, text=\textsc{SExtractor}, description={Source Extractor photometry program \citep{1996A&AS..117..393B}}, first={\textsc{SExtractor} \citep{1996A&AS..117..393B}}}
\newglossaryentry{swarp}{name=SWarp, text=\textsc{SWarp}, description={SWarp \citep{2002ASPC..281..228B}}, first={\textsc{SWarp} \citep{2002ASPC..281..228B}}}
\newglossaryentry{astrometry.net}{name=astrometry.net, text=\textsc{astrometry.net}, description={\textsc{astrometry.net} \citep{2010AJ....139.1782L}} first={\textsc{astrometry.net} \citep{2010AJ....139.1782L}}}

\newglossaryentry{astrodrizzle}{name=AstroDrizzle, text=\textsc{AstroDrizzle}, description={AstroDrizzle \citep{2012drzp.book.....G}}, first={\textsc{AstroDrizzle} \citep{2012drzp.book.....G}}}

\newglossaryentry{idl}{name=IDL,text={\textsc{idl}}, description={Interactive Data Language}}
\newglossaryentry{makee}{name=MAKEE,text=\textsc{makee}, description={MAuna Kea Echelle Extraction by Tom Barlow available}}
\newglossaryentry{minuit}{name=MINUIT,text={\textsc{minuit}}, description={collection of numerical optimization tools \citep{James:1975dr}}}
\newglossaryentry{migrad}{name=MIGRAD,text={\textsc{migrad}}, description={numerical gradient optimization tools - part of \gls{minuit}}}
\newglossaryentry{dolphot}{name=DOLPHOT, text=\textsc{dolphot}, description=photometry package for HST, first=\textsc{dolphot} \citep{2000PASP..112.1383D}}
\newglossaryentry{synphot}{name=synphot, text={\textsc{synphot}}, description={synthetic photometry package from STSCI}, first={\textsc{synphot}\protect\footnote{\textsc{synphot} is a product of the Space Telescope Science Institute, which is operated by AURA for NASA.}}}
\newglossaryentry{chianti}{name=CHIANTI, text=CHIANTI, description= CHIANTI Database 7.1, first =CHIANTI 7.1 \citep{1997A&AS..125..149D,2012ApJ...744...99L}}
\newglossaryentry{synpp}{name=SYNPP, text=SYN++, description= SYN++ software, first =SYN++ \citep{2011PASP..123..237T}}
\newglossaryentry{tardis}{name=TARDIS, text=\textsc{tardis}, description= TARDIS MC code, first = {\textsc{tardis} \citep{2014MNRAS.440..387K}}}

\newglossaryentry{artis}{name=ARTIS, text=\textsc{artis}, description= ARTIS MC code, first = \textsc{artis} \citep{2009MNRAS.398.1809K}}
\newglossaryentry{cmfgen}{name=CMFGEN, text=\textsc{cmfgen}, description=CMFGGEn radiative transfer code, first = \textsc{cmfgen} \citep{1998ApJ...496..407H}}
\newglossaryentry{sedona}{name=SEDONA, text=\textsc{sedona}, description= Sedona MC code, first = \textsc{sedona} \citep{2006ApJ...651..366K}}
\newglossaryentry{phoenix}{name=PHOENIX, text=\textsc{phoenix}, description= PHOENIX radiative transfer code, first = \textsc{phoenix} \citep{1999jcoam.109...41h,1998ApJ...495..370B,1997ApJ...483..390H,1996ApJ...462..386H,1997ApJ...490..803H}}
\newglossaryentry{mlmc}{name=MLMC, text=ML93, description= Mazzali Lucy Monte Carlo, first ={Mazzali \& Lucy (1993, ML93) code}}
\newglossaryentry{starkit}{name=STARKIT, text=\textsc{starkit}, description= TARDIS MC code, first = {\textsc{starkit} \citep{wolfgang_kerzendorf_2015_28016}}}

\newglossaryentry{pyne}{name=PYNE, text=\textsc{pyne}, description= PYNE code, first = {\textsc{pyne} \citep{Scopatz2012a}}}
\newglossaryentry{multinest}{name=MULTINEST, text=\textsc{MultiNest}, description=MultiNest, first={\textsc{MultiNest} \citep{2009MNRAS.398.1601F}}}
\newglossaryentry{wsynphot}{name=WSYNPHOT, text=\textsc{wsynphot}, description=Wsynphot, first={\textsc{wsynphot}\protect\footnote{\protect\url{https://github.com/wkerzendorf/wsynphot}}}}
\newglossaryentry{specutils}{name=SPECUTILS, text=\textsc{specutils}, description=specutils, first={\textsc{specutils} \protect\footnote{\protect\url{https://github.com/astropy/specutils}}}}
\newglossaryentry{ads}{name=ADS ,description=ADS, first={NASA Astrophysics Data System (ADS) \citep{2000A&AS..143...41K}}}

\newglossaryentry{2mass}{name=2MASS,description={Two Micron All Sky Survey \citep{2006AJ....131.1163S}}, first={Two Micron All Sky Survey \citep{2006AJ....131.1163S}}}
\newglossaryentry{wiserep}{name=\textsc{WISeREP}, description={Weizmann Interactive Supernova data REPository \citep{2006AJ....131.1163S}}, first={\textsc{WISeREP} \citep{2012PASP..124..668Y}}}
\newglossaryentry{nomad}{name=NOMAD,first={Naval Observatory Merged Astrometric Dataset \citep[NOMAD; ][]{2005yCat.1297....0Z}}, description={Naval Observatory Merged Astrometric Dataset}}
\newglossaryentry{sdss}{name=SDSS, description={Sloan Digital Sky Survey}}
\newglossaryentry{dss}{name=DSS, description={Digitized Sky Survey}}

\newglossaryentry{eso}{name=ESO, description={European Southern Observatory}, first={European Southern Observatory (ESO)}}
\newglossaryentry{eso.opc}{name=OPC, description={Observing Programmes Comittee}, first={Observing Programmes Comittee (OPC)}}
\newglossaryentry{iau}{name=IAU,description={International Astronomical Union}, first={IAU}}
\newglossaryentry{ctio}{name= CTIO, description={Cerro Tololo Inter-American Observatory}, first={Cerro Tololo Inter-American Observatory (CTIO)}}

\newglossaryentry{nsf}{name=NSF, description={National Science Foundation}, first={National Science Foundation (NSF)}}

\newglossaryentry{wifes}{name=WIFES, text=\textsc{WiFeS}, first={\textsc{WiFeS} \citep{2007Ap&SS.310..255D}},  description={Wide Field Spectrograph - \gls{ifu} mounted on the 2.3\,m telescope at Siding Spring Observatory}}

\newglossaryentry{scp}{name=SCP, description={Supernova Cosmology Project, led by Saul Perlmutter}, first={Supernova Cosmology Project (SCP)}}
\newglossaryentry{hzsns}{name=HZSNS, description={High Z Supernova Search, led by Brian Schmidt}, first={High Z Supernova Search (HZSNS)}}

\newglossaryentry{vlt}{name=VLT,description={Very Large Telescope located on Cerro Paranal (Chile)}, first={Very Large Telescope (VLT)}}
\newglossaryentry{flames}{name=FLAMES,description={Multi-object, intermediate and high resolution spectrograph mounted on the  \gls{vlt}}}

\newglossaryentry{hires}{name=HIRES, description={High Resolution Echelle Spectrometer mounted on the Keck Telescope}, first={High Resolution Echelle Spectrometer \citep[HIRES;][]{1994SPIE.2198..362V}}}

\newglossaryentry{lris}{name=LRIS,description={Low Resolution Imaging Spectrometer mounted on the Keck Telescope}, first={Low-Resolution Imaging Spectrometer \citep[LRIS;][]{Oke95}}}

\newglossaryentry{decam}{name=DECam, description={DECam is a high-performance, wide-field CCD imager mounted at the prime focus of the Blanco 4-m telescope at \gls{ctio}.}, first={Dark Energy Camera \citep[DECam; ][]{2012PhPro..37.1332D,2015AJ....150..150F}}}

\newglossaryentry{essence}{name=ESSENCE,description={The `Equation of State: SupErNovae trace Cosmic Expansion' project \citep[ESSENCE;][]{2002AAS...201.7809G}}, first={`The Equation of State: SupErNovae trace Cosmic Expansion' \citep[ESSENCE;][]{2002AAS...201.7809G}}}
\newglossaryentry{ifu}{name=IFU,description={Optical instrument combining spectrographic and imaging capabilities, used to obtain spatially resolved spectra}, first={Integral Field Unit (IFU)}, firstplural={Integral Field Units (IFUs)}}

\newglossaryentry{besancon}{name=Besan\c{c}on Model, description={Model of stellar population synthesis of the Galaxy, including kinematics.}}

\newglossaryentry{int}{name=INT,description={Isaac Newton 2.5\,m Telescope}, first={Isaac Newton 2.5\,m Telescope (INT)}}

\newglossaryentry{chandra}{name=Chandra,description={Chandra \xray\ Observatory (space-based)}}
\newglossaryentry{hst}{name=HST,description={Hubble Space Telescope}}
\newglossaryentry{hst.wfpc2}{name=WFPC2,description={Wide-Field Planetary Camera 2 mounted on the \gls{hst}}, first={Wide-Field Planetary Camera 2 (WFPC2)}}
\newglossaryentry{hst.acs}{name=ACS,description={Advanced Camera for Surveys mounted on the \gls{hst}}, first={Advanced Camera for Surveys (ACS)}}
\newglossaryentry{hst.wfc3}{name=WFC3,description={Wide-Field Camera 3 mounted on the \gls{hst}}, first={Wide-Field Camera 3 (WFC3)}}
\newglossaryentry{hst.cte}{name=CTE, description={charge transfer efficiency (CTE)}, first={charge transfer efficiency \citep[CTE; see ][for a description]{2009acs..rept....1C}}}

\newglossaryentry{snls}{name=SNLS,description={Supernova Legacy Survey \citep{2003AAS...203.8209P}}, first={Supernova Legacy Survey \citep[SNLS;][]{2003AAS...203.8209P}}}
\newglossaryentry{dass}{name=DASS, description={Digitized Astronomy Supernova Survey \citep{1975PASP...87..565C}}, first={Digitized Astronomy Supernova Survey \citep[DASS;][]{1975PASP...87..565C}}}
\newglossaryentry{bait}{name=BAIT, description={Berkley Automatic Imaging Telescope \citep{1993PASP..105.1164R}}, first={Berkley Automatic Imaging Telescope \citep[BAIT;][]{1993PASP..105.1164R}}}
\newglossaryentry{kait}{name=KAIT, description={Katzman Automatic Imaging Telescope \citep{2001ASPC..246..121F}}, first={Katzman Automatic Imaging Telescope \citep[KAIT;][]{2001ASPC..246..121F}}}
\newglossaryentry{loss}{name=LOSS, description={Lick Observatory Supernova Search  \citep{2000AIPC..522..103L}}, first={Lick Observatory Supernova Search \citep[LOSS;][]{2000AIPC..522..103L}}}
\newglossaryentry{ctss}{name=CTSS,description={Cal\'{a}n/Tololo Supernova Survey \citep{1993AJ....106.2392H}}, first={Cal\'{a}n/Tololo supernova survey \citep[CTSS;][]{1993AJ....106.2392H}}}

\newglossaryentry{ptf}{name=PTF, description={Palomar Transient Factory \citep{2009PASP..121.1334R}}, first={Palomar Transient Factory \citep[PTF;][]{2009PASP..121.1334R}}}
\newglossaryentry{batse}{name=BATSE, description={Burst and Transient Source Experiment mounted on the Compton Gamma Ray Observatory}, first={Burst and Transient Source Experiment (BATSE)}}
\newglossaryentry{bepposax}{name=BeppoSAX, description={\xray\ satellite named in honor of Giuseppe "Beppo" Occhialini}}
\newglossaryentry{rosat}{name=ROSAT, description={short for R\"{o}ntgensatellit}, first={ROSAT}}
\newglossaryentry{hete2}{name=HETE2, description={High Energy Transient Explorer}, first={High Energy Transient Explorer (HETE)}}
\newglossaryentry{ska}{name=SKA, description={Square Kilometre Array}, first={Square Kilometre Array (SKA)}}
\newglossaryentry{swift}{name=Swift, description={Swift Gamma-Ray Burst Mission}}

\newglossaryentry{gnirs}{name=GNIRS, description={Gemini Near InfraRed Spectrograph mounted on the Gemini North Telescope}}
\newglossaryentry{gmosn}{name=GMOS, description={Gemini Multi Object Spectrograph mounted on the
 Gemini North Telescope}, first={GMOS \citep[Gemini Multi Object Spectrograph;][]{2004PASP..116..425H}}}

\newglossaryentry{vla}{name=VLA, description={Very Large Array radio telescope located in North America}, first={Very Large Array (VLA)}}
\newglossaryentry{evla}{name=EVLA, description={Extended Very Large Array radio telescope located in North America}, first={Extended Very Large Array (EVLA)}}
\newglossaryentry{skymapper}{name=SkyMapper, description={SkyMapper telescope \citep{2007PASA...24....1K}}, first={SkyMapper \citep{2007PASA...24....1K}}}
\newglossaryentry{panstarrs}{name=PanSTARRS, description={Panoramic Survey Telescope \& Rapid Response System \citep{2004SPIE.5489...11K}}, first={Panoramic Survey Telescope \& Rapid Response System \citep[PanSTARRS;][]{2004SPIE.5489...11K}}}
\newglossaryentry{ps1dr1}{name=PS1~DR1, description={Panoramic Survey Telescope \& Rapid Response System \citep{2004SPIE.5489...11K} }, first={Panoramic Survey Telescope \& Rapid Response System \citep[PanSTARRS;][]{2004SPIE.5489...11K} DR1}}

\newglossaryentry{lsst}{name=LSST, description={Large Synoptic Survey Telescope}, first={Large Synoptic Survey Telescope \citep[LSST;][]{2006AAS...209.8604P}}}
\newglossaryentry{ppmxl}{name=PPMXL, description={PPMXL Catalog of Positions and Proper Motions on the ICRS \citep{2010AJ....139.2440R}}}
\newglossaryentry{gaia}{name=GAIA, description={Global Astrometric Interferometer for Astrophysics \citep{2001A&A...369..339P}}, first={Global Astrometric Interferometer for Astrophysics \citep[GAIA;][]{2001A&A...369..339P}}}
\newglossaryentry{ligo}{name=LIGO, description={Laser Interferometer Gravitational Wave Observatory}, first={Laser Interferometer Gravitational Wave Observatory \citep[LIGO;][]{1992Sci...256..325A}}}
\newglossaryentry{aligo}{name=Advanced LIGO, description={Advanced LIGO}, sort=ligo2}
\newglossaryentry{lisa}{name=LISA, description={Laser Interferometer Space Antenna \citep{1994ESAJ...18..219J}}, first={Laser Interferometer Space Antenna \citep[LISA;][]{1994ESAJ...18..219J}}}

\newglossaryentry{irc}{name=IRC, text={IRC}, description={infrared catastrophe}, first={infrared catastrophe \citep[IRC;][]{1980PhDT.........1A}}}

\newglossaryentry{sn}{name=Supernova, text={SN}, plural={SNe}, description={exploding star}, nonumberlist=true, first={supernova (SN)}, firstplural={supernovae (SNe)}}
\newglossaryentry{snia}{name=Type~Ia (SN~Ia), text={SN~Ia}, description={Thermonuclear explosion of a white dwarf - spectra show no hydrogen but a strong silicon line},first={Type~Ia supernova (SN~Ia)}, firstplural={Type Ia supernovae (SNe~Ia)}, plural={SNe~Ia}, parent=sn, nonumberlist=true}
\newcommand{\sneia}{\glspl*{snia}\xspace}

\newglossaryentry{branchnormal}{name={branch-normal}, text=\textit{Branch-normal}, description={Large homogeneous class of Type Ia Supernovae, defined in \citet{1993AJ....106.2383B}}, first={\textit{Branch-normal} SNe Ia \citep{1993AJ....106.2383B}}, parent=snia}
\newglossaryentry{91t}{name={91T-like}, description={Luminous class of Type Ia supernovae similar to \sn{1991}{T} \citep{1992AJ....103.1632P}} , first={91T-like}, parent=snia}
\newglossaryentry{91bg}{name={91bg-like}, description={Faint class of Type Ia supernovae similar to \sn{1991}{bg} \citep{1992AJ....104.1543F}}, first={91bg-like}, parent=snia}
\newglossaryentry{02cx}{name={02cx-like}, description={Peculiar class of Type Ia supernovae similar to \sn{2002}{cx} \citep{2003PASP..115..453L}}, first={02cx-like \sneia\ \citep{2003PASP..115..453L}}, parent=snia}

\newglossaryentry{snibc}{name=Type~Ib/c, text={SN~Ib/c}, description={Collapse of the core of a massive star -  spectrum shows no hydrogen and no silicon line},first={Type~Ib/c supernova (SN~Ib/c)}, firstplural={Type~Ib/c supernovae (SNe~Ib/c)}, plural={SNe~Ib/c}, parent=sn}

\newglossaryentry{snib}{name=Type~Ib, text={SN~Ib}, description={Spectrum shows no hydrogen and no silicon, but helium line},first={Type Ib supernova (SN~Ib)}, firstplural={Type~Ib supernovae (SNe~Ib)}, plural={SNe~Ib}, parent=snibc}

\newglossaryentry{snic}{name=Type~Ic, text={SN~Ic}, description={Spectrum shows no hydrogen, no silicon and no helium line},first={Type~Ic supernova (SN~Ic)}, firstplural={Type~Ic supernovae (SNe~Ic)}, plural={SNe~Ic}, parent=snibc}

\newglossaryentry{snii}{name=Type~II, text={SN~II}, description={Collapse of the core of a massive star - spectrum shows strong hydrogen line},first={Type~II supernova (SN~II)}, firstplural={Type~II supernovae (SNe~II)}, plural={SNe~II}, parent=sn}

\newglossaryentry{sniib}{name=Type~IIb, text={SN~IIb}, description={Spectrum shows hydrogen and helium lines},first={Type~IIb supernova (SN~IIb)}, firstplural={Type~IIb supernovae (SNe~IIb)}, plural={SNe~IIb}, parent=snii}

\newglossaryentry{sniip}{name=Type~II~Plateau (Type IIP), text={SN~IIP}, description={Lightcurve shows plateau},first={Type~IIP supernova (SN~IIP)}, firstplural={Type~II Plateau supernovae \citep[SNe~IIP;][]{1979A&A....72..287B}}, plural={SNe~IIP}, parent=snii}

\newglossaryentry{sniil}{name=SN~II~Linear, text={SN~IIL}, description={Lightcurve shows no plateau, but linear decline},first={Type~IIL supernova (SN~IIL)}, firstplural={Type~II~Linear supernovae \citep[SNe~IIL;][]{1990MNRAS.244..269S}}, plural={SNe~IIL}, parent=snii}

\newglossaryentry{sniin}{name=Type II narrow-lined (Type IIn), description={Spectrum shows narrow lines},first={Type~II~narrow-lined supernova (SN IIn)}, firstplural={Type~IIn supernovae (SNe~IIn)}, plural={SNe~IIn}, parent=snii}

\newglossaryentry{snr}{name=Remnant (SNR), text=SNR, description={Remnant left visible post-explosion}, first={supernova remnant (SNR)}, firstplural={supernova remnants (SNRs)}, parent=sn}

\newglossaryentry{dtd}{name=DTD,description={delay time distribution - expected supernova rate over time after a brief outburst of starformation},first={delay time distribution (DTD)}, firstplural={delay time distributions (DTDs)}, plural=DTDs}

\newglossaryentry{hvg}{name=HVG,description={high velocity gradient - Type Ia supernovae with a fast evolution of photospheric velocity},first={high velocity group (HVG)}, firstplural={high velocity groups (HVGs)}, plural=HVGs, parent=snia}

\newglossaryentry{lvg}{name=LVG,description={low velocity gradient - Type Ia supernovae with a slow evolution of photospheric velocity},first={low velocity group (LVG)}, firstplural={low velocity groups (LVGs)}, plural=LVGs, parent=snia}

\newglossaryentry{wd}{name=white dwarf (WD), text=WD, description={White Dwarf - extremely dense stellar remnant}, first={white dwarf (WD)}}
\newglossaryentry{onemgwd}{name= Oxygen/Neon (ONe), text={ONe-WD},description={Oxygen/Neon White Dwarf}, first={oxygen/neon White Dwarf (ONe-WD)}, parent=wd}
\newglossaryentry{cowd}{name=carbon/oxygen (CO), text={CO-WD}, description={carbon/oxygen white dwarf}, first={carbon/oxygen white dwarf (CO-WD)}, firstplural = {carbon/oxygen white dwarfs (CO-WDs)}, parent=wd}

\newglossaryentry{sds}{name=SD-Scenario,description={single-degenerate scenario (single white dwarf accreting from non-degenerate companion)}, first={single-degenerate scenario (SD-scenario)}}

\newglossaryentry{dds}{name=DD-Scenario, description={double degenerate scenario (merging of two white dwarfs)}, first={double-degenerate scenario (DD-scenario)}}

\newglossaryentry{sss}{name=SSS, text={supersoft \xray\ source}, description={supersoft \xray\ source - believed to be emitted by nuclear fusion on a white dwarf's surface}}

\newglossaryentry{amcvn}{name=AM CVn, description={AM Canum Venaticorum star \citep[white dwarf accreting hydrogen poor matter from a companion star; see ][]{2005ASPC..330...27N}}}

\newglossaryentry{rlof}{name=RLOF, description={Roche Lobe Overflow (see \citet{1971ARA&A...9..183P} for a more detailed description)}, first={Roche-lobe overflow (RLOF)}}

\newglossaryentry{mchan}{name={Chandrasekhar mass~}, text={Chandrasekhar~mass}, symbol={\ensuremath{M_\textrm{Chan}}}, plural={Chandrasekhar~masses}, description={Mass when the core of a star collapses due to insufficient degeneracy pressure - for a white dwarf $\approx1.38\,M_\odot$ see \citet{1931ApJ....74...81C}}, first={Chandrasekhar~mass \citep[$M_\textrm{Chan}=1.38\,M_\odot$;][]{1931ApJ....74...81C}}, sort=mchan}

\newglossaryentry{w7}{name={W7 model},description={W7 model \citep{1984ApJ...286..644N}},first = {W7 model \citep{1984ApJ...286..644N}}}

\newglossaryentry{stats.pdf}{name=PDF, description={Probability Density Function}, first={Probability Density Function}}

\newglossaryentry{dpr}{name=DPR, description={Distributed Peer Review}, first={Distributed Peer Review (DPR)}}

%% file: author_natastro.tex
\author{%
Wolfgang~E.~Kerzendorf$^{1, 2, 3}$, 
Ferdinando Patat$^{4}$,
Dominic Bordelon$^{4}$,
Glenn van~de~Ven$^{4,5}$,
Tyler A. Pritchard$^{3}$
}

%% file: affil_natastro.tex
\begin{affiliations}
\item Department of Physics and Astronomy, Michigan State University, East Lansing, MI 48824, USA
\item Department of Computational Mathematics, Science, and Engineering, Michigan State University, East Lansing, MI 48824, USA
\item Center for Cosmology and Particle Physics, New York University, 726 Broadway, New York, NY 10003, USA
\item European Southern Observatory, K. Schwarzschildstr. 2, 85748 Garching b. M\"unchen, Germany
\item Department of Astrophysics, University of Vienna, T\"urkenschanzstrasse 17, 1180 Vienna, Austria
\end{affiliations}

%% file: deepthought_dpr1_main.tex

\section{\label{sec:intro}Introduction}

All large, ground- and space-based astronomical facilities serving wide communities, like the European Southern Observatory (ESO), the Atacama large Millimeter Array, The Hubble Space Telescope, and the Gemini Observatory, face a similar problem. In many cases the number of applications they receive at each call exceeds 1000, posing a serious challenge to run an effective selection process through the classic peer-review paradigm, which assigns proposals to pre-allocated panels with fixed compositions. Although, in principle, one could increase the size of the time allocation committee (TAC), this creates logistic and financial problems which practically limit its maximum size, making this solution not viable.

Since the referees only have a limited amount of time to perform their task, the heavy load (which at ESO typically exceed 70 proposals per referee; priv. comm. ESO OPC office) has severe consequences on the quality of the review and the feedback that is provided to the applicants. This contributes to increasing levels of frustration in the community and to the loss of credibility in the whole selection process. In addition, although difficult to quantify, this will have consequences for the scientific output of the facilities.
Different measures were considered by the various facilities to alleviate the load on the reviewers. This includes quite drastic solutions, like the one deployed by \gls{nsf} to limit the number of applications \citep{Mervis1328}. 

In this context, one of the most innovative propositions was put forward by 
The concept is simple: by submitting a proposal the principal investigator accepts to review $n$ proposals submitted by peers, and to have their proposal reviewed by $n$ peers. Also, by submitting $m$ proposals, they accept to review $n \times m$ proposals, hence virtually limiting the number of submissions. We will indicate this concept as \gls{dpr}.

The Gemini Observatory deployed the \gls{dpr} for its Fast Turnaround channel \citep{2019AAS...23345503A}, which is capped to 10\% of the total time. The \gls{nsf} also explored this possibility with a pilot study in 2013, in which each PI was asked to review 7 proposals submitted by peers \citep{Parinaz2013, Mervis2014}. The NSF pilot was based on 131 applications submitted by volunteers within the Civil, Mechanical and Manufacturing Innovation Division. The \gls{nsf} did not publish a report following the study. 
A similar pilot experiment was carried out in 2016 by the National Institute of Food and Agriculture (see {\url{https://nifa.usda.gov/resource/distributed-peer-review-pilot-foundational-program}, consulted on April 9, 2019.), but also in this case the results were not published.



We report on an experiment that employed the \gls{dpr} at \gls{eso} during Period 103 (call for proposals issued on 30 August, 2018) in parallel with the regular \gls{eso.opc}. We mirrored the deployment of the \gls{dpr} implementation at Gemini as an example, and enhanced the process using \gls{nlp} and machine learning for referee selection \citep[a different method of using \gls{nlp} for proposal reviews can be found in][]{2017AJ....153..181S}. This experiment also added a feedback for individual reviews. 

The experiment was designed to test if there is a measurable difference between the \gls{dpr} and \gls{eso.opc}, if the algorithm for referee selection performed well, and if what referee attributes influenced the quality of the referee report (as judged by the feedback to a review). 

In Section~\ref{sec:method_overview}, we describe the general setup of the experiment. Section~\ref{sec:analysis.main} is devoted to the statistical analysis and divided into Section~\ref{sec:analysis.main.comparison_opc_dpr} for the comparison between the DPR and OPC outcomes, Section~\ref{sec:analysis.knowledge_vectors} for the description of the literature matching, and Section~\ref{sec:analysis.helpfulness} for the evaluation of the helpfulness of reviews by the proposers. In Section~\ref{sec:concl} we summarize and conclude this work.

The supplementary information (SI) gives a more detailed description of the time allocation process at ESO (SI Section~1), a discussion of the demographics of the experiment (SI Section~2), an extended analysis (SI Section~3), and several datasets related to the experiment (see SI Appendices). We will refer the reader to the supplementary material where appropriate.

\section{Experiment Overview}
\label{sec:method_overview}

We followed a general outline of \gls{dpr} as described in \citet{2009A&G....50d..16M}, but differences were introduced in several key areas. Specifically, we test two different referee selection methods: 1) the first selection method emulates the way that reviews are currently assigned to members of the \gls{eso.opc} to make a comparison to the OPC evaluation of the same proposals. 2) an automated machine-learning method to assign referees to proposals \citep[based on the \dt-knowledge discovery method; see][]{2019JApA...40...23K}. We believe that the advantages of this method are that it scales easily with the number of proposals and that, due to the automated construction, it might circumvent biases in self-efficacy \citep[e.g., based on gender - see][]{ehrlinger2003chronic}. Finally, we also asked the participants to assess the reviews of their proposal in order to understand what influences a constructive review and potentially reward helpful referees in the future. 

We have tested a \gls{dpr} scheme on a voluntary basis for the ESO period P103. The outcome of this experiment had no influence on the telescope allocation. The DPR program for ESO P103 recruited 172 volunteers with each submitting one participating proposal (this is 23\% of the distinct PIs in P103). These proposals were evaluated in the DPR process as well as with the general \gls{eso.opc} methods. 

For our experiment, the group of proposers and referees were the same. We selected eight referees for each proposal using several rules (see the Methods Section~\ref{sec:reviewer_selection} material for an overview of the two selection processes used). 

The reviewers accessed the assigned proposals through a web application and were given two weeks to assess them. A detailed description of the review options are described in the methods section (see Section~\ref{sec:method_overview}).

The proposal quartile and the eight unmodified comments were displayed to the proposer. We finally asked the proposer to evaluate the helpfulness of the comments (with details in see Section~\ref{sec:method_overview}).

\begin{figure*}
   \includegraphics[width=\textwidth]{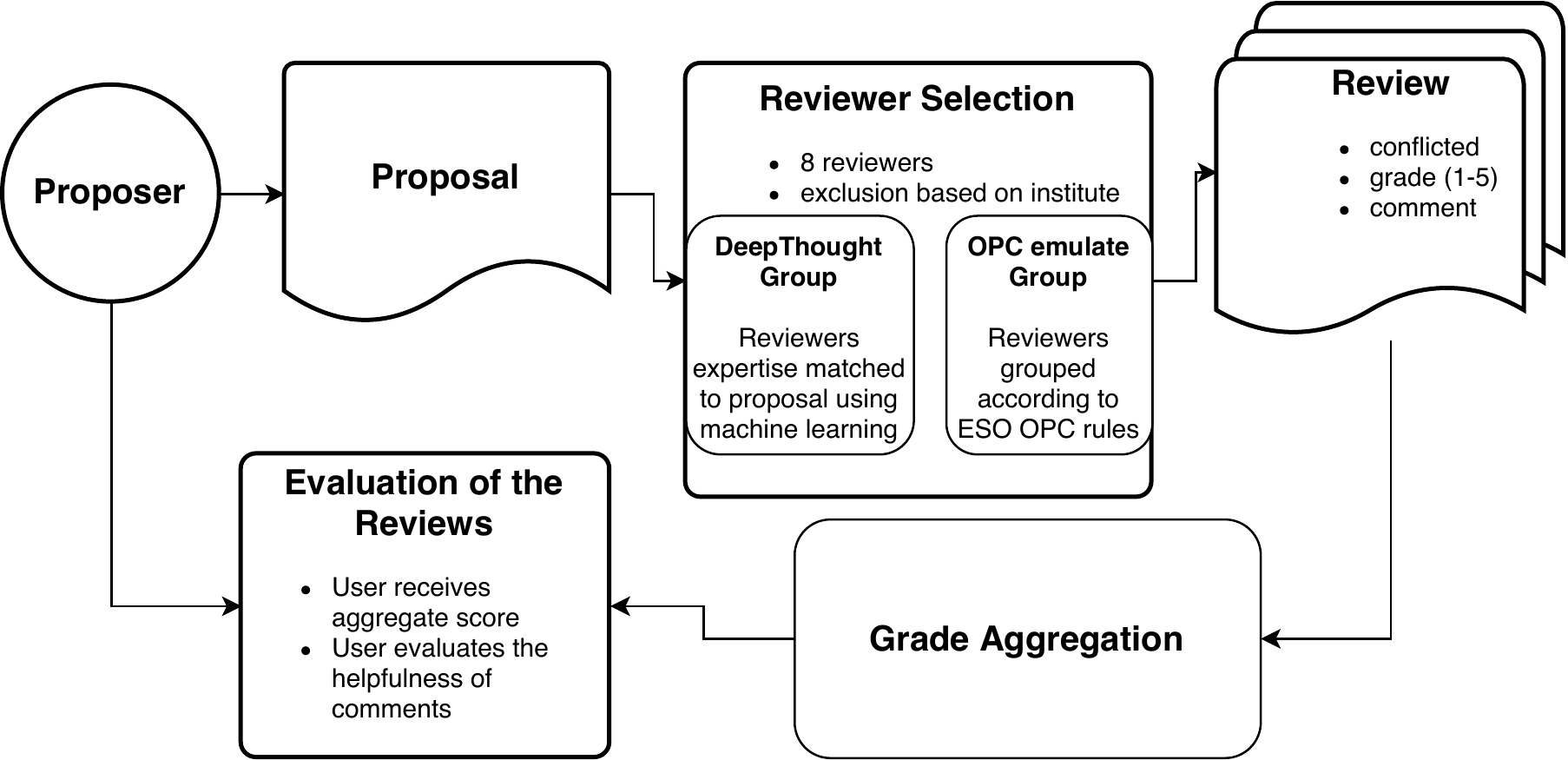}
   \caption{\textbf{Overview of the enhanced distributed peer-review process.} The graph shows the flow of the proposal starting with the proposer in the upper left-hand corned. The proposal then goes into the reviewer selection (with the two groups marked), via the review and finally to the grade-aggregation. The loop closes with the proposer giving feedback on the usefulness of the review. }
   \label{fig:general_overview}
\end{figure*}

Figure~\ref{fig:general_overview} summarizes the process as a flow-chart and the detailed description of the process can be found in the Method Section~\ref{sec:method_description}.

\section{Analysis \& Results}
\label{sec:analysis.main}
We received complete reviews for 167 out of 172 reviewers (97.1\%) at the deadline (with these 5 proposals excluded from the further analysis). The individual grades from each review were converted into a global rank for each proposal (with details found in the supplementary material) and translated to quartiles ($A, B, C, D$). 

There are a myriad of statistics available that can be interesting to apply to the dataset (some of them are explored in the supplementary material). We therefore encourage the community to make use the anonymized dataset to run independent analyses(available at \dataurl). Our main focus in this study is to address three questions: 1) How different is the DPR from a more traditional \gls{eso.opc} review? 2) How well can our algorithm predict expertise for a proposal? 3) What reviewer properties influence the helpfulness of their referee report?

\subsection{Comparison of the \glsdesc{dpr} to the
\glsdesc{eso.opc}}
\label{sec:analysis.main.comparison_opc_dpr}
The proposals used in the DPR experiment were also reviewed through the regular ESO OPC channel. This allows a comparison between the outcomes of the two processes. However, these differ by construction in many aspects, and so a one-to-one comparison is not possible. This is because in the DPR experiment:

\begin{enumerate}
\item there is no \textit{a priori} scientific-seniority selection;
\item the proposals are typically reviewed by $N_r>6$ referees (while $N_r=3$ in the pre-meeting OPC process);
\item the number of proposals per referee is much smaller;
\item the set of proposals common to different reviewers is much smaller;
\item there is no triage; an early removal of some proposals before the OPC meeting
\item there is no face-to-face discussion.
\end{enumerate}


A robust way of quantifying the consistency between two different panels reviewing the same set of proposals is that of the quartile agreement fraction introduced in \citet[hereafter P18; Section 9.2]{2018PASP..130h4501P}. Following that concept, one can compute what we will indicate as the panel-panel (p-p) \gls{qam}. The generic \gls{qam} element $M_{i,j}$ is the fraction of proposals ranked by the first panel in the $i$-th quartile of the grade distribution, that were ranked in the $j$-th quartile by the second panel.
If we indicate with $A_i$ and $B_j$ the events "a proposal is ranked by panel A in quartile $i$" and "a proposal is ranked by panel B in quartile $j$", the \gls{qam} elements represent the conditional probability:

\begin{displaymath}
M_{i,j}=P(A_i | B_j) = \frac{P(A_i \cap B_j)}{P(A_j)}.
\end{displaymath}

For a completely aleatory process $P(A_i \cap B_j)=P(A_i) P(B_j)$, and therefore all the terms of the \gls{qam} would be equal to 0.25, while for a full correlation, all terms would be null, with the exception of the diagonal terms, which would be equal to 1. We note that the matrix elements are not independent from each other as, by definition:

\begin{displaymath}
\sum_i M_{i,j} \equiv \sum_j M_{i,j} \equiv 1.
\end{displaymath}


For the purposes of the main part of this paper, we will compare the internal agreement of the DPR panels with that of the OPC (pre-meeting). Section~\ref{si-sec:analysis.opcomp.agfrac} in the supplementary information gives further comparison between the OPC and DPR statistics.

For doing this, we will bootstrap the DPR data extracting a number of sub-sets of three randomly chosen referees.

This choice for the DPR set is particularly interesting as it is directly comparable to the results presented in P18. The procedure is as follows: We first make a selection of the proposals having at least 6 reviews (164). For each of them we randomly select two distinct (i.e. non-intersecting) sub-sets of $N_r$=3 grades each, from which two average grades are derived.
These are used to compute the agreement fractions between the two sub-panels. The process is repeated a large number of times
and the average (p-p) \gls{qam} is finally obtained.

\begin{figure*}
    \includegraphics[width=0.5\textwidth]{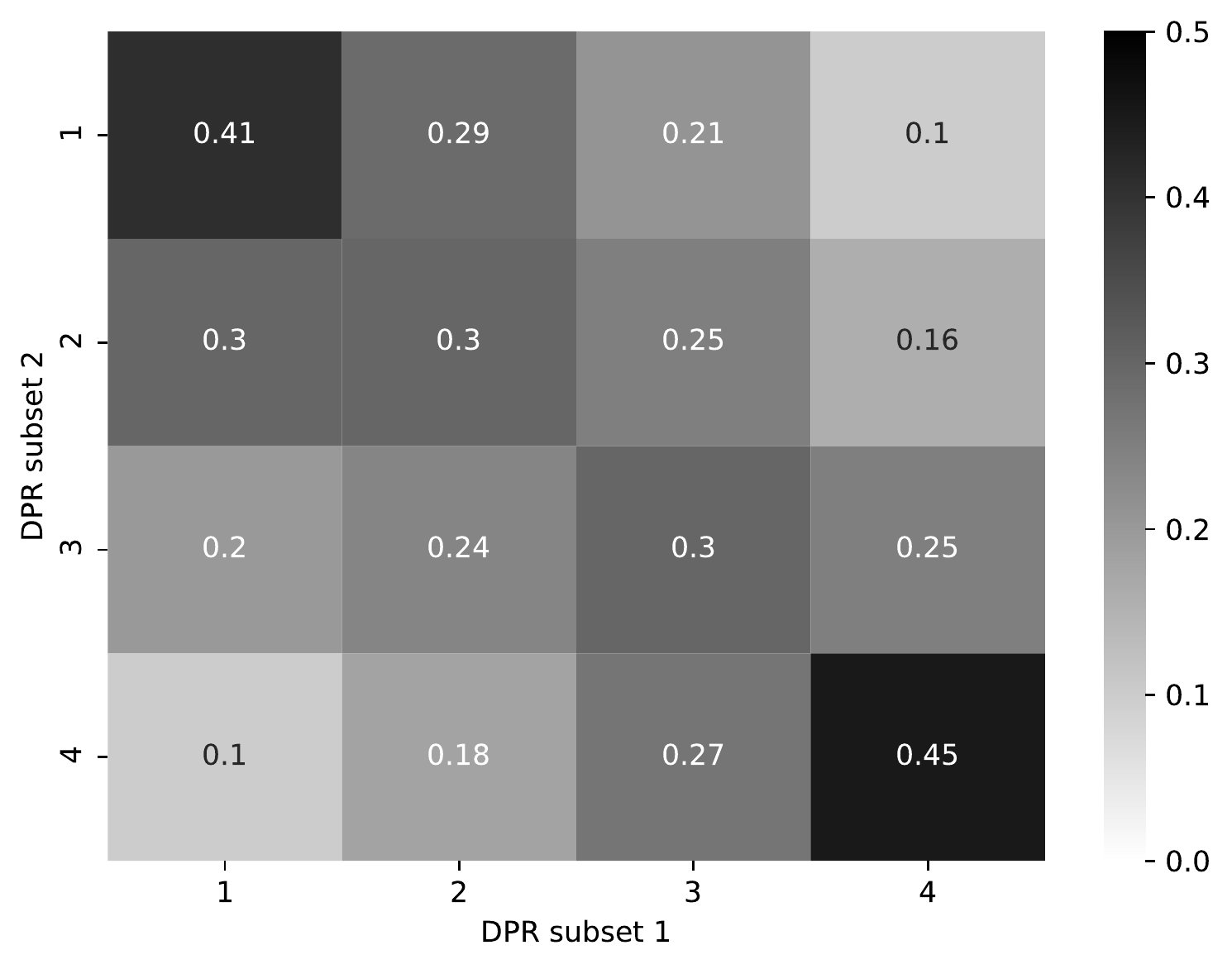}
    \includegraphics[width=0.5\textwidth]{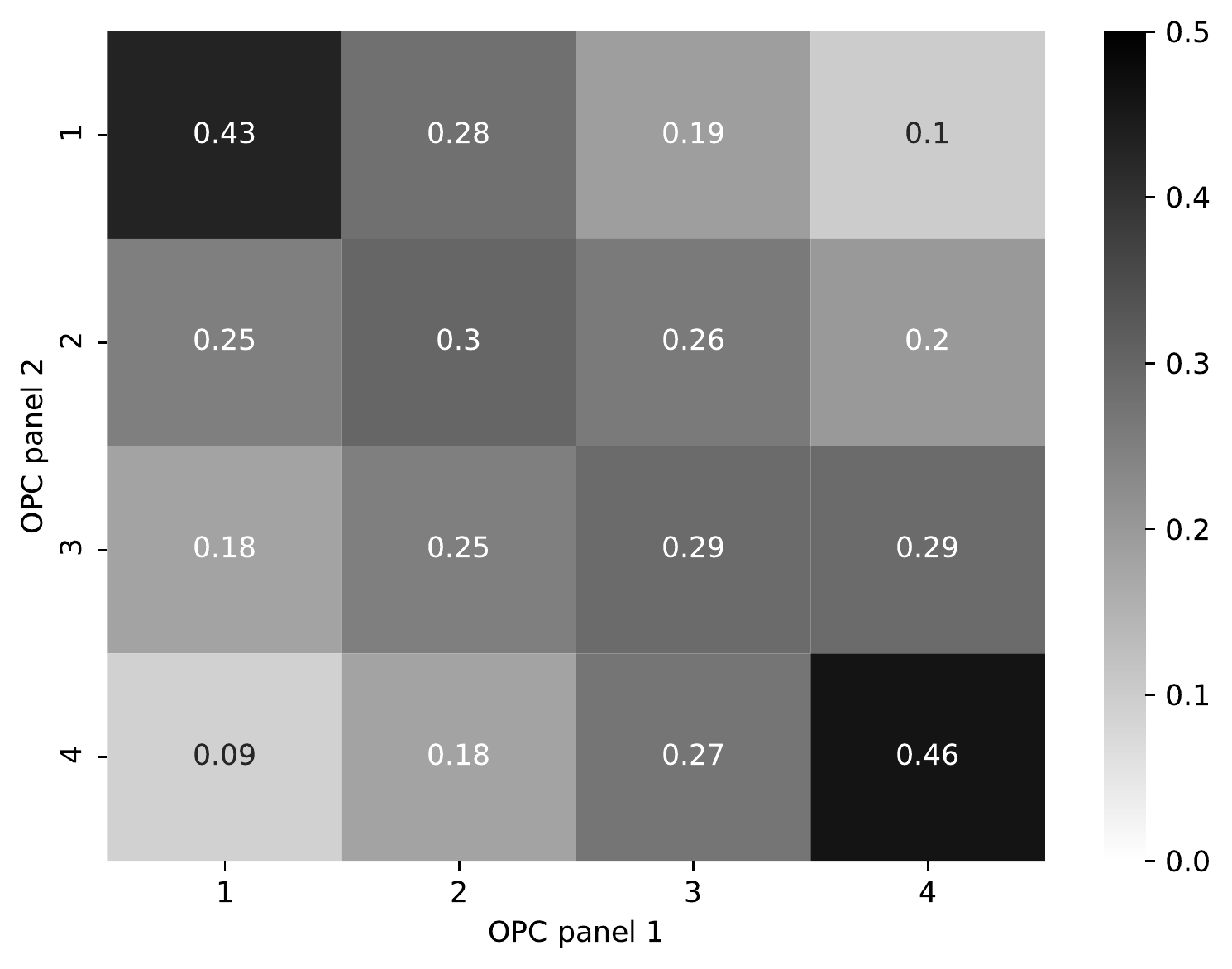}
    \caption{\textbf{Comparison of reviewer agreement between both DPR and the classic process} Comparison between the panel-panel for the OPC subsets (left, from P18 Table 3) and the DPR subsets (right). Each panel shows the probability the second panel (or DPR subset) is grading a specific proposal given the response by the first panel. The probability is highest on the diagonal which suggests there is some correlation between different reviewers grading -- albeit a relatively weak one.}
    \label{fig:qam_comparison}
\end{figure*}

In Figure~\ref{fig:qam_comparison}, we compare the \gls{qam} of our subsets with that of the OPC pre-meeting panels. The latter was derived for the OPC process for $N_r$=3 sub-panels (P18, Table~3 therein).
In both cases the first-quartile agreement is about 40\% ($k$=0.21), while for the second and third quartile this is $\sim$30\%. The top-bottom quartile agreement is 10\% ($k=-$0.60). The conclusion is that, in terms of self-consistency, the DPR review behaves in the same way as the pre-meeting OPC process. The two review processes are characterized the same level of subjectivity.


\subsection{Domain knowledge inference}
\label{sec:analysis.knowledge_vectors}
\begin{figure}[h!]
   \includegraphics[width=\columnwidth]{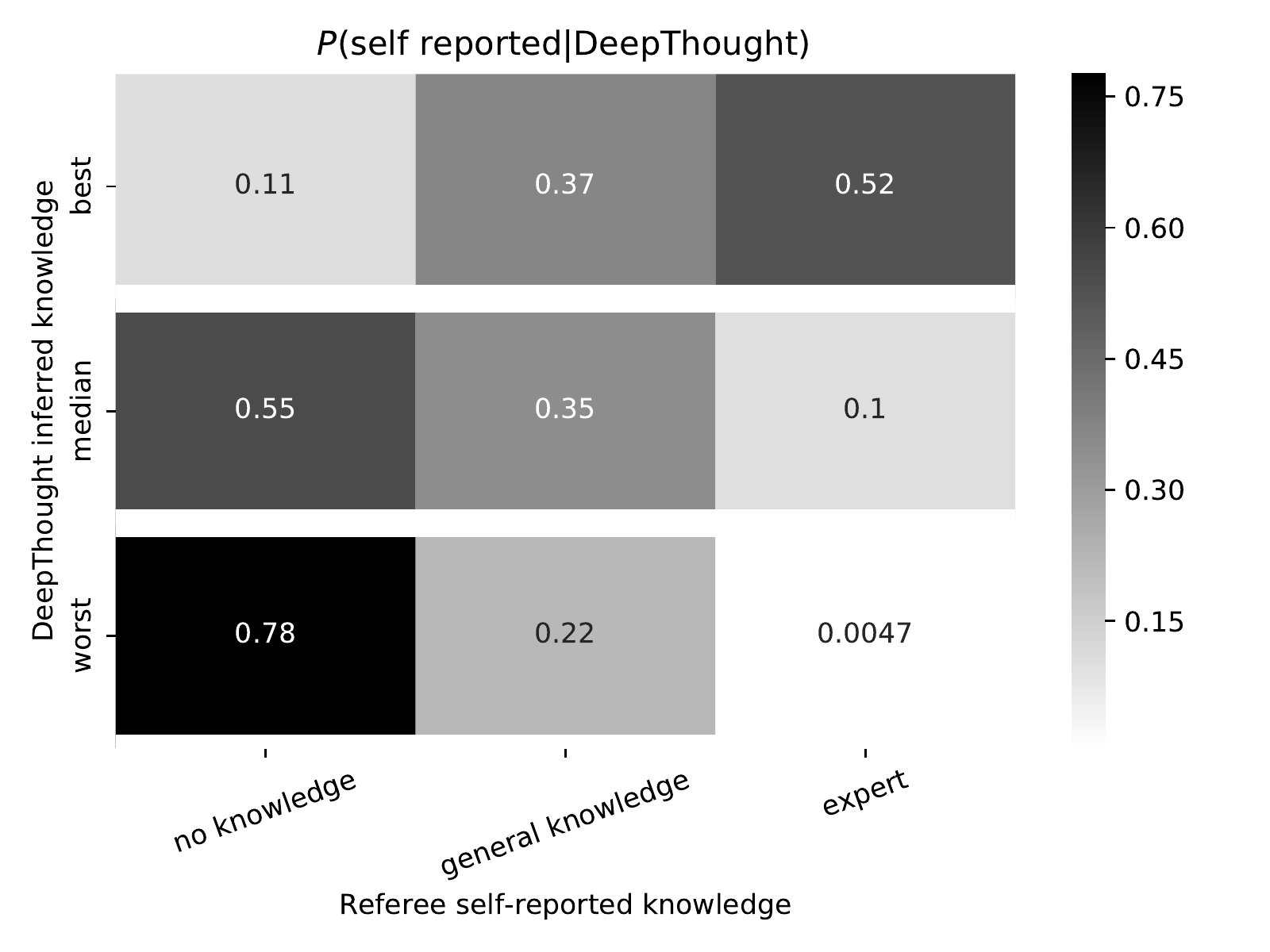}
   \caption{\textbf{Accuracy of DeepThought expertise predciction} Conditional probability $P(\textrm{self reported}|\textsc{DeepThought})$ for the various combinations of perceived and \dt inferred knowledge. The self-reported knowledge came directly from being a required field to answer in the refereeing proposal. The match provided by deepthought was calculated purely algorithmically.}
   \label{fig:dt_vs_ref_knowledge}
\end{figure}
Another aim of the DPR experiment is to infer a referee's domain knowledge for a given proposal using machine learning. ``Expertise'' is, unfortunately, no objective quantity. However, it is reasonable to assume that the self-judgement of expertise (self-efficacy) is a good measure that might approximate such a quantity.

Given:
\begin{itemize}
    \item ``self reported'' as the self reported domain knowledge 
    \item \textsc{DeepThought} as the \textsc{DeepThought} inferred domain knowledge 
\end{itemize}
For our experiment, we calculate the joint probability $P(\textrm{Self Reported}|\textsc{DeepThought})$ using Bayes theorem:

\begin{align}
   P(\textrm{Self Reported}&|\textsc{DeepThought}) =\\ 
   &\frac{P(\textsc{DeepThought}|\textrm{Self Reported})P(\textrm{Self Reported})}{P(\textsc{DeepThought})}\nonumber
\end{align}
\begin{align}
   P(\textrm{Self Reported}|&\textsc{DeepThought}) =\\ 
   &\frac{P(\textsc{DeepThought})\cap(\textrm{Self Reported})}{P(\textsc{DeepThought})}\nonumber
\end{align}

Figure~\ref{fig:dt_vs_ref_knowledge} shows the correlation between self reported knowledge (see the detailed description of the experiment in the supplementary material) and our predicted DeepThought inferred knowledge (see supplementary material for a detailed description of the method). 

We reiterate that we are not comparing to the true domain knowledge but to the self-reported knowledge. We find that DeepThought will predict the opposite of the self-reported knowledge  in only $\approx10\%$ of the cases (predicting expert with self-reported ``no knowledge'' and vice versa). We emphasize the $\approx 80\%$ success rate of predicting ``no knowledge''. These numbers show a high success rate in removing whose expertise does not overlap with the proposal.

\input{helpfulness.tex}

\section{Summary and Conclusions}
\label{sec:concl}

The main advantages of the DPR paradigm (coupled to the \dt approach) over the classic panel concept can be listed as follows. 

\paragraph{Advantages:}

\begin{itemize}
    \item it allows a much larger statistical basis (each proposal can be easily reviewed by 8-10 scientists), enabling robust outlier rejection;
    \item it removes possible biases generated by panel member nominations;
    \item the larger pool of scientists allows a much better coverage in terms of proposal--expertise matching;
    \item the smaller number of proposals per reviewer allows for more careful work and more useful feedback;
    \item coupled to the DeepThought approach for proposal--referee matching, it is suited to be made semi-automatised; it also gives an objective criteria for `expertise' removing biases in self-reporting;
    \item it removes the concept of panel, which adds rigidity to the process;
    \item it addresses the problem of maximising the proposal--referee match while maximising the overlap in the evaluations, which is a typical issue in pre-allocated panels \citep[see][and references therein]{Cook2005};
    \item the lack of a face-to-face meeting greatly simplifies the logistics and the costs, making it attractive for small, budget-limited facilities;
    \item the absence of the meeting prevents strong personal opinions from having a pivotal influence on the process;
    \item it involves a larger part of the community, increasing its democratic breadth;
    \item all applicants are exposed to the typical quality of the proposals. This allows them to better understand if their request is not allocated time by placing it in a much wider context, and helps improving their proposal-writing skills \citep[comment by Arash Takshi: \textit{``The ability to see what my competitors were doing filled a blind spot for me. Now I know that if I don't get funded, it's because of the quality of the other proposals, not something I did wrong.''}][]{Mervis2014};
    \item it trains the members of the community without additional effort.
\end{itemize}

\paragraph{Disadvantages:}

\begin{itemize}
    \item The lack of a meeting does not allow the exchange of opinions and the possibility of asking and answering questions to/from the peers;
    \item exposition of proposal content to a larger number of individuals (167 vs. 78 in the real case of the DPR experiment) increases the risk of confidentiality issues.
\end{itemize}

The two major disadvantages can, however, be easily addressed: barring the fact that its effectiveness remains to be demonstrated and quantified (see above), the social, educational and networking aspects of the face-to-face meeting should not be undervalued. In this respect, we notice that the resources freed by the DPR approach can be used by the organizations for education and community networking (training on proposal writing, fostering collaborations, ...). Another possibility to enable the interaction between the reviewers is to allow them to up-vote or down-vote the comments by other reviewers (the Science journal employs such an approach) which could be used to exclude comments and grades that were down-voted by a very significant fraction of the other referees. 


We conclude that the participating community has reacted extremely positively to this (see Section~\ref{si-sec:feedback_statistics} in the supplementary material). The presented approach to infer expertise works very well (see Figure~\ref{fig:dt_vs_ref_knowledge}). On an individual level, the behavior of the DPR referees conforms to the statistical description of the regular OPC referees (P18), and there is no statistically significant evidence that junior reviewers systematically deviate from this (see Section~\ref{sec:analysis.main.comparison_opc_dpr}).
The introduction of the possibility to rate the helpfulness of comments provides a new avenue to potentially reward helpful referees and train referees in general on giving useful feedback.

We encourage other organization to run similar studies, to progress from a situation in which the classic peer-review is adopted notwithstanding its limitations in the lack of better alternatives. As scientists, we firmly believe in experiments, even when these concern the way we select the experiments themselves.

\section{Acknowledgements}

This paper is fruit of independent research and is not to be considered as expressing the position of the European Southern Observatory on proposal review and telescope time allocation procedures and policies.

The authors wish to express their gratitude to the 167 volunteers who participated in the DPR experiment, for their work and enthusiasm.
The authors are also grateful to Markus Kissler-Patig for passionately promoting the DPR experiment following his experience at Gemini; to ESO's Director General Xavier Bar\c{c}ons and ESO's Director for Science Rob Ivison for their support; and to Hinrich Sch\"utze for several suggestions in the \gls{nlp} process. The authors would like to thank Jim Linnemann for help with some of the statistics tests.

We would like to thank the two anonymous referees and Anna Severin for providing constructive criticism that improved the paper immensly. 

W.~E.~K. is part of SNYU and the SNYU group is supported by the NSF CAREER award AST-1352405 (PI Modjaz) and the NSF award AST-1413260 (PI Modjaz). W.~E.~K was also supported by an ESO Fellowship
and the Excellence Cluster Universe, Technische Universit\"at
Munchen, Boltzmannstrasse 2, D-85748 Garching, Germany for part of this work. 
W.~E.~K would also like to thank the Flatiron Institute. 
GvdV acknowledges funding from the European Research Council (ERC) under the European Union's Horizon 2020 research and innovation programme under grant agreement No 724857 (Consolidator Grant ArcheoDyn).
\newpage
\begin{methods}
\input{deepthought_method.tex}
\end{methods}

%% file: helpfulness.tex

\subsection{Rating the helpfulness of review comments\label{sec:analysis.helpfulness}}

After the review process, we asked the proposers to evaluate the ``helpfulness'' of the review comments. A total of 136 reviewers provided feedback.

The review usefulness distribution shows a steady rise, with a
sudden drop-off at the `very helpful' bin, as shown in all panels of
Fig. 4. About 55\% of the users rated the comments in the `helpful'
and `very helpful' bins.

To check what factors might influence the ability to write helpful
comments we use the statistical method given at the beginning of
this section.

The reviewer's expertise is expected to have an influence on the
helpfulness of comments. Figure 4a,b shows the influence of both
self-reported knowledge and DeepThought-inferred knowledge on
the helpfulness of the comment. The probabilities are very similar
between the self-reported and inferred knowledge. We highlight
that experts seemingly very rarely give unhelpful comments and
that non-experts rarely give very helpful comments.

The last test is to see how the comment’s helpfulness is being
evaluated given the ranking of the proposal within the quartiles
$P(\textrm{helpful comment}|\textrm{proposal quartile})$. This shows a similar distribution to the other panels in Fig. 4. There are some small differences. Comments for proposals from the second to the top quartile often were perceived as relatively helpful. Comments on proposals in the last quartile were rarely ranked as very helpful \citep[][finds a similar effect]{van1999effect}.

We checked whether seniority has an influence on the ability to create helpful comments. Figure 4c shows some correlation
between the seniority and the ability for the referee to give helpful
comments. Most interesting is the apparent inability of graduate
students to give very helpful comments. This might be a training
issue and can be resolved by exposing the students to schemes
such as DPR.

We have also asked about the helpfulness of the comment in our
general feedback (Supplementary Information). The distribution of
comment usefulness follows the distribution of helpfulness for individual comments relatively closely (see statistics in Supplementary
Information).
The comments given in the DPR compare very favourably with
the OPC (see details in Supplementary Information).

\begin{figure*}[ht!]
   \includegraphics[width=\columnwidth]{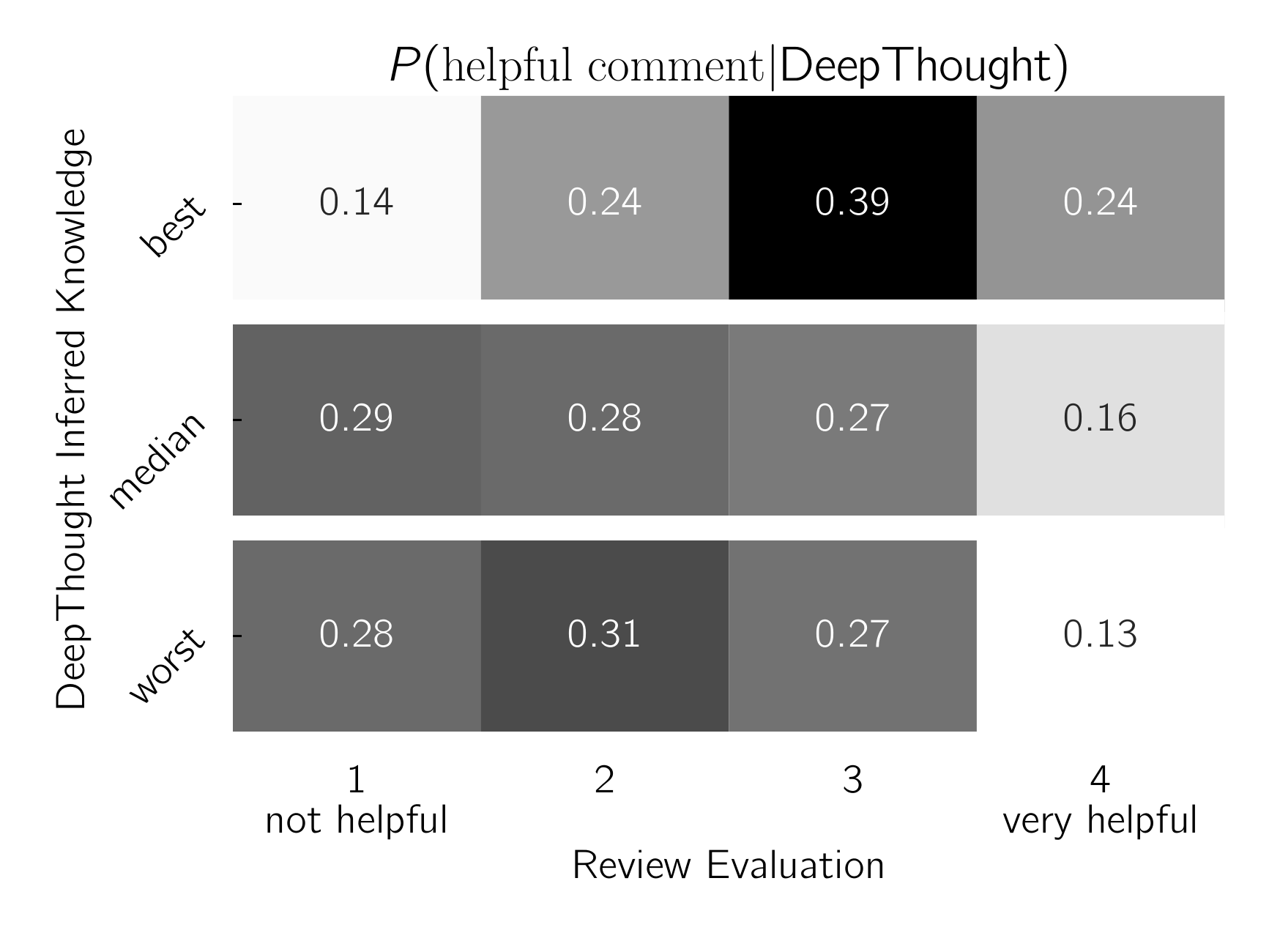}
   \includegraphics[width=\columnwidth]{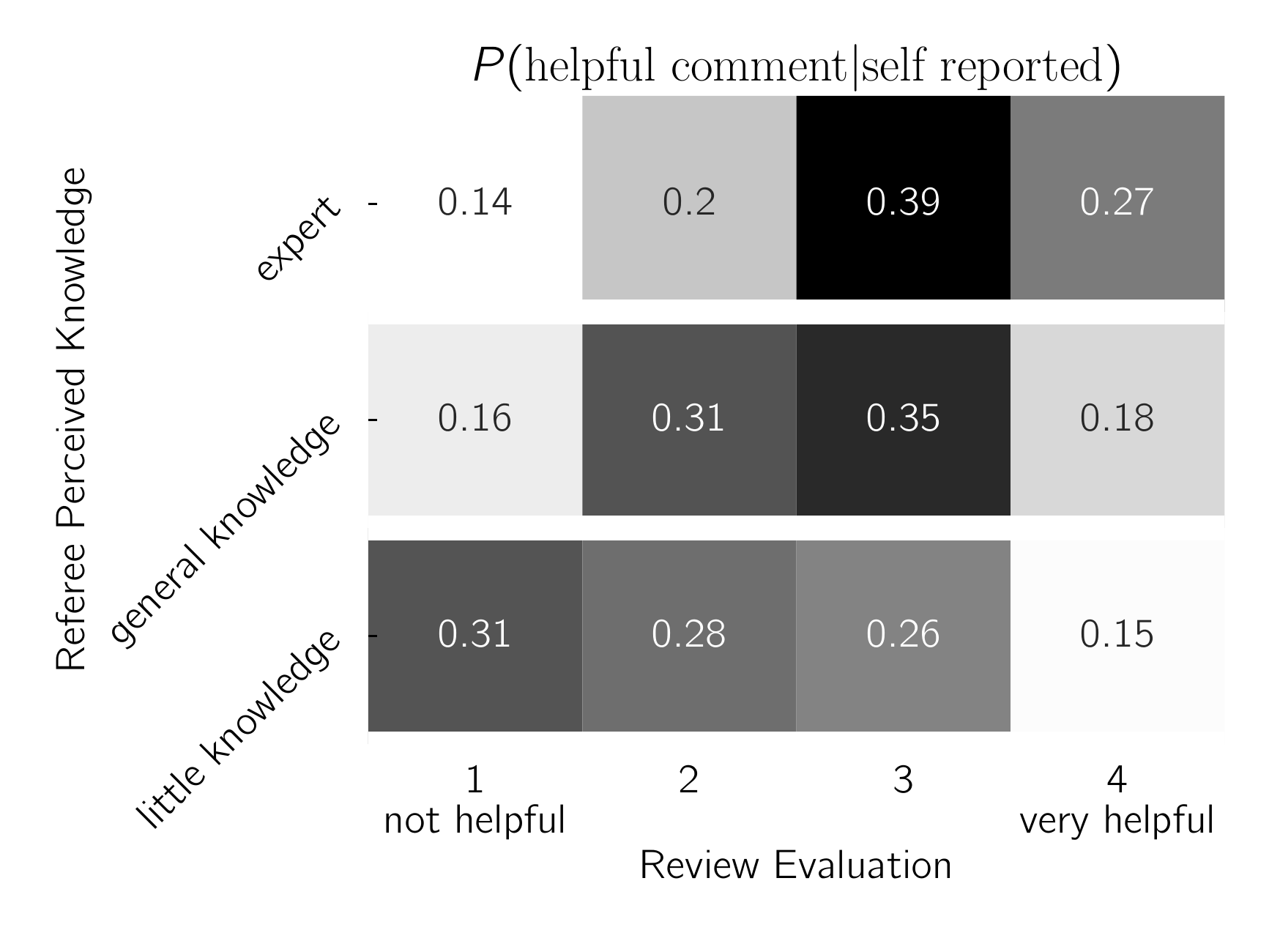}
   \includegraphics[width=\columnwidth]{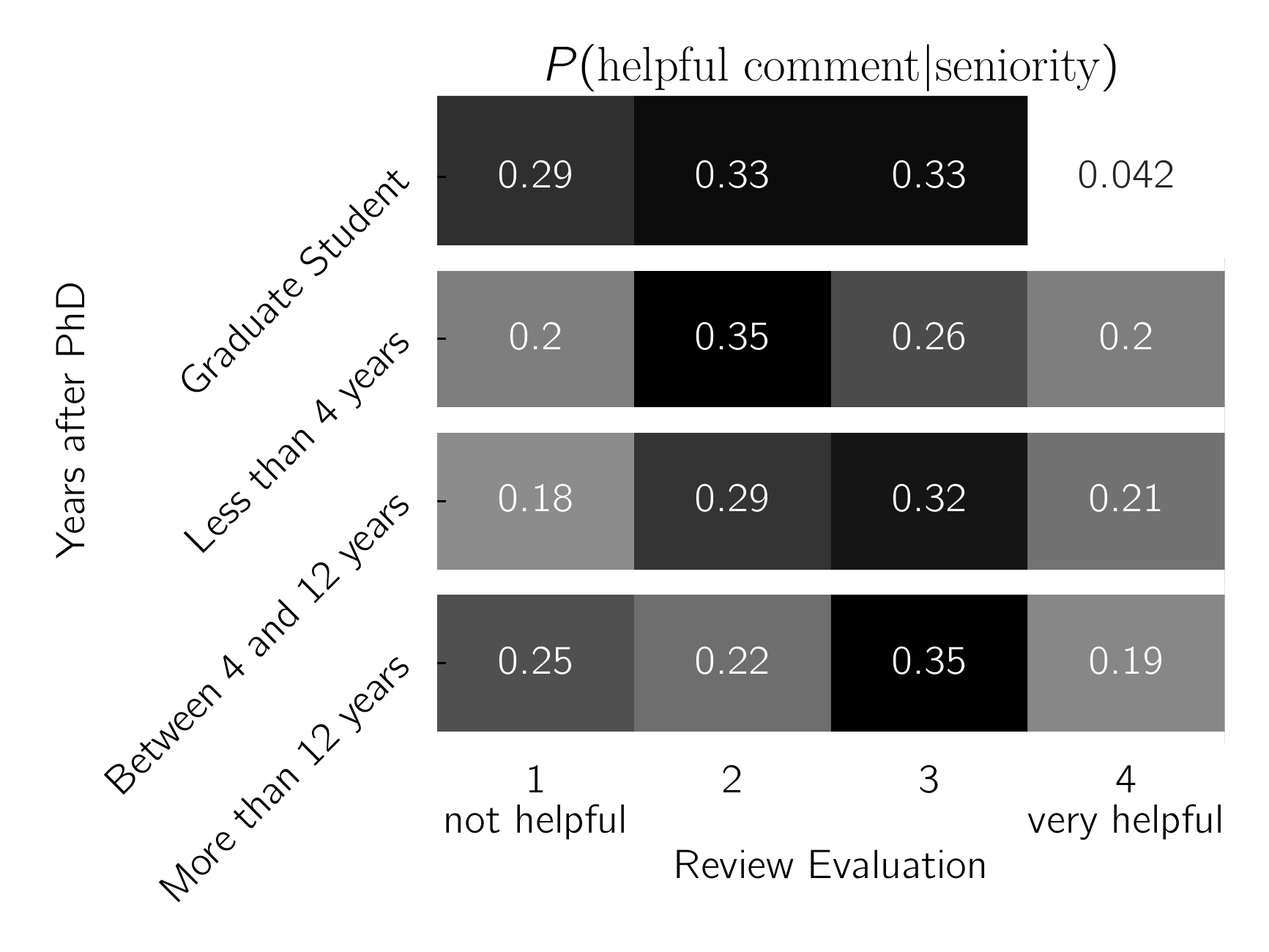}
   \includegraphics[width=\columnwidth]{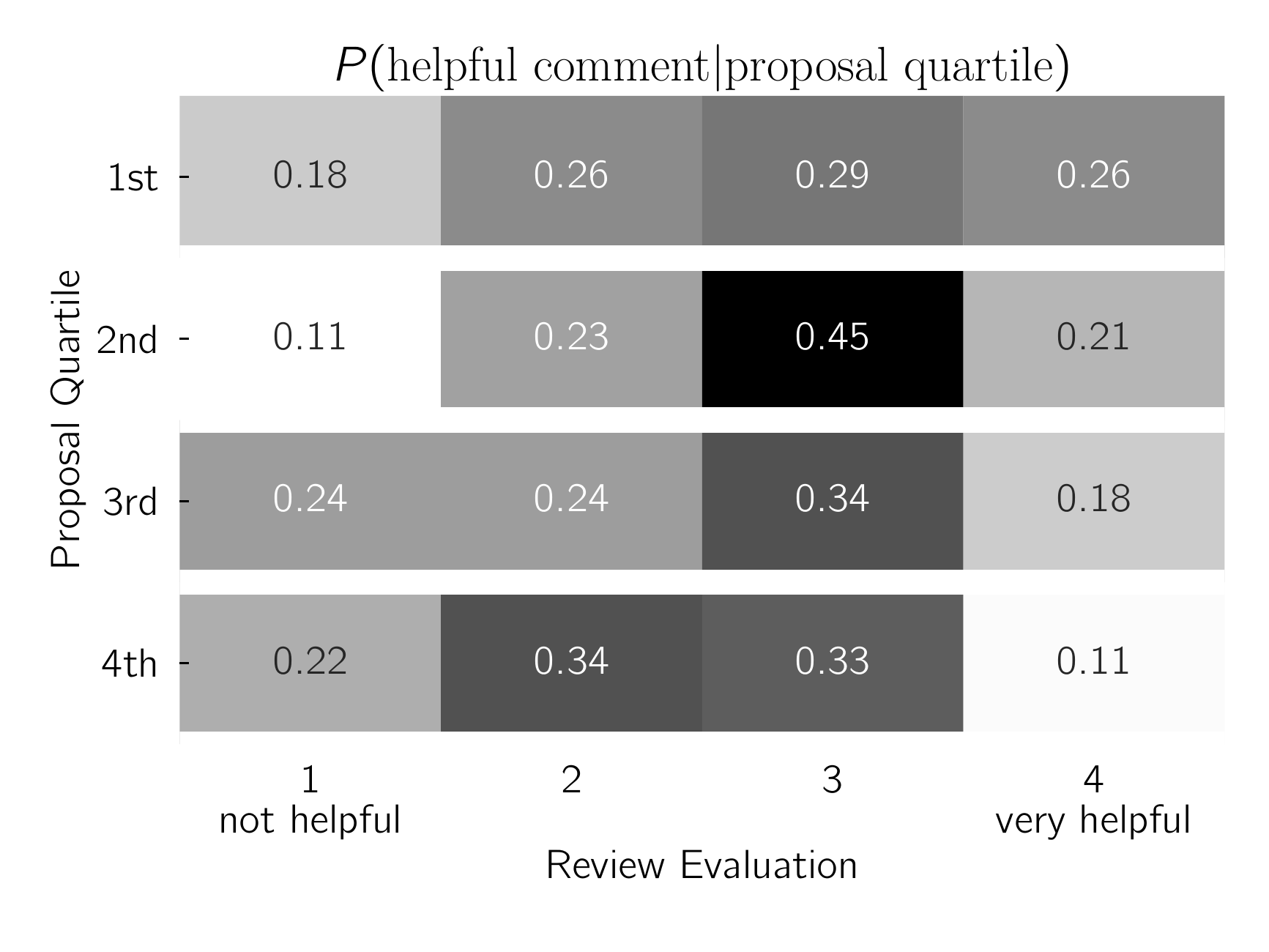}

   \caption{\textbf{Influence of various factors on the perceived quality of the review} We show a number of factors that can influence the (perceived) quality of the review that was graded by the proposers in the feedback step. \textbf{Upper Row:} Evaluation of the helpfulness of the comment given expertise (inferred by  \dt on the left panel and self reported on the right panel). 
   \textbf{Lower Left:} We show the conditional probability $P(\textrm{helpful comment}|\textrm{gender})$ for the various combinations of helpful comment rating and reported gender.
   \textbf{Lower Right:}We show the conditional probability $P(\textrm{helpful comment}|\textrm{proposal quartile})$ for the various combinations of helpful comment rating and proposal quartile.}
   \label{fig:review_review_overview}
\end{figure*}

%% file: deepthought_method.tex
\section{Description of the DeepThought DPR experiment}
\label{sec:method_description}

For an overview of the process see Section~\ref{sec:method_overview}. In the next sections, we will give a detailed description of the process of the DeepThought DPR experiment.

\subsection{Reviewer exclusion and selection}
\label{sec:reviewer_selection_general}

Reviewer selection is a core part of the experiment. We separated this step into reviewer exclusion and reviewer selection. We tested two different strategies for reviewer selection: one based on a standard methodology (called ``OPC emulate'' [OE]), the other based on a machine-estimated of the domain-specific knowledge (called ``\textsc{DeepThought}'' [DT]), which are described in detail in Section~\ref{sec:method.deepthought}.

We describe the reviewer exclusion and selection using the abstract concept of a matrix. The matrix has a row for each proposer and a column for each referee. In our case, the referees and proposers are the same set, and thus we have a square matrix.

Our exclusion matrix was constructed in a way that would mark a referee ineligible to review a proposal where any of the investigators were from the same institute as the referee
\begin{equation}
   A_\textrm{exclusion} = 
   \bordermatrix{~ & \textrm{p}_1 & \textrm{p}_2 & \cdots & \textrm{p}_n\cr
   \textrm{referee}_1 & 1 & 1 & \cdots & 0 \cr
   \textrm{referee}_2 & 0 &1 & \cdots & 0\cr
   \vdots           & 1 & 0 & \cdots & 0 \cr
   \textrm{referee}_n & 0 & 0 & \cdots & 1 \cr},
   \label{eq:exclusion_matrix}
\end{equation}

where $\textrm{p}_n$ stands for proposer$_{n}$, 1 indicates conflict and 0 no conflict.

For the OPC-emulate group, we also constructed an additional matrix (that was combined to the previous exclusion matrix using the logical or) that marks a referee ineligible to review a proposal that was submitted to the same unit telescope as the referee's submitted proposal. 

We constructed a reviewing matrix that marks a reviewer-proposal combination with 1 (and the rest of the matrix with 0). 
We then used a round-robin selection process to iterate through the referees. For each referee, we use the exclusion matrix to determine eligible proposals and then the specific selection criterion for each of the groups (outlined in Section~\ref{sec:reviewer_selection_general}) to assign one of the remaining available proposals (taking into account the exclusion matrix). This process was repeated until all referees had been assigned eight proposals. If the process failed before completion, it was restarted with a different random number seed until we found a solution. A solution matrix needs to have all row sums and column sums equal to eight. For future projects, we strongly suggest to research algorithms from operations research (or combinatorial optimization) that have been optimized for the given process.

\subsection{Reviewer selection methodology}
\label{sec:reviewer_selection}

We separated the volunteer base into two groups for our two experiments. The first group was 60 randomly chosen volunteers out of the 172. This group was assigned proposals that would closely emulate the current way ESO assigns proposals in the \gls{eso.opc} (which is a variant of the common time-allocation strategy present in the astronomy community).

The second group of the remaining 112 volunteers was assigned by predicting their expertise of a proposal based on their publication history using machine learning. 

\subsubsection{OPC emulate Group}

We aim with this selection process to emulate the OPC process. The members of the OPC assign themselves to expert groups in four categories

\begin{itemize}
   \item A - Cosmology and Intergalactic Medium
   \item B - Galaxies
   \item C - Interstellar Medium, Star Formation, and Planetary systems
   \item D - Stellar Evolution
\end{itemize}

Multiple panels are then constructed for each subgroup (depending on the number of proposals for each subgroup).

We attempt to emulate this process by constructing four groups (A, B, C, and D) of 15 referees. Each of these groups only reviews the proposals in their group. Thus each referee will only see proposals within the same category that they proposed in.

We construct an exclusion matrix for each of the subgroups (see Equation~\ref{eq:exclusion_matrix}) and then proceed with the review selection (see Section~\ref{sec:reviewer_selection_general}) where at each selection step we simply randomly select any eligible proposal. 

The total number of reviews from this process was 480. 

\subsubsection{DeepThought Group}
\label{sec:method.deepthought}

The general idea behind this selection process is to use the published papers of each participant to predict how knowledgeable they were for each proposal. This made extensive use of the dataset and techniques presented in \citet{2019JApA...40...23K}. This required identifying their publications, construct knowledge-vectors for each referee, construct proposal vectors from the submitted latex document, construct a knowledge matrix for each of the combination of referee and proposal, and use this matrix in the selection process. 

\paragraph{Name disambiguation}

The first part of this process was to uniquely identify participants in ADS to infer their publications. \citet{milojevic2013accuracy} has shown using the last name and the first initials only 6.1\% of author's identities are contaminated (either due to splitting or merging) which is sufficient for the statistical requirements of our experiment.

We then used the \textsc{Python} package \textsc{ads} (available at \url{https://ads.readthedocs.io}) to access the ADS API to search for the participant's papers (and their arxiv identifiers) without regard for position of authorship. We excluded participants (moved them to the OPC emulate group) that have less than 3 papers (9 participants) or more than 500 papers (4 participants). 

\paragraph{Knowledge Vectors}

\citet{2019JApA...40...23K} shows in Section~4 the construction of vectors from publication (using a technique called TFiDF). We used the document vectors from \citet{2019JApA...40...23K} given of all publications identified for each participant in the previous step. These document  vectors were summed up and then normalized. We call such a vector sum for each referee a ``knowledge vectors''. 

\paragraph{Proposal vectors}

We use the machinery described in \citet{2019JApA...40...23K} to process several sections of the latex representation ('Title', 'Abstract', 'ScientificRationale', 'ImmediateObjective') of the submitted proposal. These were then converted to normalized document vectors which we refer to as ``proposal vectors''. 

\paragraph{Knowledge Matrix}

We then construct a knowledge matrix similar to the exclusion matrix and fill each of its elements with the dot-product between the proposal vector and referee knowledge vector (cosine distance; see Equation~\ref{eq:knowledge_matrix}).

\begin{equation}
   A_\textrm{knowledge} = 
   \bordermatrix{~ & \textrm{p}_1 & \textrm{p}_2 & \cdots & \textrm{p}_n\cr
   \textrm{referee}_1 & 0.8 & 0.4 & \cdots & 0.1 \cr
   \textrm{referee}_2 & 0.5 & 0.9 & \cdots & 0.5\cr
   \vdots           & \vdots & \vdots & \ddots &\vdots  \cr
   \textrm{referee}_n & 0.6 & 0.2 & \cdots & 0.7 \cr},
   \label{eq:knowledge_matrix}
\end{equation}
where $\textrm{p}_n$ stands for proposer$_{n}$.

As opposed to the OPC-emulate case, we do not assign proposals randomly to the referees during the selection step in the referee selection process (see Section~\ref{sec:reviewer_selection_general}). The proposals are picked according to the following algorithm with different steps for the first four, subsequent two and last two proposals picked for each referee:

\begin{enumerate}
   \item from the available proposals choose the one with the highest cosine distance for the first four proposals assigned to each referee
   \item from the available proposals choose the proposal closest to the median of all cosine scores for that particularly referee for the next two proposals
   \item from the available proposals choose the one with the lowest cosine distance for the last two proposals assigned to each referee
\end{enumerate}

The process was repeated with a different random if there were no eight suitable proposals available for each referee. Depending on the number of constraints and participants it took on the order of three times too find a suitable solution. 

\subsection{\label{sec:review}Review process}

The participants were given a login to evaluate the proposals. After signing a Non Disclosure Agreement (which is identical to the one signed by the OPC members), the participants could view the proposals assigned to them. They were first given the option to indicate a conflict of interest (removing them from making an eligible vote on the proposal). Then were asked for their expertise on the proposal's topic. They were tasked to review the proposals by giving them a score (1 - 5; same as in the OPC), their assessment of their knowledge of the proposal, and a comment. 

We outline the steps in more detail in the following. The following options were given to indicate a conflict:\\[+5pt]
%
\noindent
\begin{framed}
   \it
\begin{itemize}
   \item No, I do not have a conflict.
   \item Yes, I have a close personal or professional relationship with the PI and/or team.
   \item Yes, I am a direct competitor to this proposal.
\end{itemize}
\end{framed}

\noindent\\[+5pt]
The referees were instructed to consider the following questions when evaluating a proposal:\\[+5pt] 
\noindent
\begin{framed}
   \it
\begin{itemize}
   \item Is there sufficient background/context for the non-expert (i.e., someone not specialized in this particular sub-field)?
   \item Are previous results (either by proposers themselves or in the published literature) clearly presented?
   \item Are the proposed observations and the Immediate Objectives pertinent to the background description?
   \item Is the sample selection clearly described, or, if a single target, is its choice justified?
   \item Are the instrument modes, and target location(s) (e.g., cosmology fields) specified clearly?
   \item Will the proposed observations add significantly to the knowledge of this particular field?
\end{itemize}
\end{framed}

\noindent\\[+5pt]
They were then asked to assess their expertise of the proposal:\\[+5pt] 
\noindent
\begin{framed}
   \it
\begin{itemize}
   \item This is my field of expertise.
   \item I have some general knowledge of this field.
   \item I have little or no knowledge of this field.
\end{itemize}
\end{framed}

\noindent\\[+5pt]
They were instructed to use the following general grading rules:\\[+5pt]
\noindent
\begin{framed}
   \it
\begin{itemize}
   \item \textbf{1.0} outstanding: breakthrough science
   \item \textbf{1.5} excellent: definitely above average
   \item \textbf{2.0} very good: no significant weaknesses
   \item \textbf{2.5} good: minor deficiencies do not detract from strong scientific case
   \item \textbf{3.0} fair: good scientific case, but with definite weaknesses
   \item \textbf{3.5} rather weak: limited science return prospects
   \item \textbf{4.0} weak: little scientific value and/or questionable scientific strategy
   \item \textbf{4.5} very weak: deficiencies outweigh strengths
   \item \textbf{5.0} rejected
\end{itemize}
\end{framed}

\noindent\\[+5pt]
The referees then had to write a comment with a minimum of 10 characters. 

\subsection{Grade Aggregation}
\label{sec:method.aggregation}

In the experiment design, each proposal was assigned to $N_r$=8 peers, and each PI was assigned $N_p$=8 proposals. In practice, because of the declared conflicts, proposals were evaluated by 4 to 8 referees, with $N_r\geq$6 in 95\% of the cases. For the same reason, each referee reviewed between 5 and 8 proposals, with $N_p\geq$6 in 98\% of the cases.
This guarantees statistical robustness in the grade aggregation.

In the OPC process, the grades given by the distinct referees are combined using a simple average, after applying the referee calibration. This operation is described in \citet{2018PASP..130h4501P} (Sect. 2.4 and Appendix A therein), and aims at minimizing the systematic differences in the grading scales used by the reviewers. In the current implementation, the calibration consists in a shift-and-stretch linear transformation, by which the grade distributions of the single referees are brought to have the same average and standard deviation (grades $\geq$3 are excluded from the calculations). This operation is justified by the relatively large number of proposals reviewed by each referee ($>$60), which makes the estimate of the central value and dispersion reasonably robust.

The case of the DPR is different in this respect, as a given person would have reviewed at most $N_p$=8 proposals. Especially for the dispersion, this limitation certainly weakens its statistical significance. For this reason, following the example of the Gemini Fast Turn Around channel \citep{2019AAS...23345503A}, and for the purposes of providing feedback to the users, the raw grades were combined without applying any referee calibration (the effects of calibration in the DPR experiment are presented and discussed in the main text).

\subsection{\label{sec:evaluation}Review evaluation}

After the review deadline the participants were given access to a page with the peer-reviews (most of the time seven or eight) of their proposal. The applicants are given the quartile rank as a letter (A-D; as calculated in Section~\ref{sec:method.aggregation}). For each comment, they were asked to rank its \textit{helpfulness}:\\[+5pt]
\noindent
\begin{framed}
\it
We would be grateful if you could rate each review on a scale of 1 (not helpful) to 4 (very helpful) how much this comment helps improve your proposal (positive comments like "best proposal I ever read" - can be ranked as not helpful as it does not improve the proposal further). These ratings will not be distributed further \textbf{but help us for statistical purposes.}
\end{framed}

\subsection{Questionnaire}
\label{sec:method.feedback}

Each participant in the DPR experiment was asked to fill out a questionnaire after performing the reviews and receiving the feedback on their own proposal:\\[+5pt]
\noindent
\begin{framed}
   Also, after reading the reviews, please take 10-15 minutes to fill the final questionnaire (see the link at the bottom of this page). Although it is optional, it is your chance to give us feedback on any aspect that you liked/disliked and to shape a future DPR process. It will also greatly assist us in understanding how the experiment went, learn about what works and not, which biases are still present etc. It will allow us to build better tools for you in the future.
\end{framed}

\noindent\\[+5pt]
Out of 167 participants, 140 returned a completed on-line questionnaire (83.8\%). Most of the questions were multiple choice (see supplementary material Section~\ref{si-sec:feedback_questions}) with many of the answers used in the following sections in the evaluation of the DPR experiment. 

The questionnaire also included 5 free-format questions: (i) What suggestions do you have to improve the software?, (ii) What suggestions do you have to improve the assessment criteria and/or review process?, (iii) Do you have concerns about your proposals being evaluated through distributed peer review?, (iv) Do you have any further feedback or suggestions regarding distributed peer review?, and (v) Would you like to give any further feedback and/or suggestions regarding earlier raised points on securing confidentiality, external expertise, and robustness versus bias?

Each of the 5 free-form questions was answered on average by 50 persons with a sentence or more. The answers were very helpful in specific suggestions for improvement and to get an overall feedback as summarized in the main text).

\subsection{\label{sec:data}Data Collection}

The proposals were distributed to the participated on 8 October, 2018. The participants were given until Oct 25 (17 days) until the deadline to submit the reviews. At the time of the deadline 2 of 112  participants in the \dt group and 3 of 60 participants in the OPC emulate group had not completed their reviews and were excluded from the further process (completion rate 97.1\%). We received a total of 1336 reviews (from 167 reviewers for 172 proposals).

On 30 October the 167 remaining participants were given access to their evaluated proposals and were given two weeks to provide feedback. Of these, 136 (81.4\%) completed the questionnaire.

We aim to allow further study of this dataset by the community. We also want to ensure the privacy of our participants and thus have anonymized and redacted some of the dataset. In particular, we have given the participants randomized IDs and only give some derived products from the \dt machinery (we are not sharing properties knowledge vectors as they might allow the reconstruction of the individuals).  In addition, we have removed any free text data the participants entered (such as the comments on the proposal). We have also removed all participants that did not reveal their gender. This is a very small number and might be used to de-anonymize individuals.

The anonymized dataset is available at \dataurl.

%% file: statements.tex
\section{Statements}
\paragraph{Author Contributions:} 
We use the CRT standard (see \url{https://casrai.org/credit/}) for reporting the author contributions:

\begin{itemize}
    \item \textbf{Conceptualization:} Kerzendorf, Patat, van de Ven
    \item \textbf{Data curation:} Kerzendorf, Patat
    \item \textbf{Formal Analysis:} Kerzendorf, Patat
    \item \textbf{Investigation:} Kerzendorf, Patat
    \item \textbf{Methodology:} Kerzendorf, van de Ven, Pritchard
    \item \textbf{Software:} Kerzendorf, Bordelon
    \item \textbf{Supervision:} Kerzendorf,  Patat
    \item \textbf{Validation:} Kerzendorf, Patat, van de Ven, Pritchard
    \item \textbf{Visualization:} Kerzendorf, Patat, Pritchard
    \item \textbf{Writing – original draft:} Kerzendorf, Patat, Pritchard
    \item \textbf{Writing – review \& editing:} Kerzendorf, Patat, van de Ven, Pritchard
\end{itemize}

\paragraph{Competing Interests:} The authors declare no competing interests.
\paragraph{Data Availability:}
The anonymized data is available at \dataurl.

%% file: deepthought_dpr1_si_main.tex
The Supplementary information is structured in the following way. In Section~\ref{sec:tac_eso}, we described the general process of ESO Time allocation. Section~\ref{sec:analysis.demo} gives an overview of the demographics of the experiment and discusses self-efficacy based on several criteria. Section~\ref{sec:analysis.supplementary} provides further analysis of the data. Appendix~\ref{sec:feedback_questions} gives the exact wording of the questions including some statistics. Appendix~\ref{sec:feedback_statistics} provides additional statistics for the feedback survey.

\input{time_allocation_eso.tex}

\input{demographics.tex}
\section{Supplementary Analysis\label{sec:analysis.supplementary}}
\input{self-efficacy.tex}

\input{expertise_influence_review.tex}
\input{gender-statistics.tex}
\input{long_dpr_opc.tex}

%% file: time_allocation_eso.tex
\section{\label{sec:tac_eso}Time allocation at ESO\ }

ESO is an intergovernmental organization established in 1962, and runs one of the largest ground-based astronomical facilities world-wide. Every semester, about 900 proposals including more than 3000 distinct scientists from 50 different countries are submitted to ESO, requesting time on a large suite of telescopes: the four 8.2 m units of the Very Large Telescope (VLT), VISTA (4.2 m), 3.6 m, NTT (3.5 m), VST (2.6 m), and the Atacama Pathfinder Experiment (APEX), placed on three different sites in Chile. The total time request for Period 103 (the one relevant for the DPR experiment), was about \num{23000} hours. The over-subscription rate varies with telescopes, and for the VLT is typically around 4 (a detailed oversubscription statistics can be found at \url{https://www.eso.org/sci/observing/phase1/p96/pressure.html}).

The proposal review process at ESO is extensively described in \citet{2013ops2.book..231P}, and is summarized in \citet{2018PASP..130h4501P}. The reader is referred to those publications for more details. Here only a very brief summary is given. 

The members of the ESO TAC (the Observing Programmes Committee; OPC) who review the proposals are mostly nominated by the Users Committee, in turn composed by representatives of all member states. The nominations are reviewed by the OPC Nominating Committee, which makes a final recommendation to the Director General. As a rule, the panel members are required to have a minimum seniority level, starting with scientists at their second post-doc onward. The current implementation includes 13 Panels with 6 referees each, and cover the four scientific categories (Cosmology; Galaxy structure and evolution; Interstellar medium, Star formation and planetary systems; Stellar evolution). The panels are composed in such a way to maximize the expertise coverage (while also taking into account other constraints, like affiliations, gender, and seniority). Although some panels are created to cover special cases (e.g. interferometry), the proposals are distributed randomly within the panels of a given scientific category. Institutional conflicts are taken into account by the distribution software, while scientific conflicts are declared by the referees during the initial phases of the review.

The process is composed of two steps: 1) asynchronous at home, and 2) synchronous face-to-face. In the first step (which is called pre-meeting), the proposals are assigned to 3 referees, who are asked to review and grade the proposals using a scale from 1 to 5.\footnote{Before Period 102 all non-conflicted members of a panel (up to 6) would review all proposals assigned to their panel. As of Period 102, in order to alleviate the work-load, this was reduced to 3.} The pre-meeting grades are used to compile a first rank list, and triage is applied to the bottom 30\% (in time) for each telescope separately.
At the meeting, the surviving proposals are discussed and graded by all non-conflicted members. The final rank lists per telescope are compiled based on the meeting grades, and the time is allocated following this final rank.

%% file: demographics.tex
\section{Demographics of the Experiment}
\label{sec:analysis.demo}

The participants were asked to voluntarily give feedback and statistics about themselves and their experience in the experiment. We received this information from 140 out of 167 final participants. 

\begin{figure}
    \includegraphics[width=0.49\textwidth]{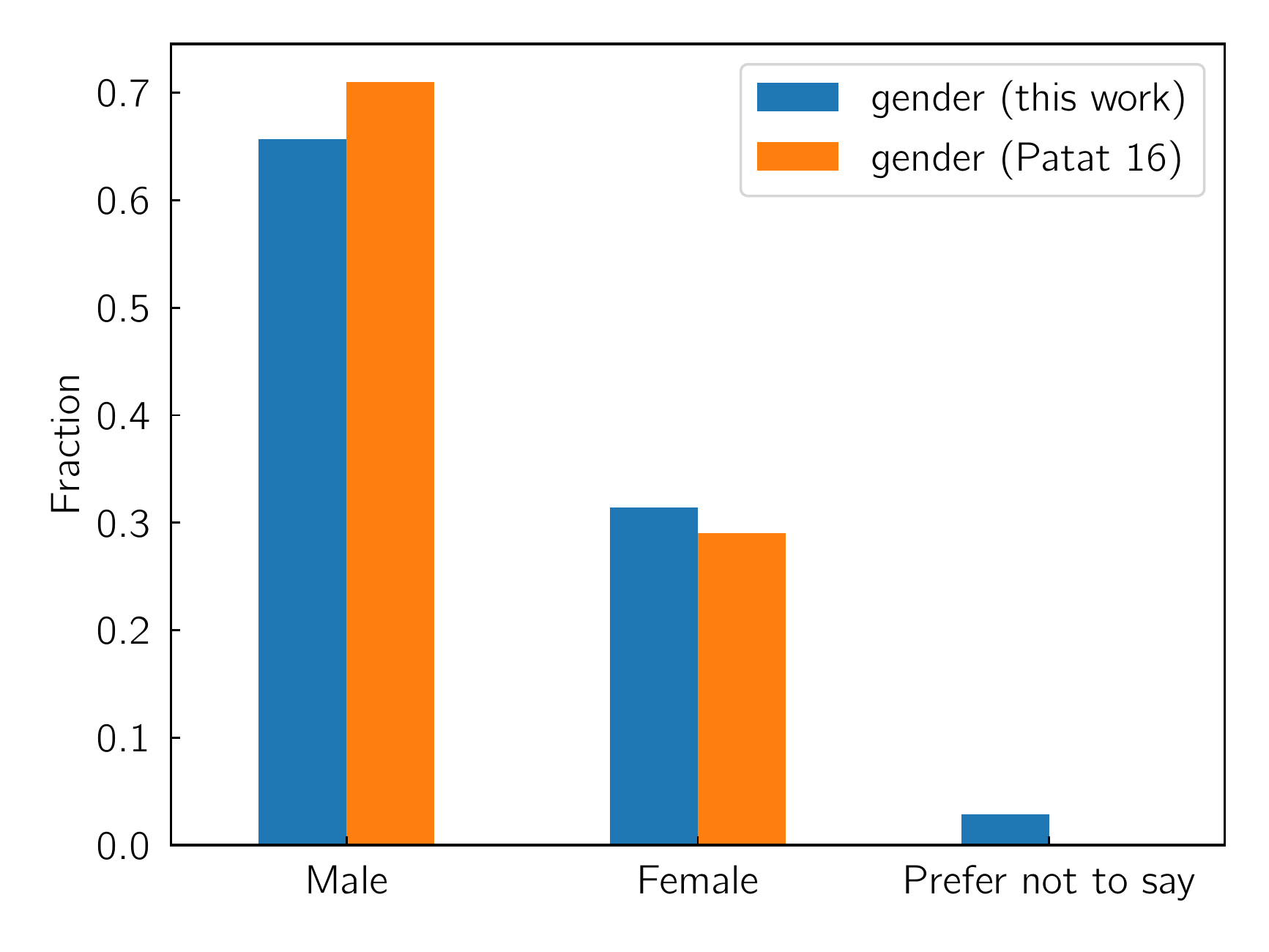}
    \caption{Gender distribution of Principal Investigators comparing this work participants and the general community.}
    \label{fig:gender_distribution}
 \end{figure}
 
 \begin{figure}
    \includegraphics[width=0.49\textwidth]{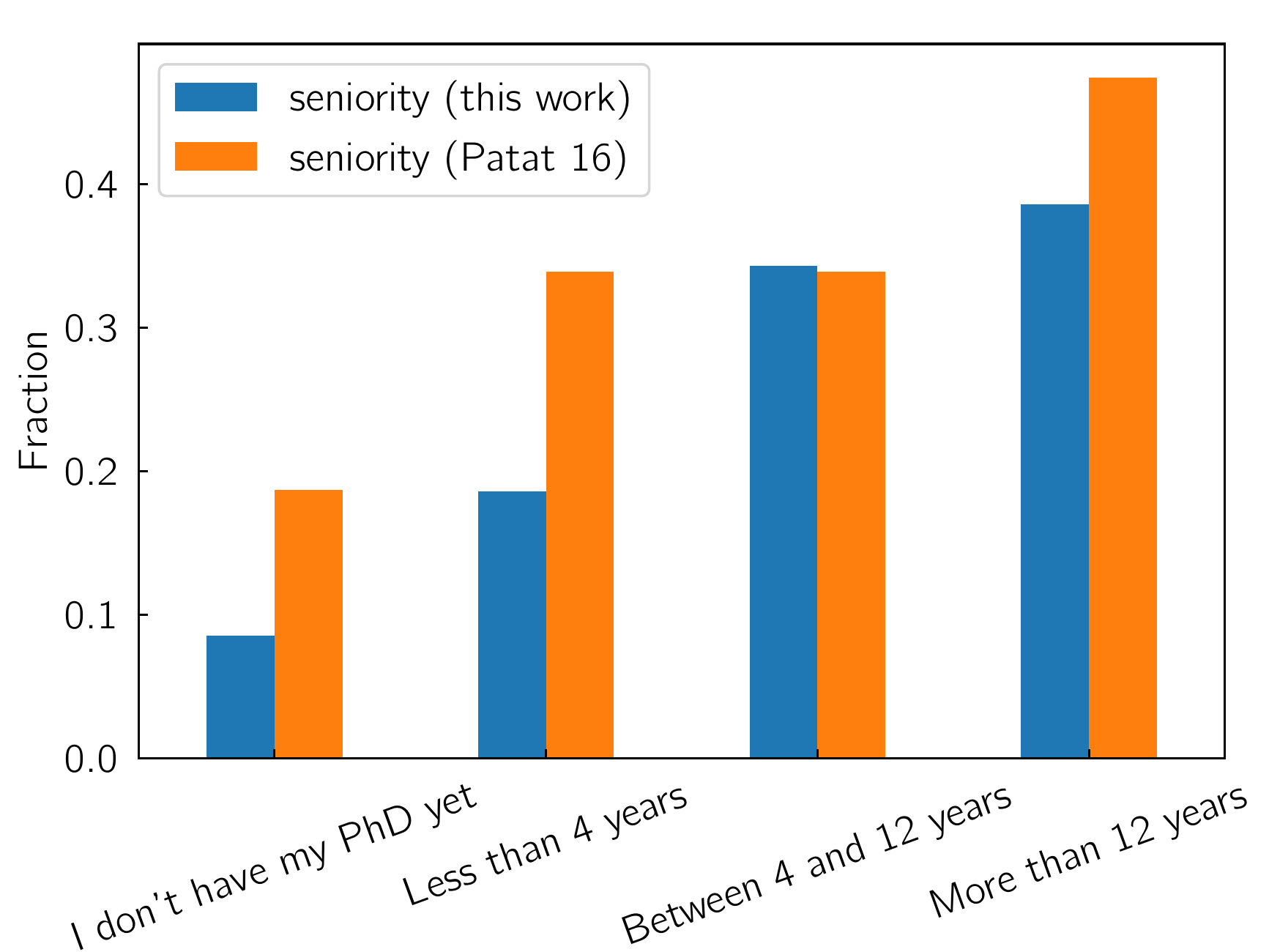}
    \caption{Career level distribution of Principal Investigators. We chose to plot the \citet{2016Msngr.165....2P} postdoc fraction for both the ``less than 4 years after PhD'' and ``between 4 and 12 years''. We chose the professional astronomer category to plot over the ``more than 12 years'' past PhD column.}
    \label{fig:stats_years_past_phd}
 \end{figure}

Figure~\ref{fig:gender_distribution} and Figure~\ref{fig:stats_years_past_phd} suggest that the self-selected participants of this study are representative of the ESO community at large. The PI gender distribution in the DPR experiment is 32.4\% (F) and 67.6\% (M), to be compared to the values derived from a sample of about \num{3000} PIs \citep{2016Msngr.165....2P}: 29.0\% (F), 71.0\% (M). In terms of scientific seniority, the DPR PIs are distributed as follows: 8.9\% (no PhD), 18.3\% ($<$4 years after PhD), 33.8\% (between 4 and 12 years after PhD), 39\% (more than 12 years after PhD). For the sample discussed in \citet{2016Msngr.165....2P} the distribution is: 18.7\% (PhD students), 33.9\% (post-docs), 47.4\% (senior astronomers). Although the two studies have a different seniority classification scheme, the overall comparison shows that the DPR includes a fair amount of junior scientists, properly sampling the underlying population.

We cannot guarantee that there are not specific attributes that lead our participants to self-selection, but if they exist they are well hidden.

\begin{figure}[h!]
    \includegraphics[width=\columnwidth]{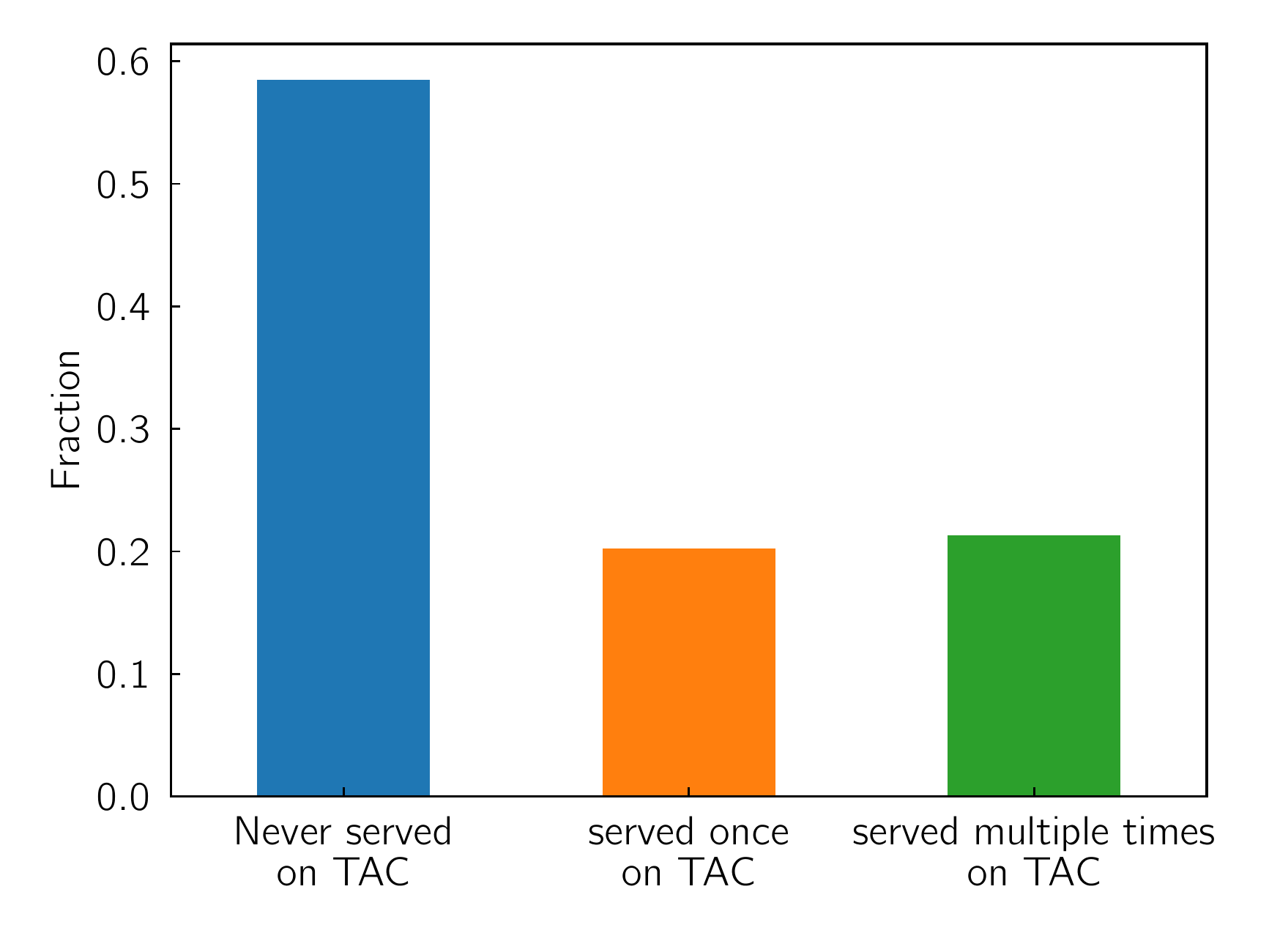}
 
    \caption{Combined fractions from the feedback questions (see Section~\ref{sec:feedback_questions}) of \textit{Have you served on another time allocation committee?} and \textit{How often have you been a panel member of the ESO Observing Programmes Committee (OPC)?}}
    \label{fig:tacs_served}
 \end{figure}

Most participants are relatively experienced submitting proposals (see Figure~\ref{fig:prop_submitted_to_eso}). However, most of them have not served on a TAC (see Figure~\ref{fig:tacs_served}). 

%% file: self-efficacy.tex
\subsection{Self-Efficacy}
\label{sec:analysis.self-efficacy}

The \dt algorithm described in this work is built to predict domain expertise. However, domain expertise is not a measurable attribute of reviewers. A close approximator might be the self-reported domain expertise but this comes with its own biases when compared with an actual -- but immeasurable -- domain knowledge. Evaluating one's ability to perform a task (in this case reviewing a proposal is called self efficacy). We can use our dataset to study the self efficacy of our participants before moving onto testing the \dt algorithm against the self-reported expertise. 

\begin{figure*}[ht!]
   \includegraphics[width=0.95\columnwidth]{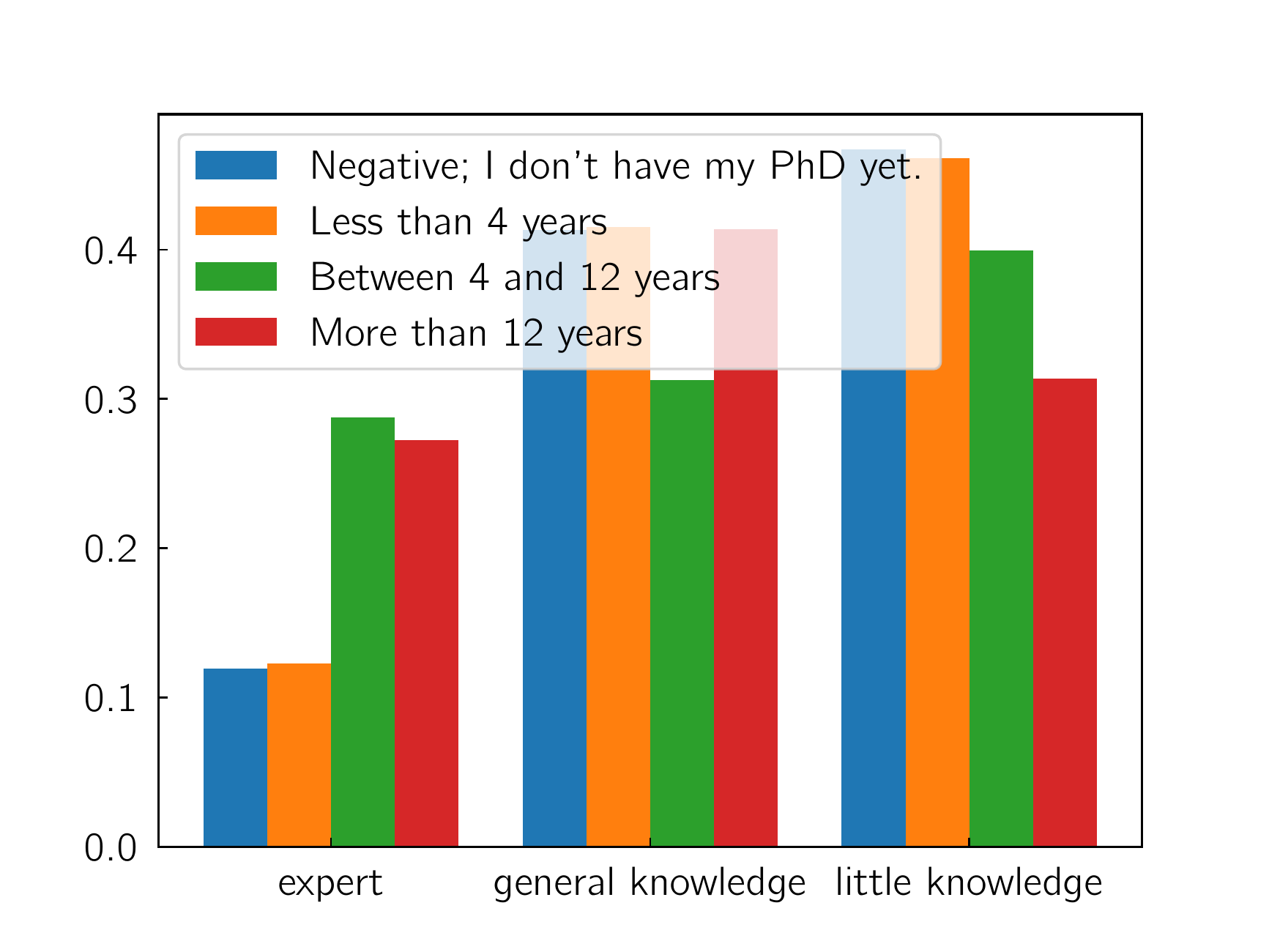}
   \includegraphics[width=0.95\columnwidth]{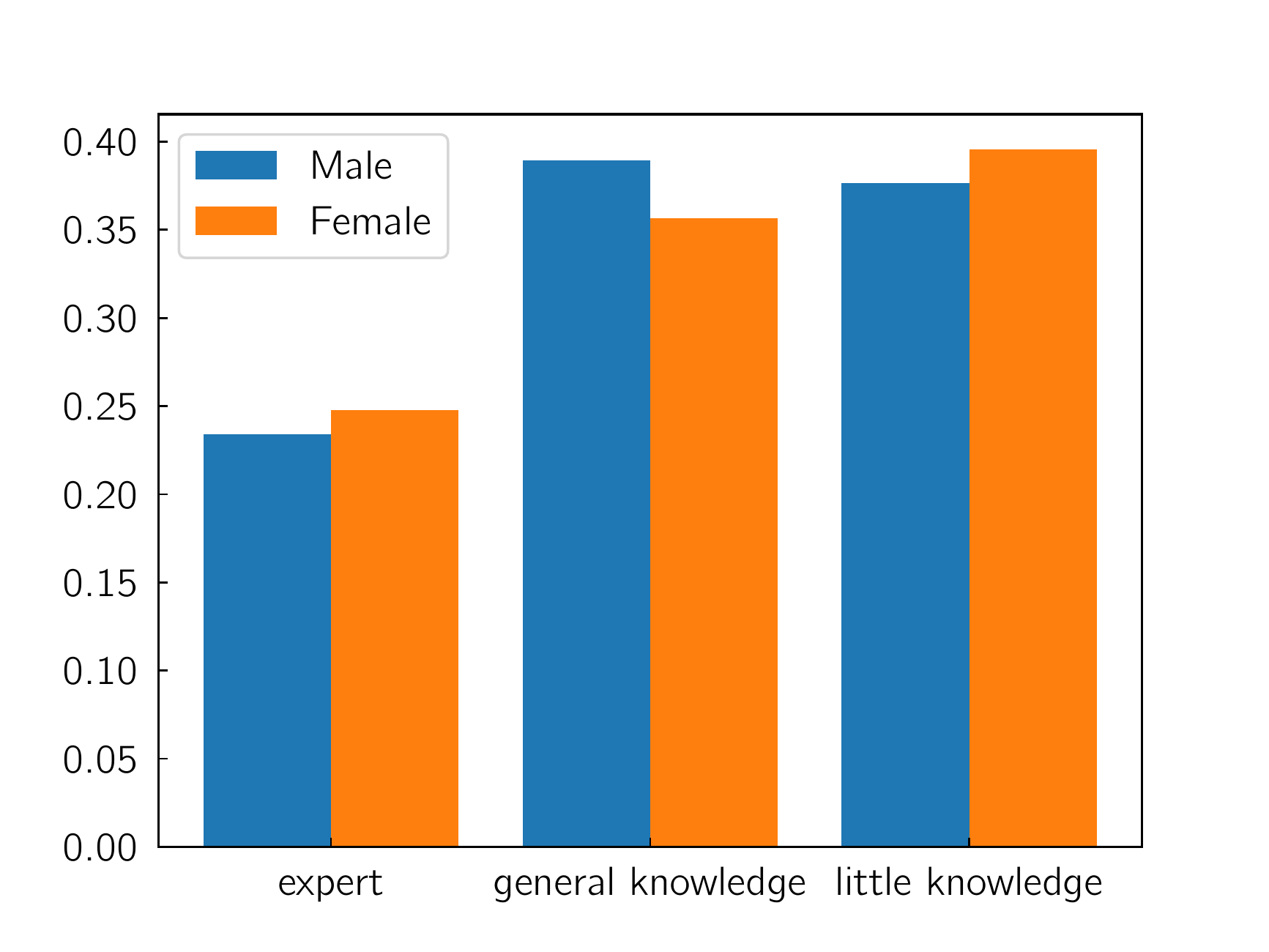}
   \caption{\label{fig:self_efficacy_sen_gender}\textbf{Left:} Self-reported domain knowledge given seniority of participants. \textbf{Right:} Self Efficacy for Male and Female reviewers.}

\end{figure*}

Seniority is expected to have an impact on expertise. Figure~\ref{fig:self_efficacy_sen_gender} shows that the self-reported domain expertise is significantly lower for junior scientists compared to senior scientists. Junior scientist, also often suggest that they have little to no knowledge about a field. This might suggests that self-reported knowledge is a useful proxy for actual knowledge. 

 \citet{huang2013gender} is a large meta-study of academic self-efficacy and shows that women tend to under-predict their performance in certain STEM fields. Figure~\ref{fig:self_efficacy_sen_gender} suggests that in the domain of astrophysics, and for post-grad individuals, the difference between the self-efficacy of men and women is relatively small.

%% file: expertise_influence_review.tex
\subsection{Expertise influence on Review}

An important question is whether expertise has an influence on the review process itself. 

Statistics shows that experts take tentatively less time to review proposals than non-experts (see supplementary material for statistics). 

\begin{figure*}[t]
   \includegraphics[width=\columnwidth]{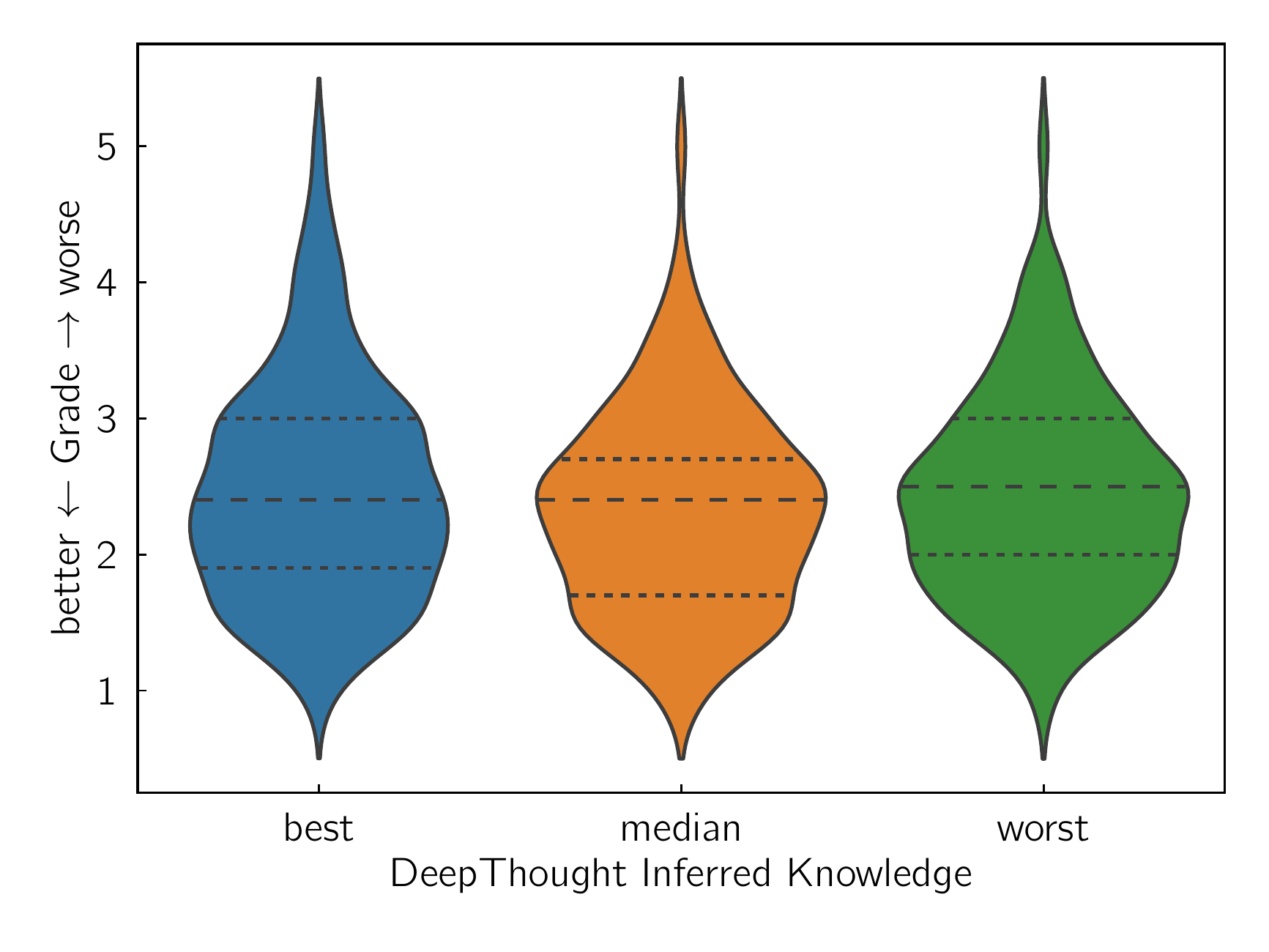}
   \includegraphics[width=\columnwidth]{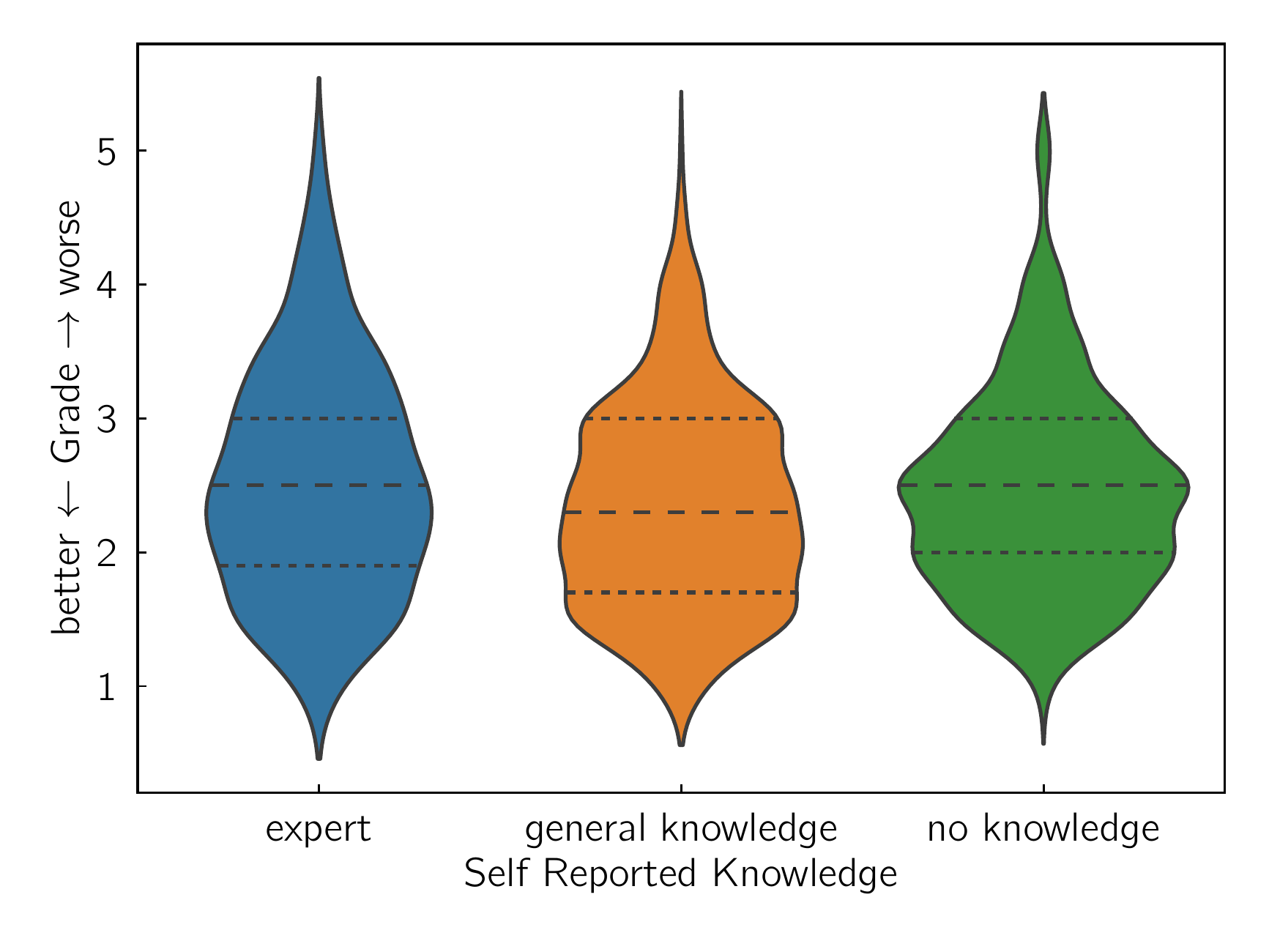}
   \caption{The distribution of grades for different groups of referees. The horizontal lines mark quartiles. \textbf{Left:} The distribution of grades for the different groups domain knowledge inferred by the \dt algorithm. \textbf{Right:} The distribution of grades for the different groups of self-reported domain knowledge.}
   \label{fig:dt_vs_ref_grade_violin}
\end{figure*}

We also study how expertise influences the grade distribution. Figure~\ref{fig:dt_vs_ref_grade_violin} shows that the difference in the shapes related to experts and referees with little to no domain knowledge is relatively small. The distributions show an extended tail towards poor grades, similarly to what is seen in Figure~4 of \citet{2018PASP..130h4501P} for the large OPC sample.

In their study on the effects of reviewer expertise on the evaluation of funding applications, \citet{2016PLoSO..1165147G} concluded that "reviewers with higher levels of self-assessed expertise tended to be harsher in their evaluations". That study was based on a sample of 1044 reviewers, i.e. 6 times larger than ours. Although our data may contain some indication of this trend, the differences are statistically marginal.

%% file: gender-statistics.tex
\subsection{Influence of gender on helpfulness of referee report}
\begin{figure}[ht!]
   \includegraphics[width=
   \columnwidth]{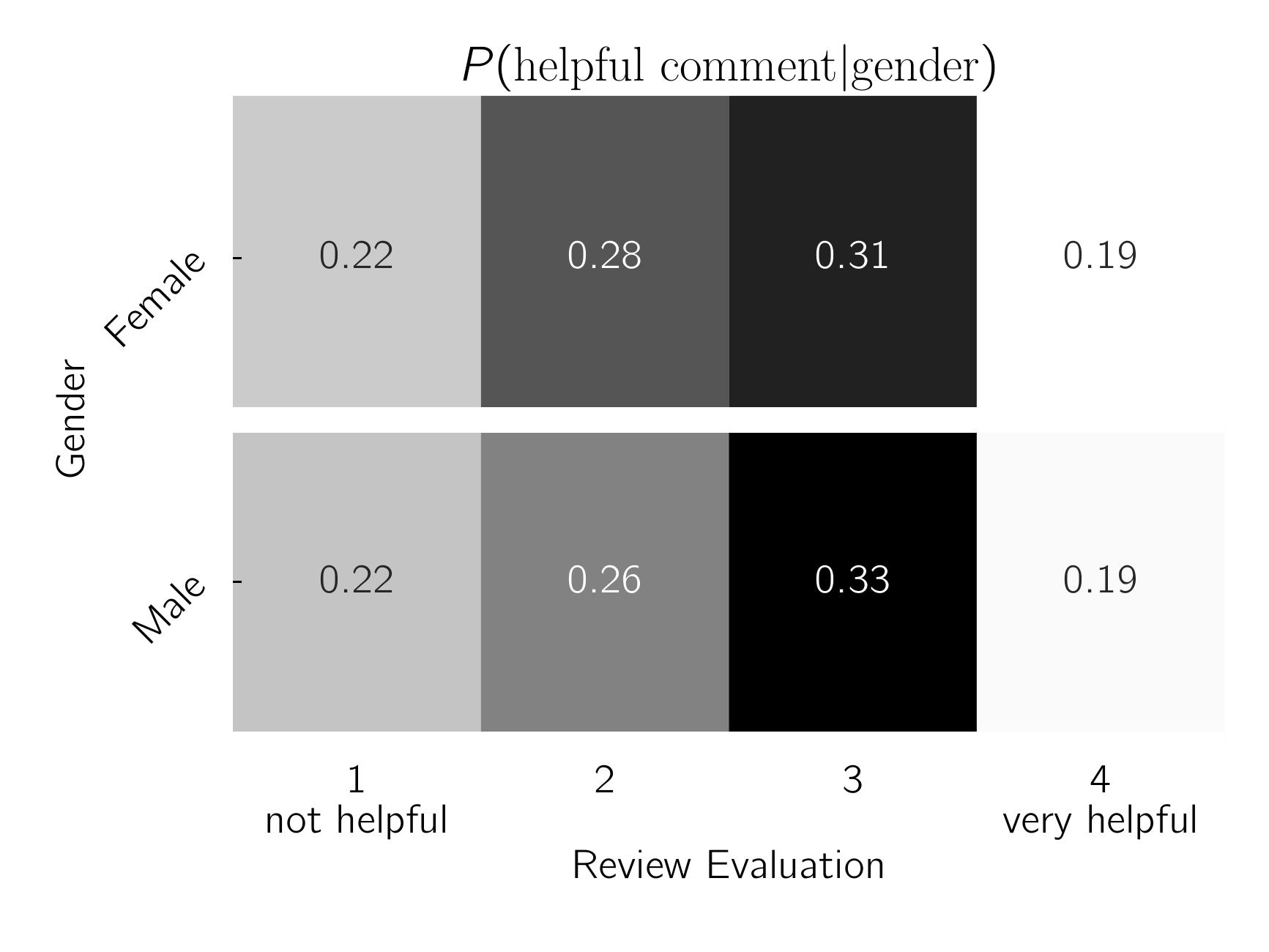}
   \caption{We show the conditional probability $P(\textrm{helpful comment}|\textrm{gender})$ for the various combinations of helpful comment rating and reported gender.}
   
   \label{fig:review_review_gender_vs_rating}
\end{figure}

We study if gender of the referee has an influence on the helpfulness of the comments in Figure~\ref{fig:review_review_gender_vs_rating} and find that there is no statistically significant difference between the helpfulness of comments by male and female referees. There were very few participants that used the option "prefer not to say" on the gender question (see Figure~\ref{fig:gender_distribution}). We thus did not report the helpfulness statistics on those participants as the statistical noise is too large to draw conclusions.

%% file: long_dpr_opc.tex
\subsection{Comparison of DeepThought Distributed Peer Review to the Observing Progr}

We will use the concept of the quartile agreement matrix (QAM) and related statistics as described in Section~3 of the main paper. We will present additional statistics such as individual referee comparison in this supplementary information.

We first  compute the agreement fractions between pairs of sub-sets of reviews within the DPR sample. In a first test, for each of the 172 proposals we extract a random pair of reviews, which are then used to produce two rank-lists, from which the QAM is derived. The process is then repeated a large number of times, leading to the construction of an average QAM and the standard deviation matrix. Since this is constructed for simple reviewer pairs, we will indicate it as the referee-referee (r-r) QAM. The result is presented in Table~\ref{tab:rrqam} for the calibrated grades (very similar values are obtained using the raw grades). The typical standard deviation of single realizations is 0.06, while the uncertainty (using poisson statistics;)on the average values is below 0.01.

\begin{table}
\caption{\label{tab:rrqam}Bootstrapped DPR r-r QAM.}
\tabcolsep 5.0mm
\centerline{
\begin{tabular}{c|cccc}
\hline
\hline
 1st referee.  & \multicolumn{4}{c}{2nd referee Quartile}\\
 quartile & 1 & 2 & 3 & 4\\
\hline
1 & 0.33 &  0.26 & 0.24 &  0.18\\
2 & 0.26 &  0.26 & 0.25 &  0.23\\
3 & 0.24 &  0.25 & 0.25 &  0.26\\
4 & 0.18 &  0.23 & 0.26 &  0.34\\
\hline\hline
\end{tabular}
}
\end{table}

The values are statistically indistinguishable from those reported in P18 (Table~2) for the OPC sample including $\sim$15,000 proposals. This shows that the average r-r agreement is identical in the two processes. On average, the ranking lists produced by two distinct referees have about 33\% of the proposals in common in their first and last quartiles. This corresponds to a Cohen's kappa coefficient $\kappa$=0.11 \citep{cohen1960}. In the central quartiles the intersection is compatible with a purely random selection. This extends to the mixed cases ($i \neq j$), with the exception of the extreme quartiles: the fraction of proposals ranked in the first quartile by one referee and in the fourth quartile by another referee is $\sim$18\%, which deviates in a statistically significant way from the random value ($\kappa=-$0.28). No meaningful difference is seen in the QAMs computed for the OPC-Emulate (OE; 60 proposals) and Deep-Thought (DT; 112 proposals) sub-samples.

As a further test, we have investigated the possible dependence on the scientific seniority level. In the feedback questionnaire we asked the participants to express it in terms of the years after PhD, specified within 4 groups (0=no PhD; 1=less than 4 years; 2=between 4 and 12; 3: more than 12). Of the 167 reviewers, 136 provided this information, which we used to divide the reviewers in two classes: junior (groups 0 and 1) and senior (groups 2 and 3). These classes roughly correspond to PhD students plus junior post-docs (37), and advanced post-docs plus senior scientists (99). We then computed the r-r QAM for the two classes. The first quartile terms are 0.22 and 0.32 for the two classes, respectively. At face value this indicates a larger agreement between senior reviewers. However, the small size of the junior class (37 people) produces a significant scatter, and therefore we do not attach too much confidence to this result.

We have defined the panel-panel (p-p) agreement fraction in the main text (see Section~3). For the OE and DT sub-samples yield statistically indistinguishable values. Finally, the numbers derived from the bootstrap procedure fully agree with the results presented in P18 for different panel sizes. For $N_r$=1, 2, and 3 the first quartile agreement fractions are 0.34, 0.37 and 0.41, respectively. These match (within the noise) the corresponding P18 values (Table~8): 0.33, 0.39 and 0.45.

The conclusion is that, in terms of self-consistency, the DPR review behaves in the same way as the pre-meeting OPC process.

\subsection{\label{sec:analysis.opcomp.agfrac}The DPR-OPC agreement fraction}     

As anticipated, the proposals used in the DPR experiment were also subject to the regular OPC review. This enables the comparison between the outcomes of the two selections, with the caveats outlined above about their inherent differences.

For a first test we used a bootstrap procedure in which, for each proposal included in the DPR, we randomly extracted one evaluation from the DPR (typically one out of 7) and one from the OPC (one out of 3), forming two ranking lists from which a QAM was computed. The operation was repeated a large number of times and the average and standard deviation matrices were constructed. This approach provides a direct indication of the DPR-OPC agreement at the r-r level, and overcomes the problem that the two reviews have a different number of evaluations per proposal (see below). The result is presented in Table~\ref{tab:dpropc1to1}. The typical standard deviation of single realizations from the average is 0.06.

\begin{table}
\caption{\label{tab:dpropc1to1}DPR-OPC (pre-meeting) r-r QAM.}
\tabcolsep 5.0mm
\centerline{
\begin{tabular}{c|cccc}
\hline
\hline
 DPR ref.  & \multicolumn{4}{c}{OPC referee quartile}\\
 quartile & 1 & 2 & 3 & 4\\
\hline
1 & 0.31 &  0.26 & 0.24 &  0.18\\
2 & 0.24 &  0.27 & 0.25 &  0.24\\
3 & 0.24 &  0.23 & 0.26 &  0.26\\
4 & 0.20 &  0.23 & 0.25 &  0.31\\
\hline\hline
\end{tabular}
}
\end{table}

This matrix is very similar to that derived within the DPR reviews (Table~\ref{tab:rrqam}), possibly indicating a DPR-OPC r-r agreement slightly lower than the corresponding DPR-DPR. A check run on the two sub-samples for the junior and senior DPR reviewers (according to the classification described in the previous section) has given statistically indistinguishable results.

As explained in the introduction, the proposals were reviewed by $N_r$=3 OPC referees in the pre-meeting phase. This constitutes a significant difference, in that the DPR ranking is typically based on $\sim$7 grades, whilst the pre-meeting OPC ranking rests on 3 grades only. With this caveat in mind, one can nevertheless compute the QAM for the two overall ranking lists.

 The result (computed using calibrated grades) is presented in Table~\ref{tab:qampre}. At face value, about 37\% of the proposals ranked in the 1st quartile by the DPR were ranked in the same quartile by the OPC, with a similar fraction for the bottom quartile. When looking at these values, one needs to consider that this is only one realization, which is affected by a large scatter, as one can deduce from the comparatively large fluctuations in the QAM. These are evident when comparing, for instance, to the average values obtained from the bootstrapping procedures described in the previous section; or considering that the average matrix is expected to be symmetric.

\begin{table}
\caption{\label{tab:qampre}DPR-OPC (pre-meeting) p-p QAM.}
\tabcolsep 5.0mm
\centerline{
\begin{tabular}{c|cccc}
\hline
\hline
 DPR  & \multicolumn{4}{c}{OPC quartile}\\
 quartile & 1 & 2 & 3 & 4\\
\hline
1 & 0.37 &  0.26 & 0.28 &  0.09\\
2 & 0.28 &  0.16 & 0.28 &  0.28\\
3 & 0.16 &  0.40 & 0.19 &  0.26\\
4 & 0.19 &  0.19 & 0.26 &  0.37\\
\hline\hline
\end{tabular}
}
\end{table}
     
To quantify the expected noise on a single panel-panel realization, we have run a set of simulations along the lines described in P18, which are based on the statistical description of the reviewers and on the True Grade Hypothesis (see P18; Section~7 therein). For the purposes of this analysis we modified the original code to allow the calculation of the QAM for pairs of panels with different sizes. We emphasize that the P18 code was designed to simulate the OPC process, i.e. it deals with panels reviewing the same set of proposals which, in turn, are reviewed by all referees in the panels. It therefore does not fully reproduce the DPR side, in which each proposal is reviewed by a virtually different panel in the \dt sub-sample. For this reason the results should be considered only as indicative. More sophisticated simulations will be presented elsewhere.

The outcome is presented in Table~\ref{tab:qampresim} (top), which also shows the standard deviation of the single realizations (bottom). For the DPR simulated panels we have used $N_r$=7, which is close to the actual average number of reviewers per proposal in the experiment. For the OPC we have set $N_r$=3.
The numerical calculations predict a top and bottom quartile agreement of about 50\%, while in the central quartiles this drops to about 30\%. 
The statistical significance of the deviations of the observed QAM from the average QAM can be quantified using the predicted dispersion. For the first and fourth quartile, the observed value (0.37) differs at the 1.3$\sigma$-level from the average value. For the central quartiles the difference is at the $\sim$1.5$\sigma$-level. 

Therefore, although lower than expected on average, the observed DPR-OPC agreement is statistically consistent with that expected from the statistical description of the pre-meeting OPC process (P18).

\begin{table}
\caption{\label{tab:qampresim}Simulated p-p QAM (top) and uncertainty (bottom) for two panels with $N_r$=3 and $N_r$=7.}
\tabcolsep 5.4mm
\centerline{
\begin{tabular}{c|cccc}
\hline
\hline
 1st pan.  & \multicolumn{4}{c}{2nd panel quartile}\\
 quartile & 1 & 2 & 3 & 4\\
\hline
1 & 0.51 &  0.28 & 0.16 &  0.06\\
2 & 0.28 &  0.31 & 0.27 &  0.15\\
3 & 0.16 &  0.27 & 0.30 &  0.28\\
4 & 0.06 &  0.15 & 0.28 &  0.51\\
\hline
1 & 0.11 &  0.10 & 0.09 &  0.06\\
2 & 0.10 &  0.10 & 0.10 &  0.09\\
3 & 0.08 &  0.09 & 0.10 &  0.10\\
4 & 0.06 &  0.08 & 0.10 &  0.11\\
\hline\hline
\end{tabular}
}
\end{table}

One important aspect to remark is that, given the large noise inherent to the process, a much larger data-set (or more realizations of the experiment) would be required to reach a statistical significance sufficiently high to make robust claims about systematic deviations.

\subsection{\label{sec:analysis.opcomp.postopc}Comparison with the post-meeting outcome}

The fact that in the real OPC process there is a face-to-face meeting constitutes the most pronounced difference between the two review schemes. In the meeting, the opinions of single reviewers are changed by the discussion, so that the grades attributed by the single referees are not completely independent from each other (as opposed to the pre-meeting phase, in which the possible correlation should depend only on the intrinsic merits of the proposal).

The effects of the meeting can be quantified in terms of the quartile agreement fractions between the pre- and post-meeting outcomes, as outlined in \citet[][hereafter P19]{Patat2019}. Based on the P18 sample, P19 concludes that the change is significant: on average, only 75\% of the proposals ranked in the top quartile before the meeting remain in the top quartile after the discussion (about 20\% are demoted to the second quartile, and 5\% to the third quartile). P19 characterizes this effect introducing the Quartile Migration Matrix (QMM). For the specific case of P103, the QMM is reported in Table~\ref{tab:qmm} for the 683 proposals that passed the triage (top), and for the sub-set of the DPR experiment (bottom).\footnote{Of the initial 172 proposals included in the DPR sample, 36 were triaged out in the OPC process.} 

\begin{table}
\caption{OPC Quartile Migration Matrix}
\label{tab:qmm}
\tabcolsep 4.5mm
\centerline{
\begin{tabular}{c|cccc}
\hline
\hline
  OPC pre-m.  & \multicolumn{4}{c}{OPC post-meeting quartile (N=683)}\\
  quartile & 1 & 2 & 3 & 4\\
\hline
1 & 0.64 &  0.25 & 0.06 &  0.05\\
2 & 0.26 &  0.37 & 0.28 &  0.08\\
3 & 0.08 &  0.26 & 0.39 &  0.26\\
4 & 0.02 &  0.11 & 0.25 &  0.62\\
\hline 
  & \multicolumn{4}{c}{DPR sub-set (N=136)}\\
\hline
1 & 0.56 &  0.32 & 0.12 &  0.00\\
2 & 0.32 &  0.32 & 0.29 &  0.06\\
3 & 0.12 &  0.26 & 0.38 &  0.24\\
4 & 0.00 &  0.09 & 0.21 &  0.71\\
\hline\hline
\end{tabular}
}
\end{table}

As anticipated, the effect of the discussion is very pronounced. In the light of these facts one can finally inspect the QAM between the DPR and the final outcome of the OPC process. This is presented in Table~\ref{tab:qamfinal}. With the only possible exception of $M_{4,4}$, which indicates a relatively marked agreement for the proposals in the bottom quartile, the two reviews appear to be almost completely uncorrelated. By means of simple Monte-Carlo calculations one can show that for two fully aleatory panels the standard deviation of a single realization around the average value (0.25) is 0.10. Therefore, the majority of the $M_{i,j}$ elements in Table~\ref{tab:qamfinal} are consistent with a stochastic process at the 1$\sigma$ level.

\begin{table}
\caption{DPR vs. OPC post-meeting QAM}
\label{tab:qamfinal}
\tabcolsep 5.1mm
\centerline{
\begin{tabular}{c|cccc}
\hline
\hline
  DPR  & \multicolumn{4}{c}{OPC post-meeting quartile (N=136)}\\
quartile & 1 & 2 & 3 & 4\\
\hline
1 & 0.26 &  0.38 & 0.24 &  0.12\\
2 & 0.24 &  0.35 & 0.24 &  0.18\\
3 & 0.32 &  0.12 & 0.29 &  0.26\\
4 & 0.19 &  0.15 & 0.24 &  0.44\\
\hline \hline
\end{tabular}
}
\end{table}

The main conclusion is that while the pre-meeting agreement is consistent with the DPR and OPC reviewers behaving in a very similar way (in terms of r-r and p-p agreements), the face-to-face meeting has the effect of significantly increasing the discrepancy between the two processes.
We remark that the sample is relatively small, and therefore the results are significantly affected by noise.

%% file: deepthought_dpr1_si_appendix.tex
\begin{appendix}
    \section{Feedback questionnaire}
    \label{sec:feedback_questions}
    The following is the multiple choice questions that were presented to the participants in the questionnaire at the end of the experiment. The bold numbers after the multiple choice answers are the number of the participants that selected this answer. 
    \input{feedback_q_overview.tex}

    \section{Feedback questions statistics and overview}
    \label{sec:feedback_statistics}
    Statistics figures of the feedback overview.
    \input{feedback_overview.tex}

\end{appendix}

%% file: feedback_q_overview.tex
\subsection{What is your gender?}
\begin{itemize}
\item Male \textbf{96}
\item Female \textbf{46}
\item Prefer not to say \textbf{5}
\item (no answer given) \textbf{0}
\end{itemize}

\subsection{How many years are you after your PhD?}
\begin{itemize}
\item Negative; I don't have my PhD yet. \textbf{12}
\item Less than 4 years \textbf{27}
\item Between 4 and 12 years \textbf{51}
\item More than 12 years \textbf{57}
\item (no answer given) \textbf{0}
\end{itemize}

\subsection{How many proposals have you submitted to ESO as PI in the past?}
\begin{itemize}
\item None; this is my first proposal as PI. \textbf{0}
\item Less than 3 proposals \textbf{14}
\item Between 3 and 10 proposals \textbf{55}
\item More than 10 proposals \textbf{78}
\item (no answer given) \textbf{0}
\end{itemize}

\subsection{How often have you been a panel member of the ESO Observing Programmes Committee (OPC)?}
\begin{itemize}
\item Never; I have not yet been on the OPC. \textbf{112}
\item I am currently and OPC panel member. \textbf{3}
\item I served as OPC panel member for one term. \textbf{12}
\item I served more than one term on the OPC. \textbf{20}
\item (no answer given) \textbf{0}
\end{itemize}

\subsection{How easy was it to navigate and use the interface to review the proposals?}
\begin{itemize}
\item Easy, no problems in usage \textbf{117}
\item Mostly easy; some improvements possible \textbf{28}
\item Rather difficult; some improvements needed \textbf{2}
\item Difficult; several problems in usage \textbf{0}
\item (no answer given) \textbf{0}
\end{itemize}

\subsection{How much time did you spend, on average, per proposal (including writing comments)?}
\begin{itemize}
\item Less than 15 minutes \textbf{0}
\item Between 15 and 30 minutes \textbf{41}
\item Between 30 and 45 minutes \textbf{62}
\item Over 45 minutes \textbf{44}
\item (no answer given) \textbf{0}
\end{itemize}

\subsection{Was the time spent for proposals for which you are an expert versus a non-expert different?}
\begin{itemize}
\item More time on proposals for which I am an expert \textbf{15}
\item About the same \textbf{62}
\item Less time on proposals for which I am an expert \textbf{69}
\item (no answer given) \textbf{1}
\end{itemize}

\subsection{How appropriate were the assessment criteria to evaluate the proposals?}
\begin{itemize}
\item Fully appropriate; the criteria were clear and easy to apply \textbf{53}
\item Mostly appropriate; the criteria were clear but not easy to apply \textbf{73}
\item Somewhat appropriate; the criteria can be clarified but were applicable \textbf{18}
\item Not appropriate; the criteria should be changed \textbf{2}
\item (no answer given) \textbf{1}
\end{itemize}

\subsection{How satisfactorily were you able to evaluate the proposals for which you were a non-expert?}
\begin{itemize}
\item Fully; I could evaluate well and fairly as a non-expert \textbf{3}
\item Mostly; I sometimes missed the expertise but was still able to evaluate fairly \textbf{53}
\item Somewhat; I struggled and might not always have been able to evaluate fairly \textbf{75}
\item Not satisfactorily; I might have unintentionally provided an unfair evaluation \textbf{16}
\item (no answer given) \textbf{0}
\end{itemize}

\subsection{How useful were the comments that you received on your proposal?}
\begin{itemize}
\item Fully; overall the comments will allow me to improve my proposed project \textbf{17}
\item Mostly; several comments will help me to strengthen my proposed project \textbf{70}
\item Somewhat; some comments might help me to strengthen my proposed project \textbf{53}
\item Not useful; the comments will not help me to improve my proposed project \textbf{6}
\item (no answer given) \textbf{1}
\end{itemize}

\subsection{How clear and appropriate were the comments that you received on your proposal?}
\begin{itemize}
\item Fully; overall the comments were clear and appropriate \textbf{36}
\item Mostly; some comments were not fully clear but mostly appropriate \textbf{72}
\item Somewhat; several comments were unclear or not fully appropriate \textbf{36}
\item Not; several comments were unclear and inappropriate \textbf{3}
\item (no answer given) \textbf{0}
\end{itemize}

\subsection{How do these raw comments compare to the edited comments from the OPC in the past?}
\begin{itemize}
\item These raw comments were better; the OPC edited comments were less helpful. \textbf{62}
\item Of similar quality \textbf{66}
\item These raw comments were less helpful; the OPC edited comments were better. \textbf{18}
\item Not applicable; I have not received OPC edited comments before. \textbf{1}
\item (no answer given) \textbf{0}
\end{itemize}

\subsection{For which types of proposals do you think distributed peer review would be beneficial? (multiple answers possible)}
\begin{itemize}
\item Short proposals, with total requested time less than 20 hours \textbf{0}
\item Regular proposals, with total requested time between 1 and 100 hours \textbf{0}
\item Large proposals, with total requested time more than 100 hours \textbf{0}
\item None \textbf{0}
\item (no answer given) \textbf{4}
\end{itemize}

\subsection{The ESO Time Allocation Working Group recommended to introduce—in addition to the existing calls for proposals—a 'Fast Track' channel with a fraction of the time allocated through (two) monthly calls. If implemented, would distributed peer review be appropriate for this time allocation? (multiple answers possible)}
\begin{itemize}
\item I would support distributed peer review being used for the 'Fast Track' calls. \textbf{105}
\item I have concerns with distributed peer review being used for the 'Fast Track' calls. \textbf{23}
\item I am indifferent. \textbf{19}
\item (no answer given) \textbf{0}
\end{itemize}

\subsection{Securing confidentiality in the process:}
\begin{itemize}
\item I am neither more nor less concerned about confidentiality issues in the DPR process than in the OPC process. \textbf{73}
\item I am more concerned about confidentiality issues in the DPR process than in the OPC process. \textbf{42}
\item I am less concerned about confidentiality issues in the DPR process than in the OPC process. \textbf{8}
\item I have no strong opinion on this point. \textbf{24}
\item (no answer given) \textbf{0}
\end{itemize}

\subsection{External expertise in the review process:}
\begin{itemize}
\item I think that the pool of eight peer reviewers (including some non-experts) are sufficient to assess the proposals. \textbf{43}
\item I think that the pool of eight peer reviewers are sufficient to assess the proposals, provided that they have some expertise in the field. \textbf{73}
\item I think that external experts (non-ESO facility users) are critical to the assessment of the proposals and must be added as reviewers. \textbf{22}
\item I have no strong opinion on this point. \textbf{8}
\item (no answer given) \textbf{1}
\end{itemize}

\subsection{Robustness of the review process against any biases:}
\begin{itemize}
\item I think that DPR review process is as robust against biases as the OPC review process. \textbf{38}
\item I think that DPR review process is more robust against biases than the OPC review process. \textbf{39}
\item I think that DPR review process is less robust against biases than the OPC review process. \textbf{25}
\item I have no strong opinion on this point. \textbf{43}
\item (no answer given) \textbf{2}
\end{itemize}

%% file: feedback_overview.tex

\begin{figure}[p]
    \includegraphics[width=0.5\textwidth]{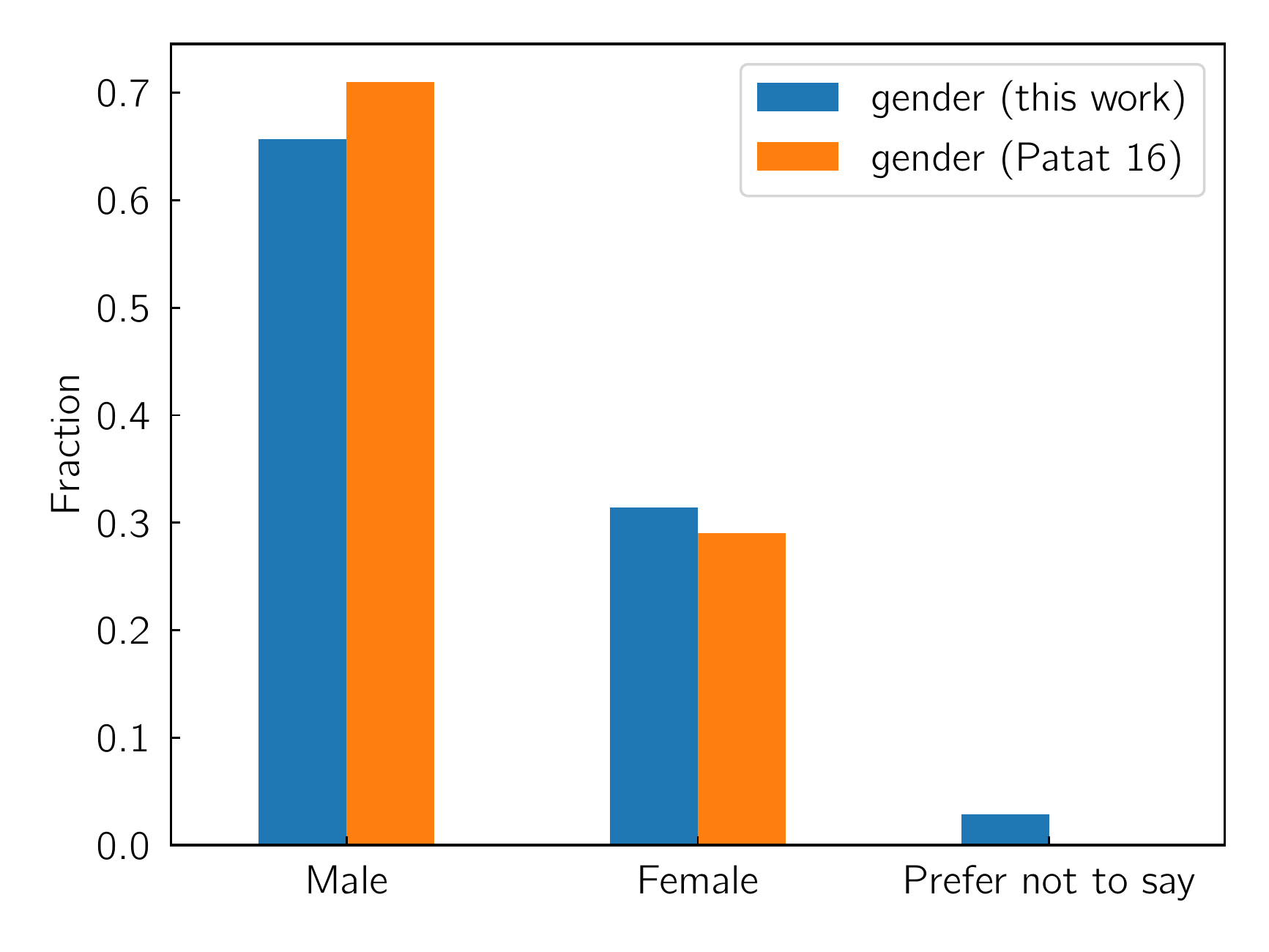}
    \caption{The referee was asked the question \textit{What is your gender?}}
    \label{fig:gender}
\end{figure}


\begin{figure}[p]
    \includegraphics[width=0.5\textwidth]{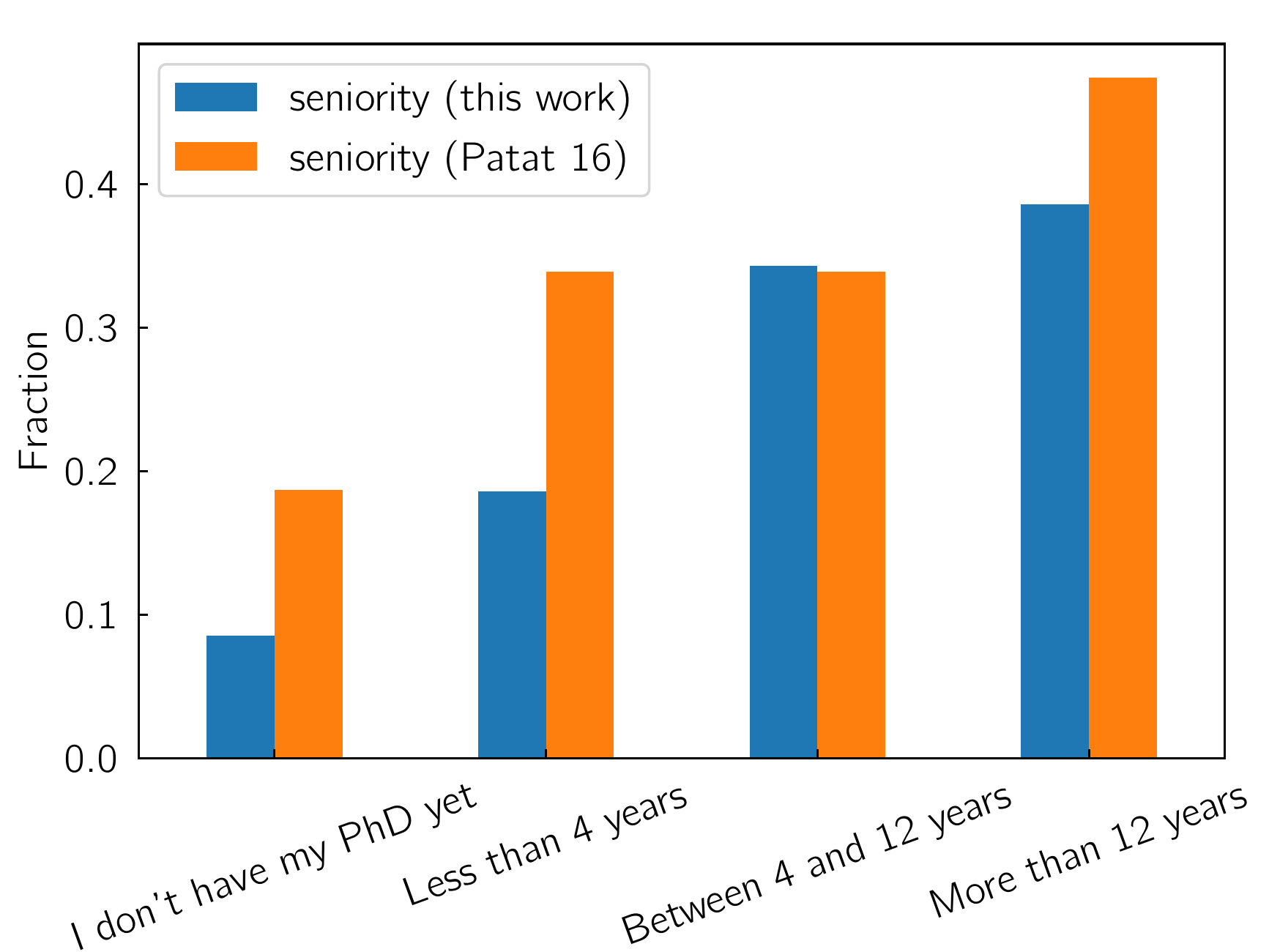}
    \caption{The referee was asked the question \textit{How many years are you after your PhD?}}
    \label{fig:years_past_phd}
\end{figure}


\begin{figure}[p]
    \includegraphics[width=0.5\textwidth]{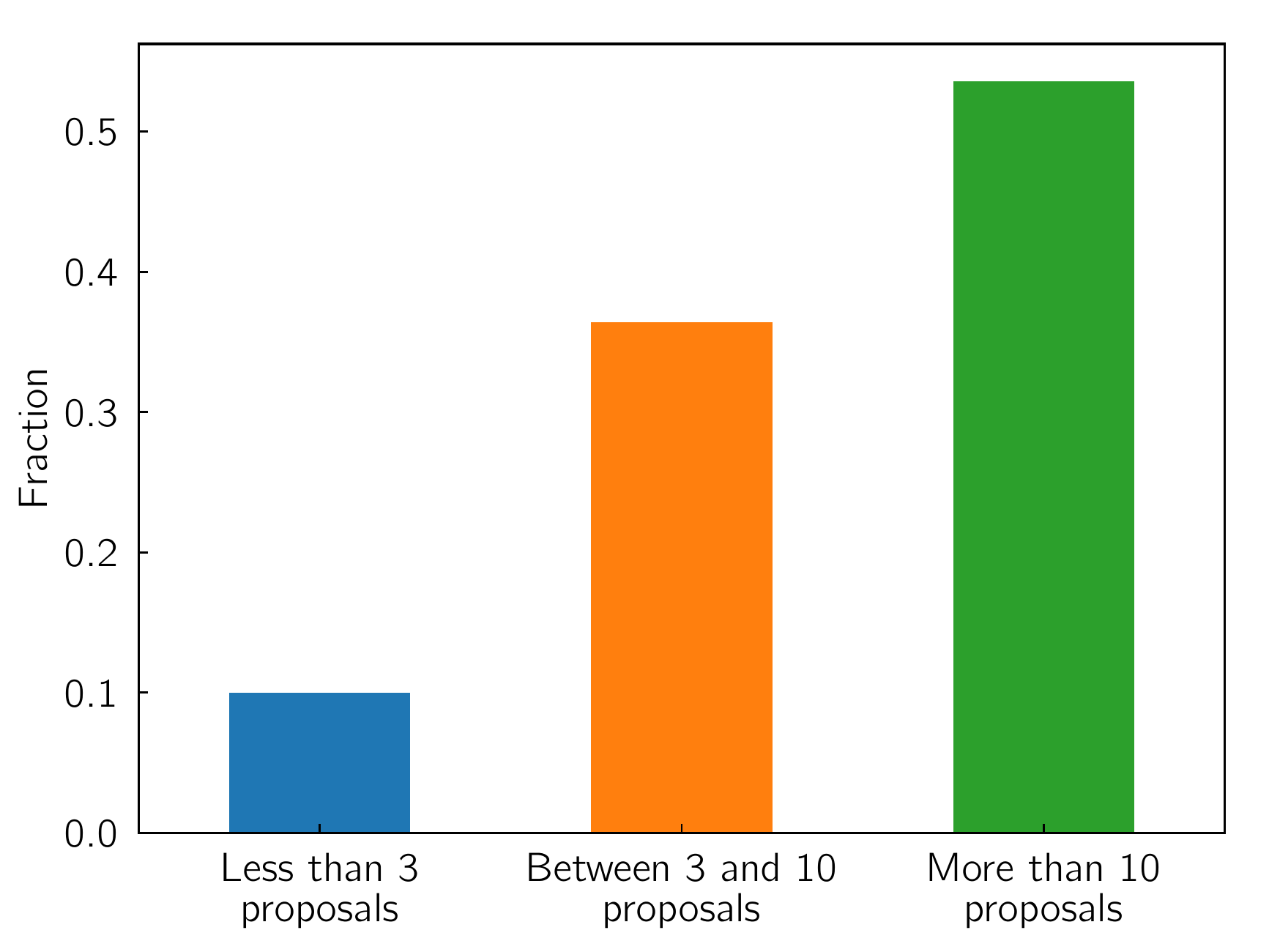}
    \caption{The referee was asked the question \textit{How many proposals have you submitted to ESO as PI in the past?}}
    \label{fig:prop_submitted_to_eso}
\end{figure}


\begin{figure}[p]
    \includegraphics[width=0.5\textwidth]{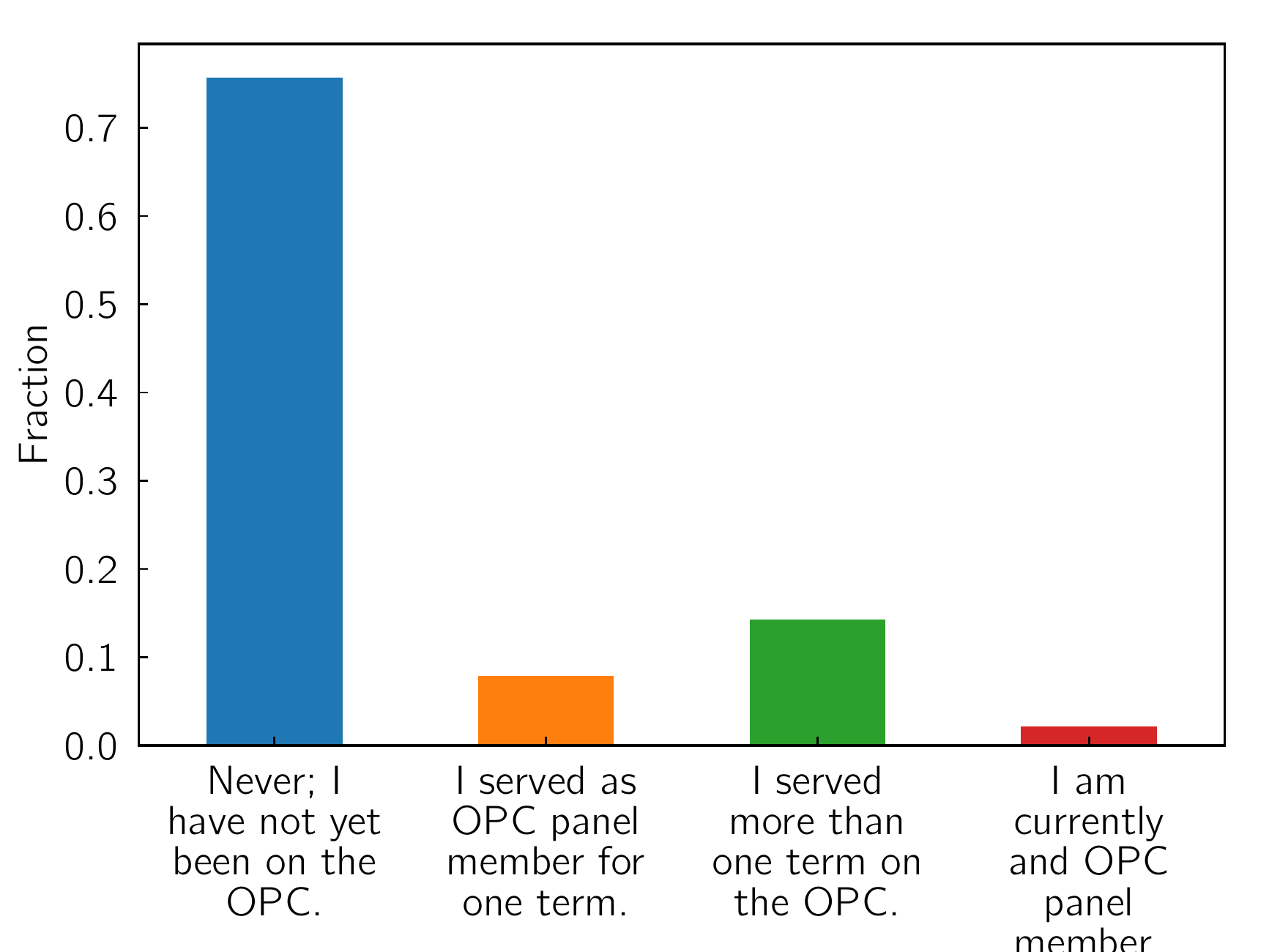}
    \caption{The referee was asked the question \textit{How often have you been a panel member of the ESO Observing Programmes Committee (OPC)?}}
    \label{fig:eso_opc_served}
\end{figure}


\begin{figure}[p]
    \includegraphics[width=0.5\textwidth]{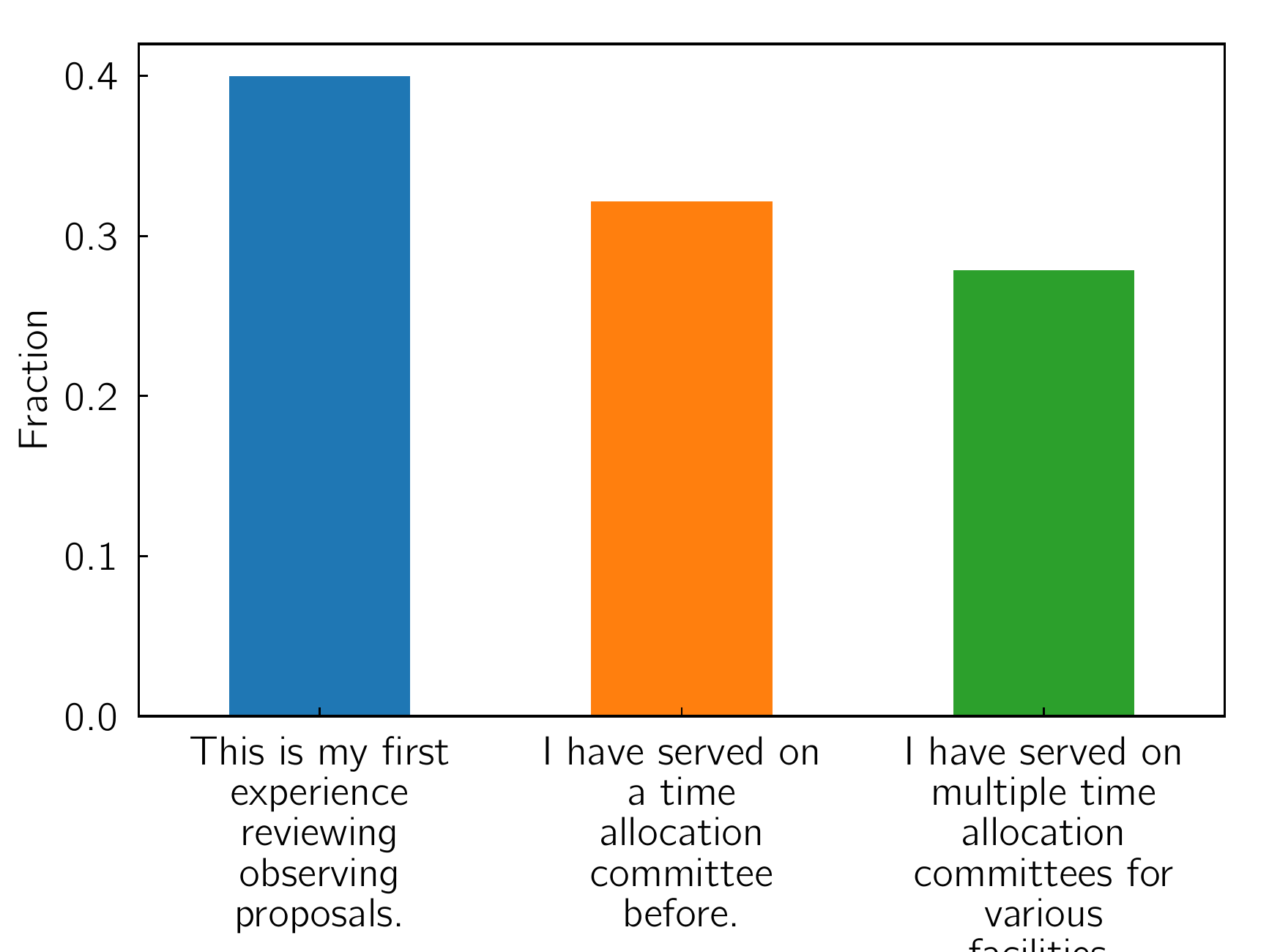}
    \caption{The referee was asked the question \textit{Have you served on another time allocation committee?}}
    \label{fig:other_tac_served}
\end{figure}


\begin{figure}[p]
    \includegraphics[width=0.5\textwidth]{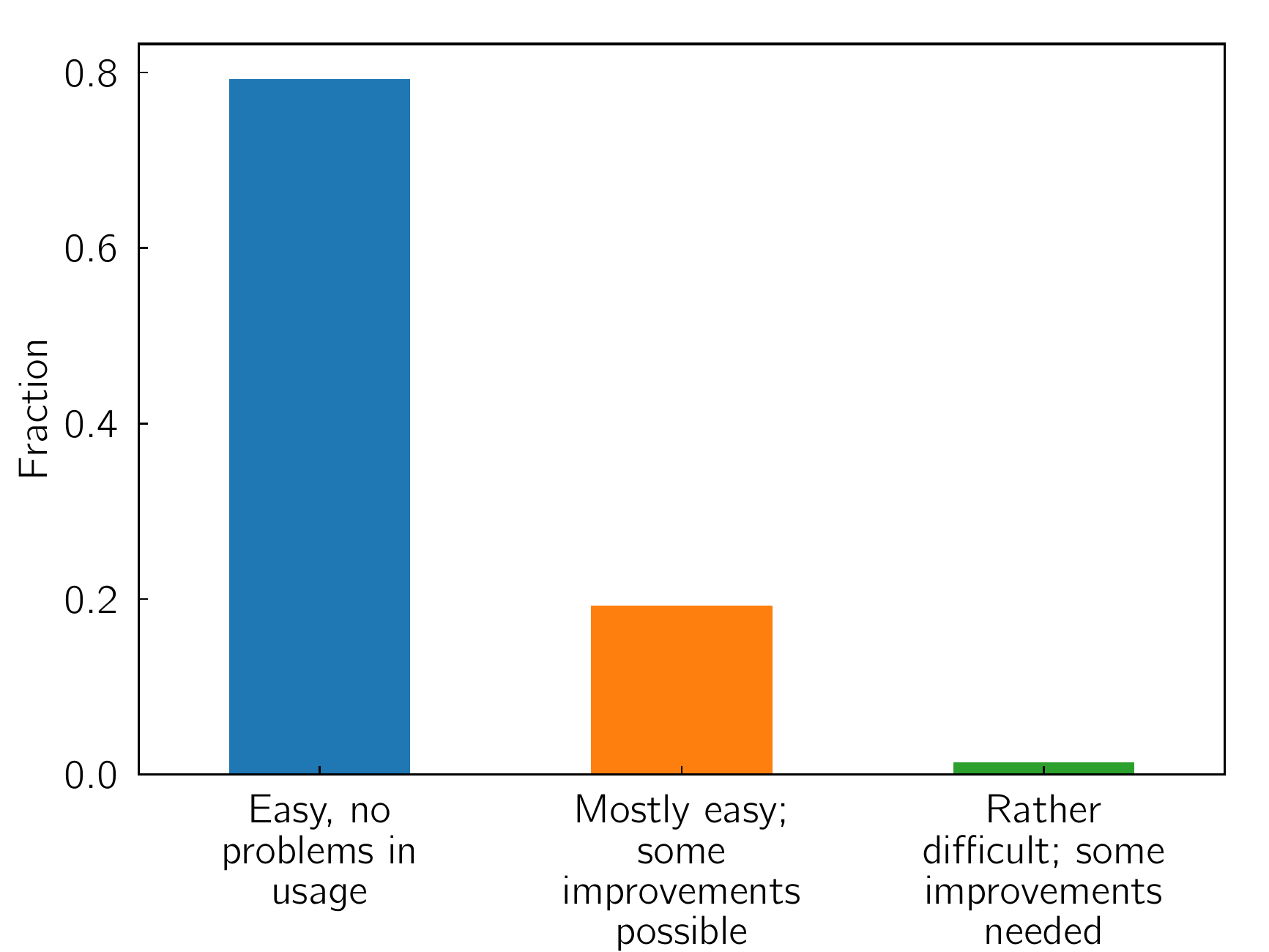}
    \caption{The referee was asked the question \textit{How easy was it to navigate and use the interface to review the proposals?}}
    \label{fig:ease_of_use}
\end{figure}


\begin{figure}[p]
    \includegraphics[width=0.5\textwidth]{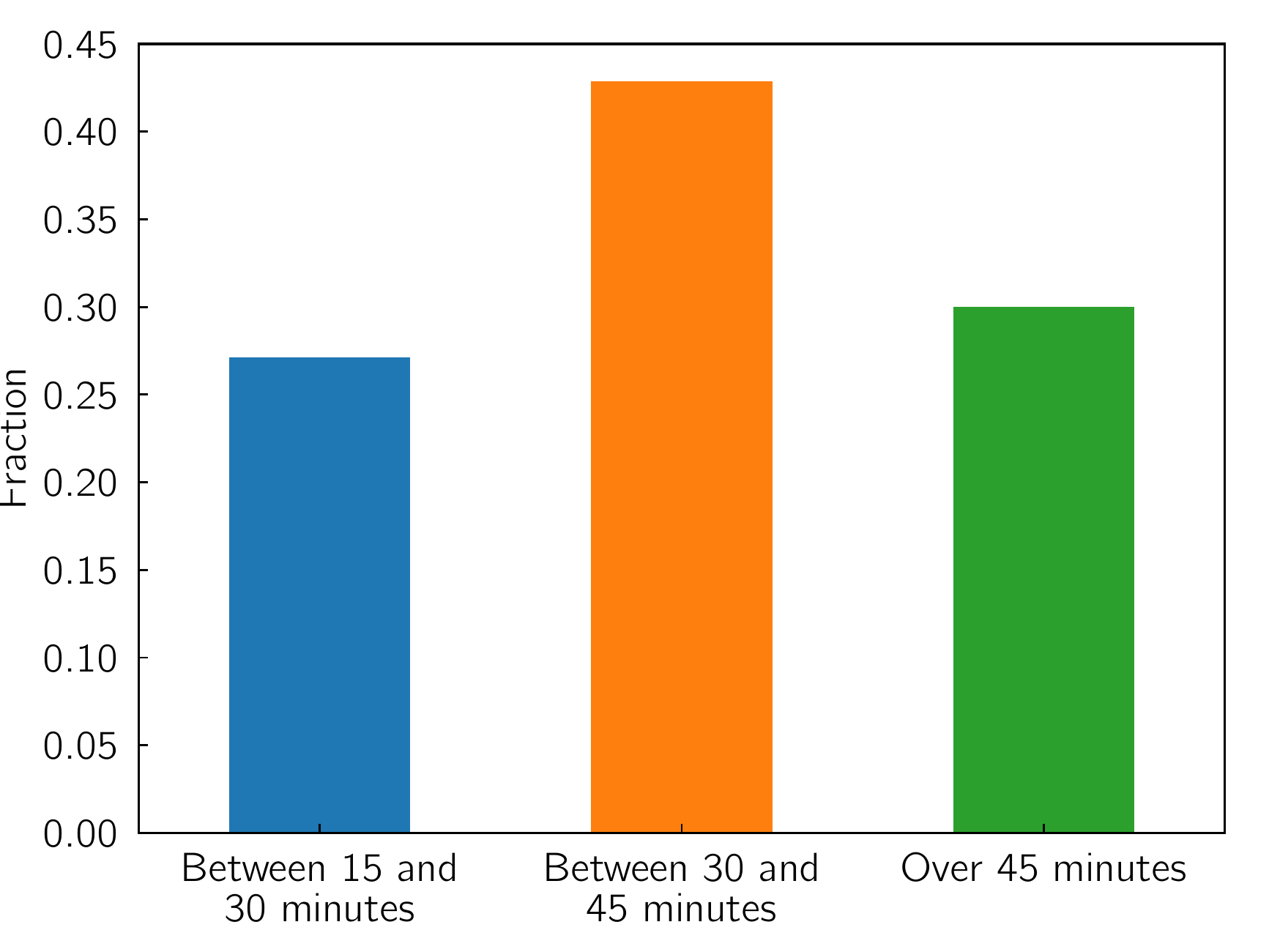}
    \caption{The referee was asked the question \textit{How much time did you spend, on average, per proposal (including writing comments)?}}
    \label{fig:time_spent}
\end{figure}


\begin{figure}[p]
    \includegraphics[width=0.5\textwidth]{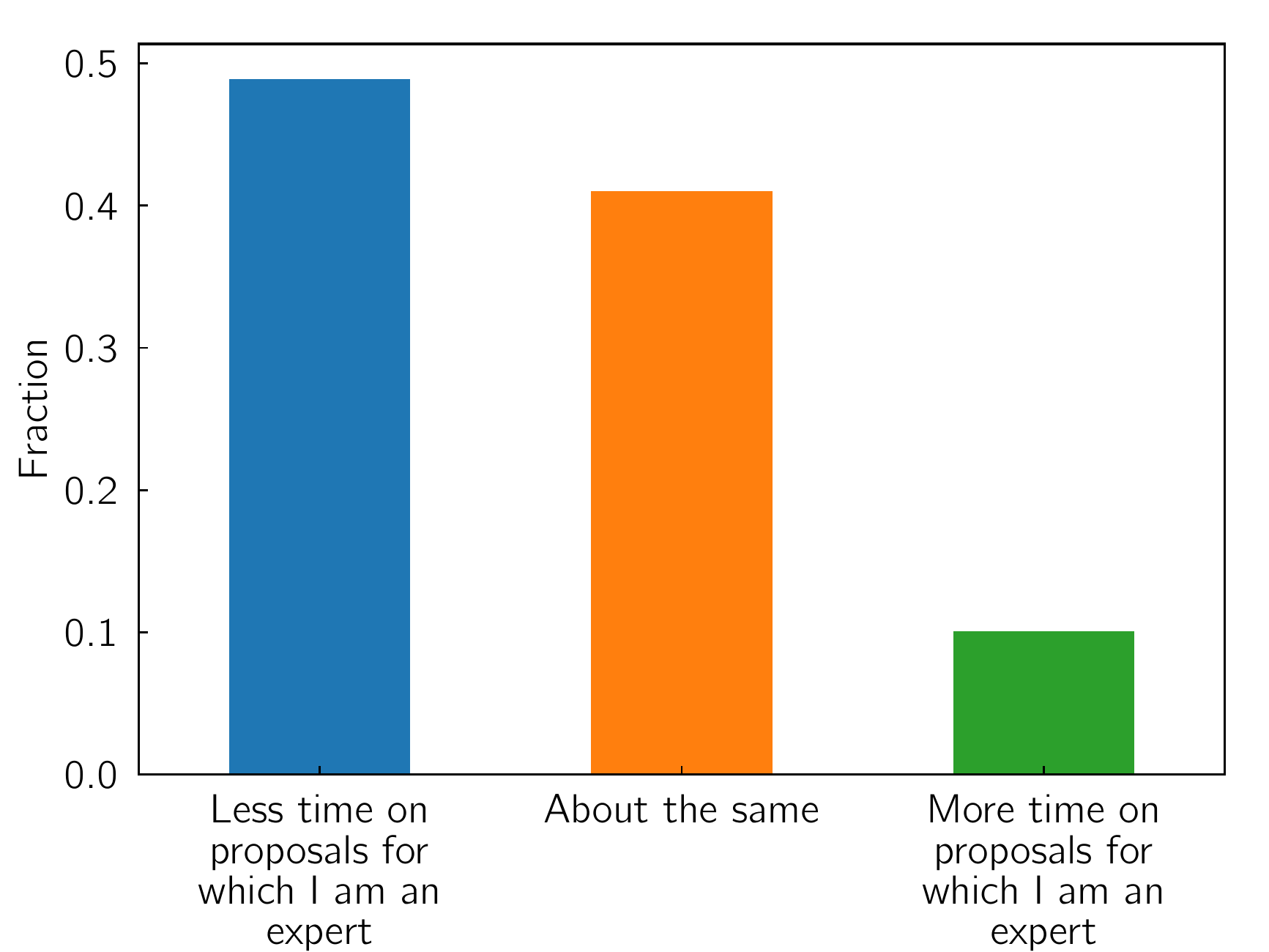}
    \caption{The referee was asked the question \textit{Was the time spent for proposals for which you are an expert versus a non-expert different?}}
    \label{fig:time_spent_expertise}
\end{figure}


\begin{figure}[p]
    \includegraphics[width=0.5\textwidth]{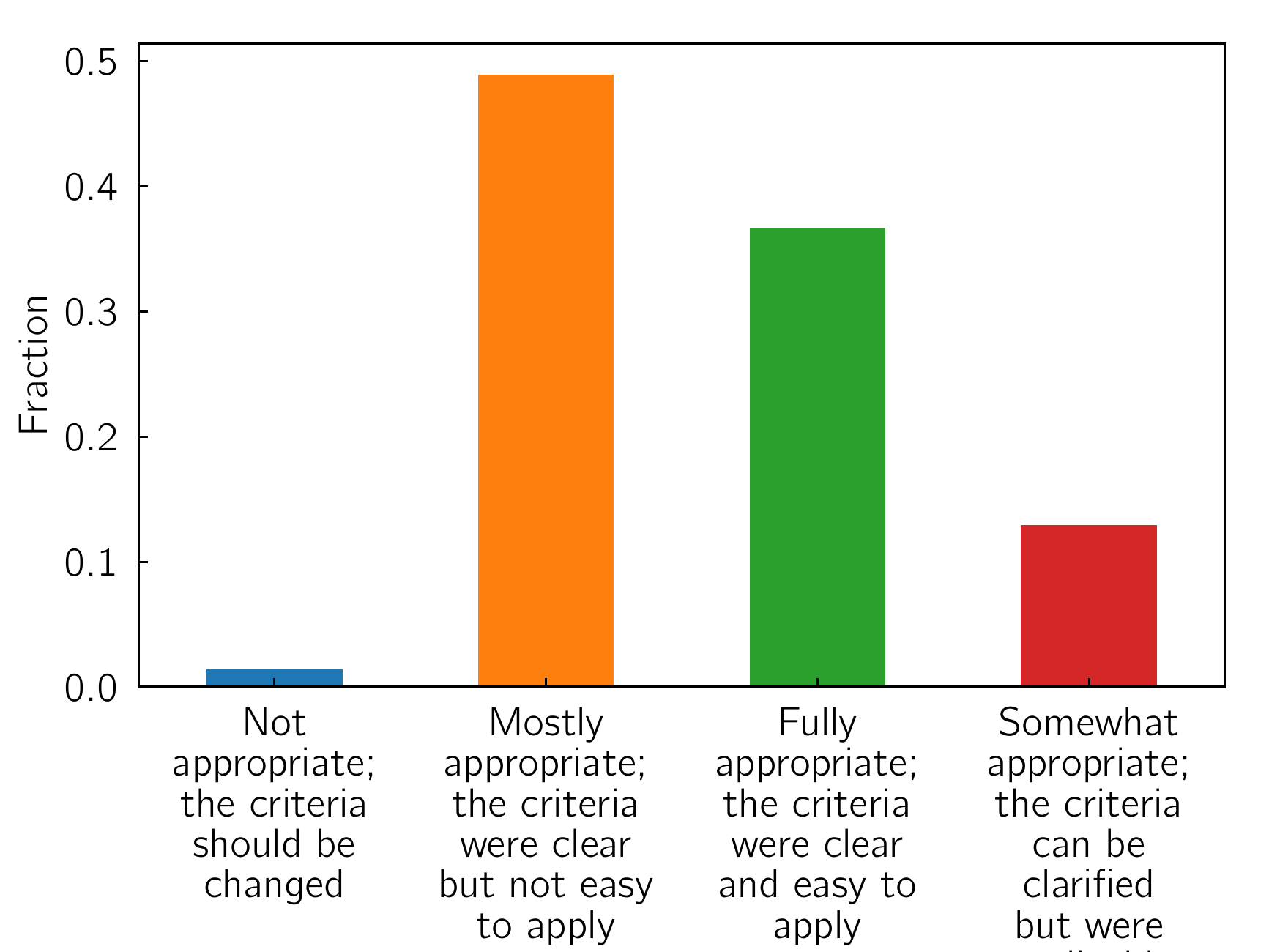}
    \caption{The referee was asked the question \textit{How appropriate were the assessment criteria to evaluate the proposals?}}
    \label{fig:appropriate_criteria}
\end{figure}


\begin{figure}[p]
    \includegraphics[width=0.5\textwidth]{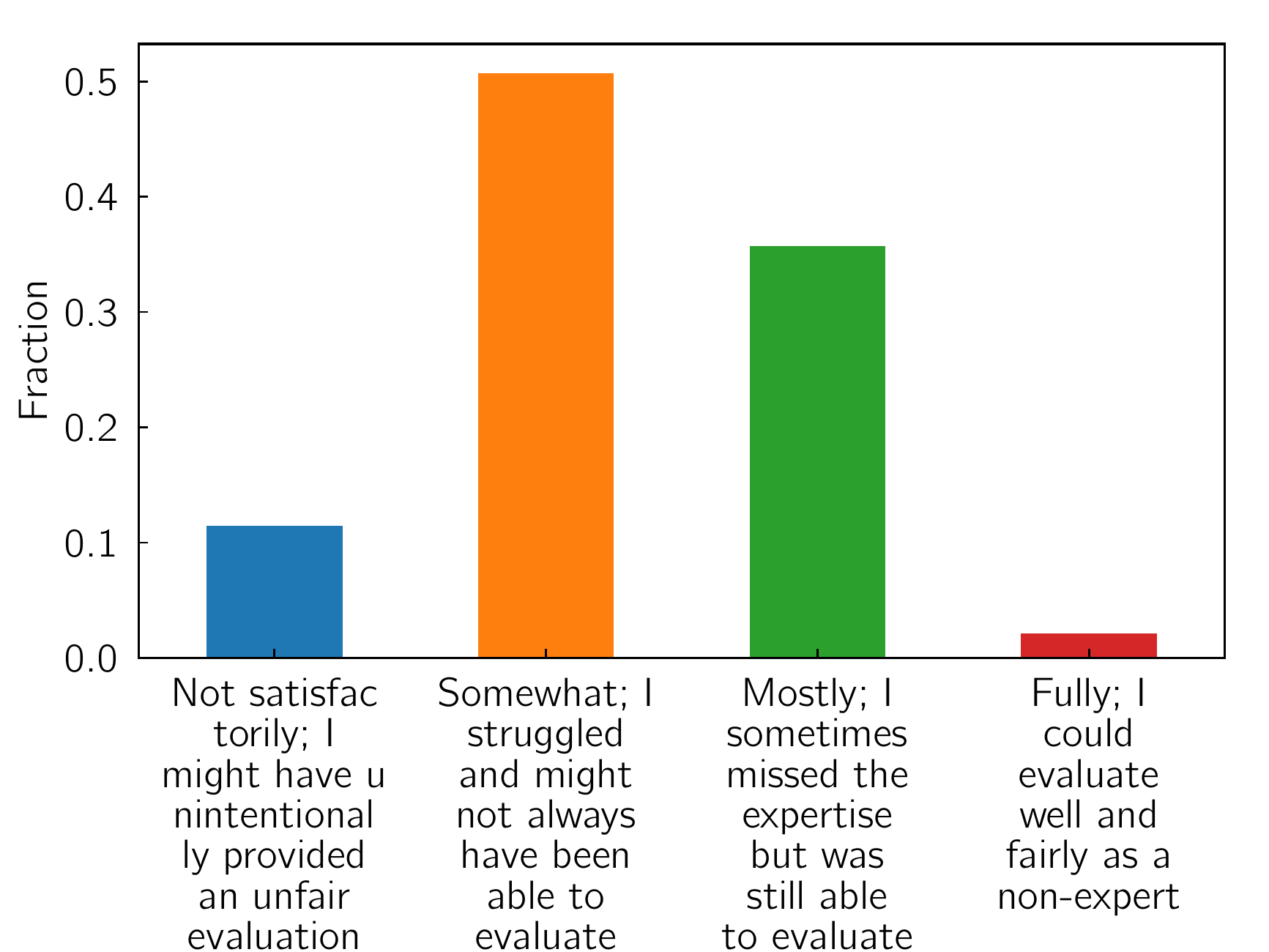}
    \caption{The referee was asked the question \textit{How satisfactorily were you able to evaluate the proposals for which you were a non-expert?}}
    \label{fig:non_expert_review_satisfaction}
\end{figure}


\begin{figure}[p]
    \includegraphics[width=0.5\textwidth]{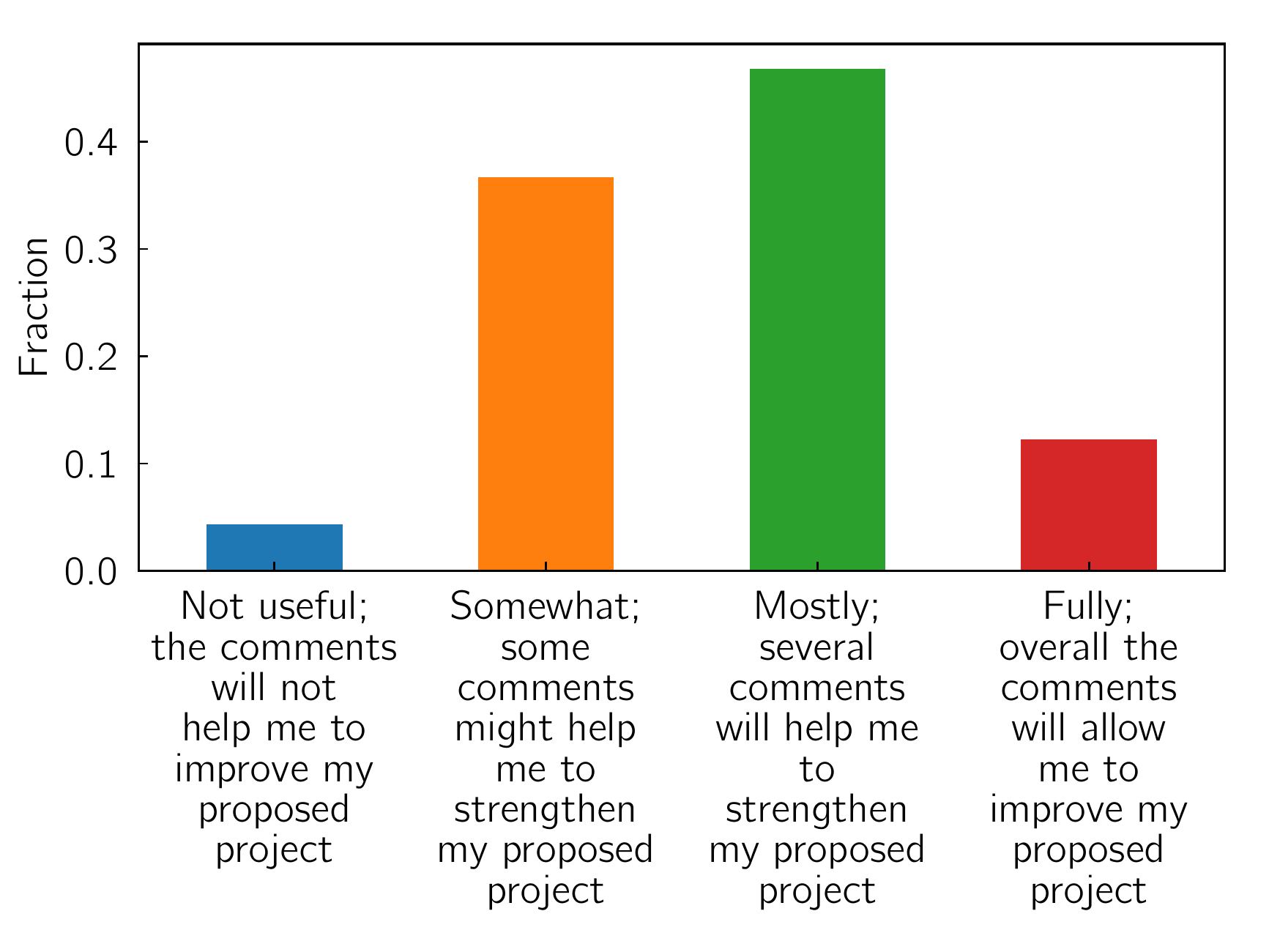}
    \caption{The referee was asked the question \textit{How useful were the comments that you received on your proposal?}}
    \label{fig:helpful_comments}
\end{figure}


\begin{figure}[p]
    \includegraphics[width=0.5\textwidth]{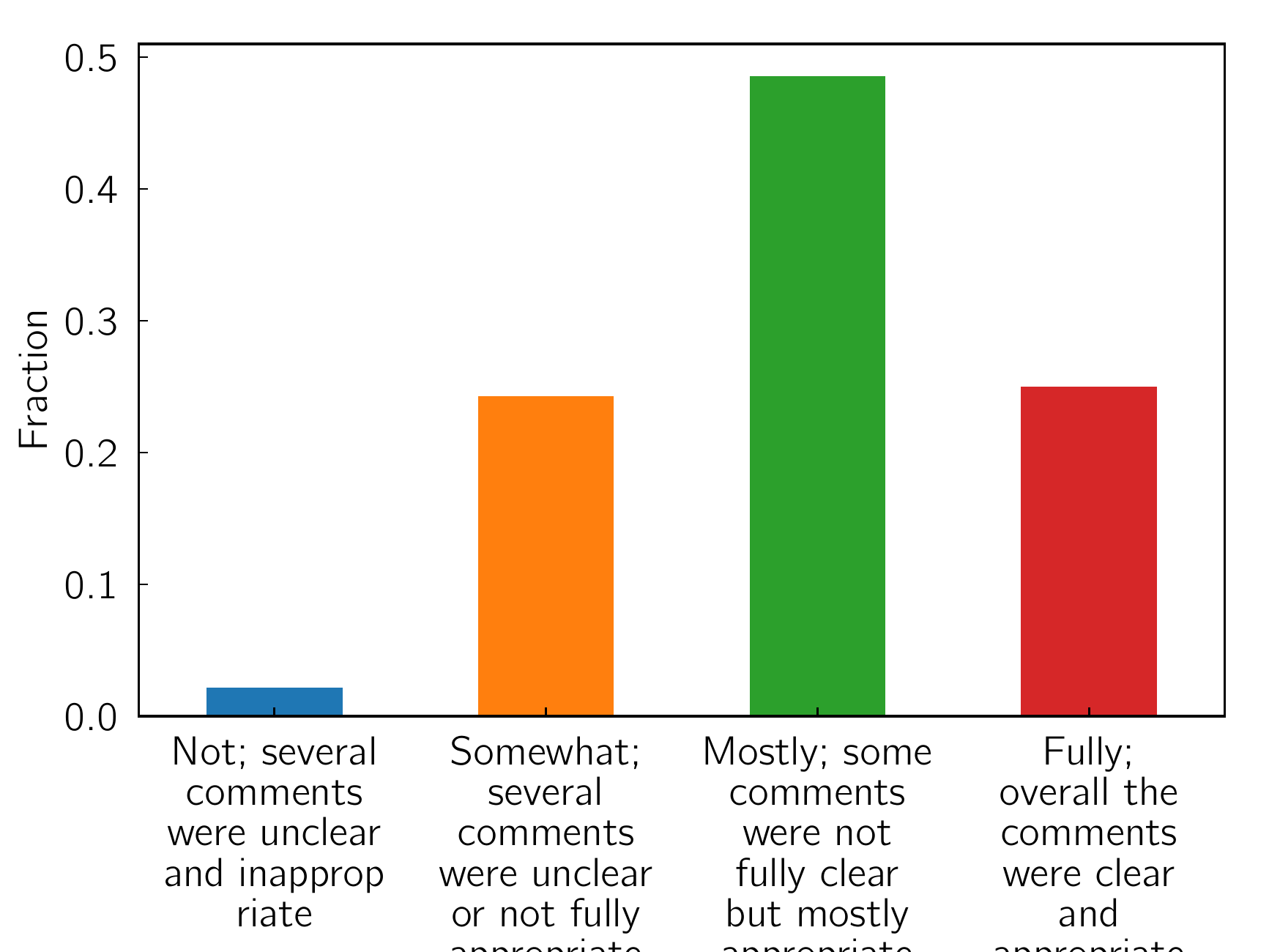}
    \caption{The referee was asked the question \textit{How clear and appropriate were the comments that you received on your proposal?}}
    \label{fig:appropriate_comments}
\end{figure}


\begin{figure}[p]
    \includegraphics[width=0.5\textwidth]{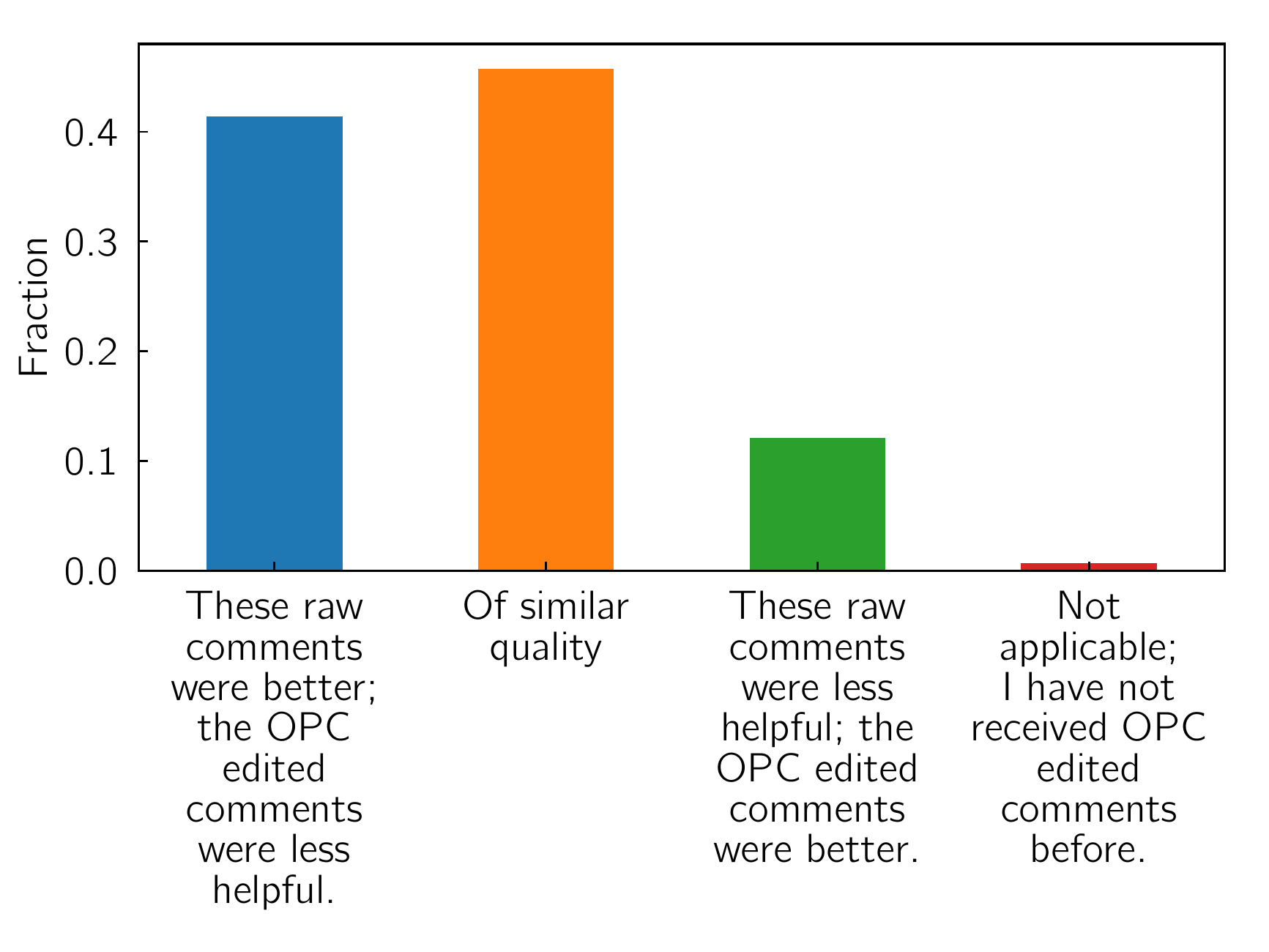}
    \caption{The referee was asked the question \textit{How do these raw comments compare to the edited comments from the OPC in the past?}}
    \label{fig:comment_opc_compare}
\end{figure}


\begin{figure}[p]
    \includegraphics[width=0.5\textwidth]{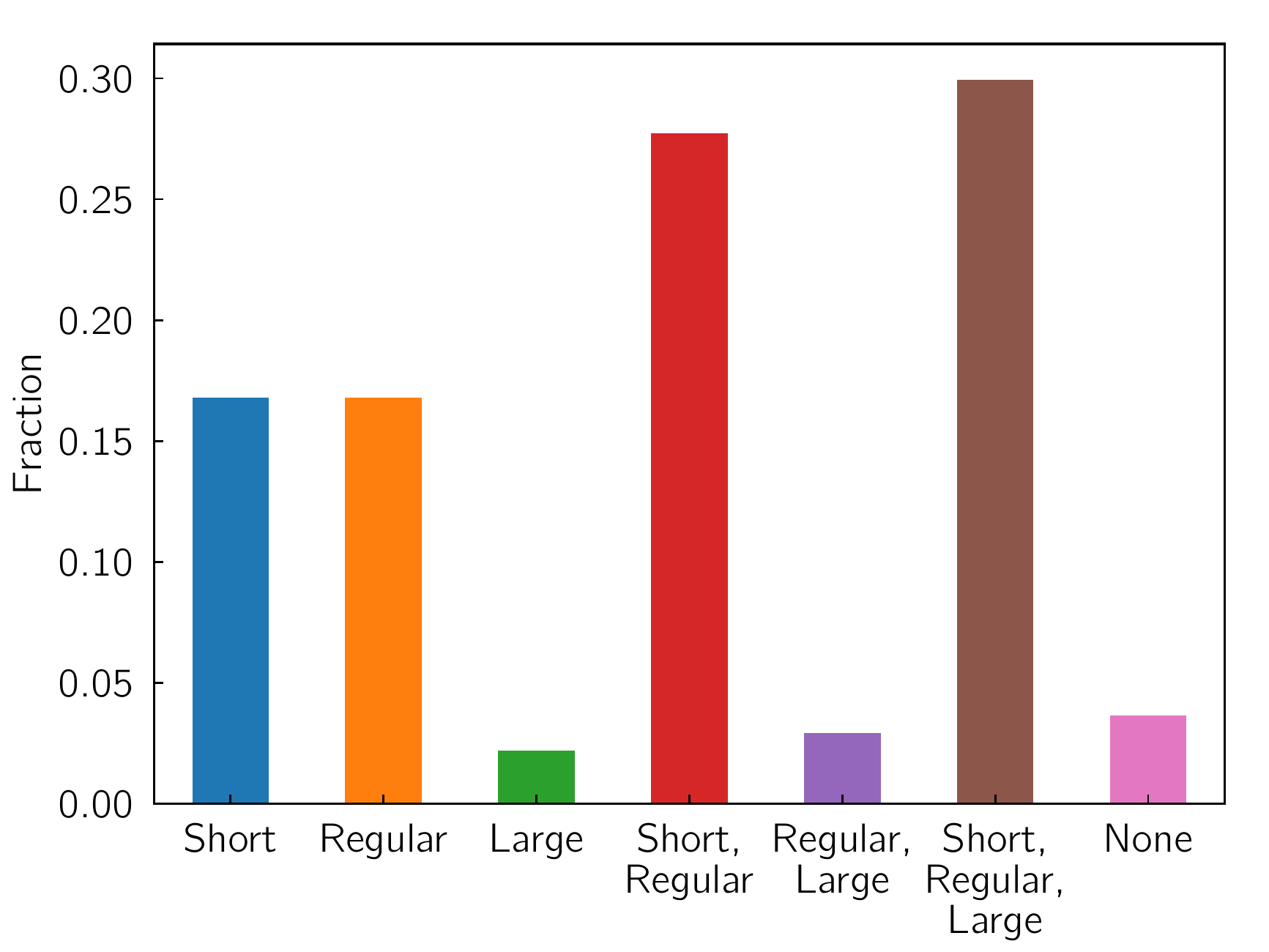}
    \caption{The referee was asked the question \textit{For which types of proposals do you think distributed peer review would be beneficial?}}
    \label{fig:application_of_dpr}
\end{figure}


\begin{figure}[p]
    \includegraphics[width=0.5\textwidth]{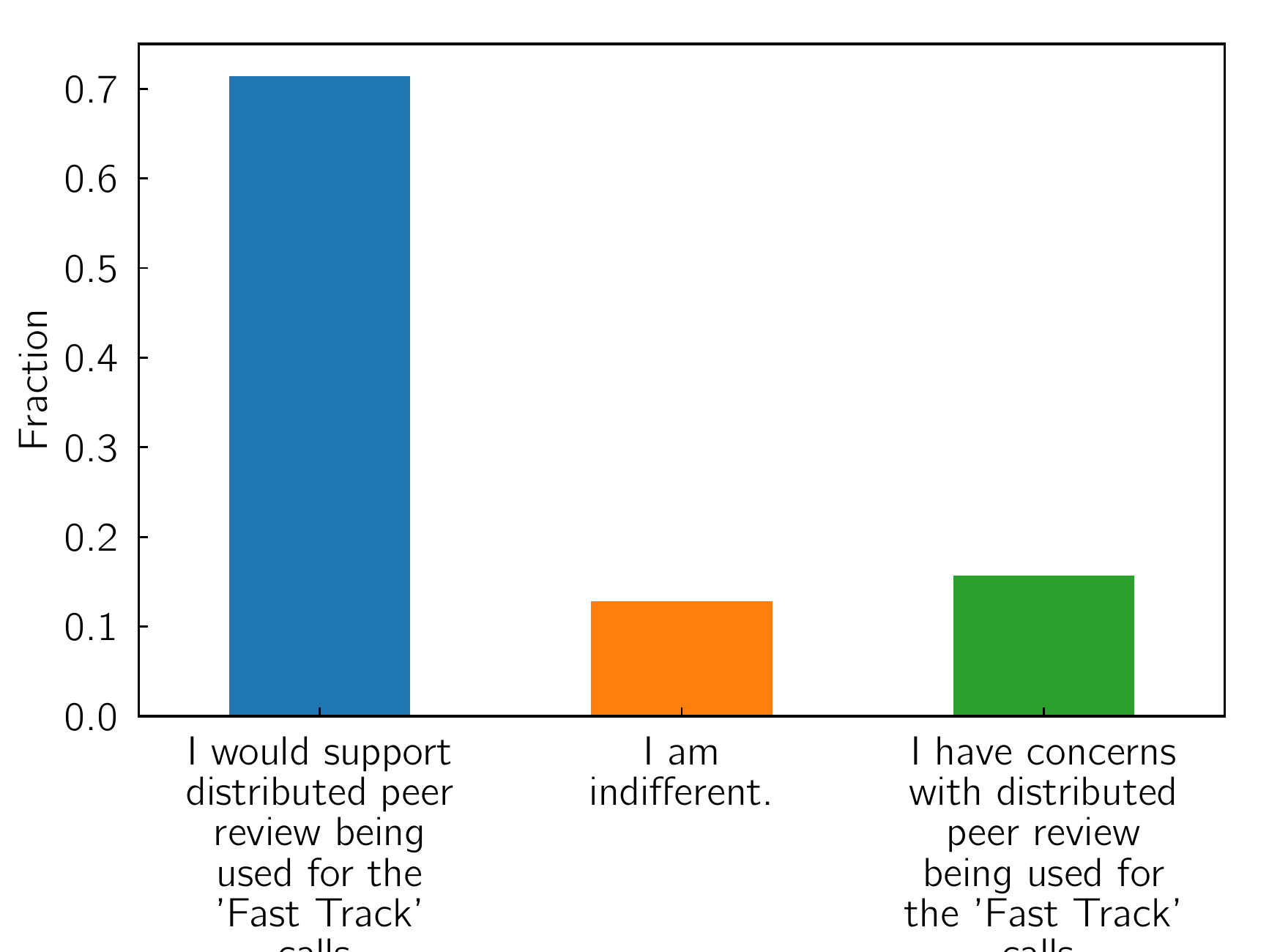}
    \caption{The referee was asked the question \textit{The ESO Time Allocation Working Group recommended to introduce—in addition to the existing calls for proposals—a 'Fast Track' channel with a fraction of the time allocated through (two) monthly calls. If implemented, would distributed peer review be appropriate for this time allocation?}}
    \label{fig:dpr_fast_track}
\end{figure}


\begin{figure}[p]
    \includegraphics[width=0.5\textwidth]{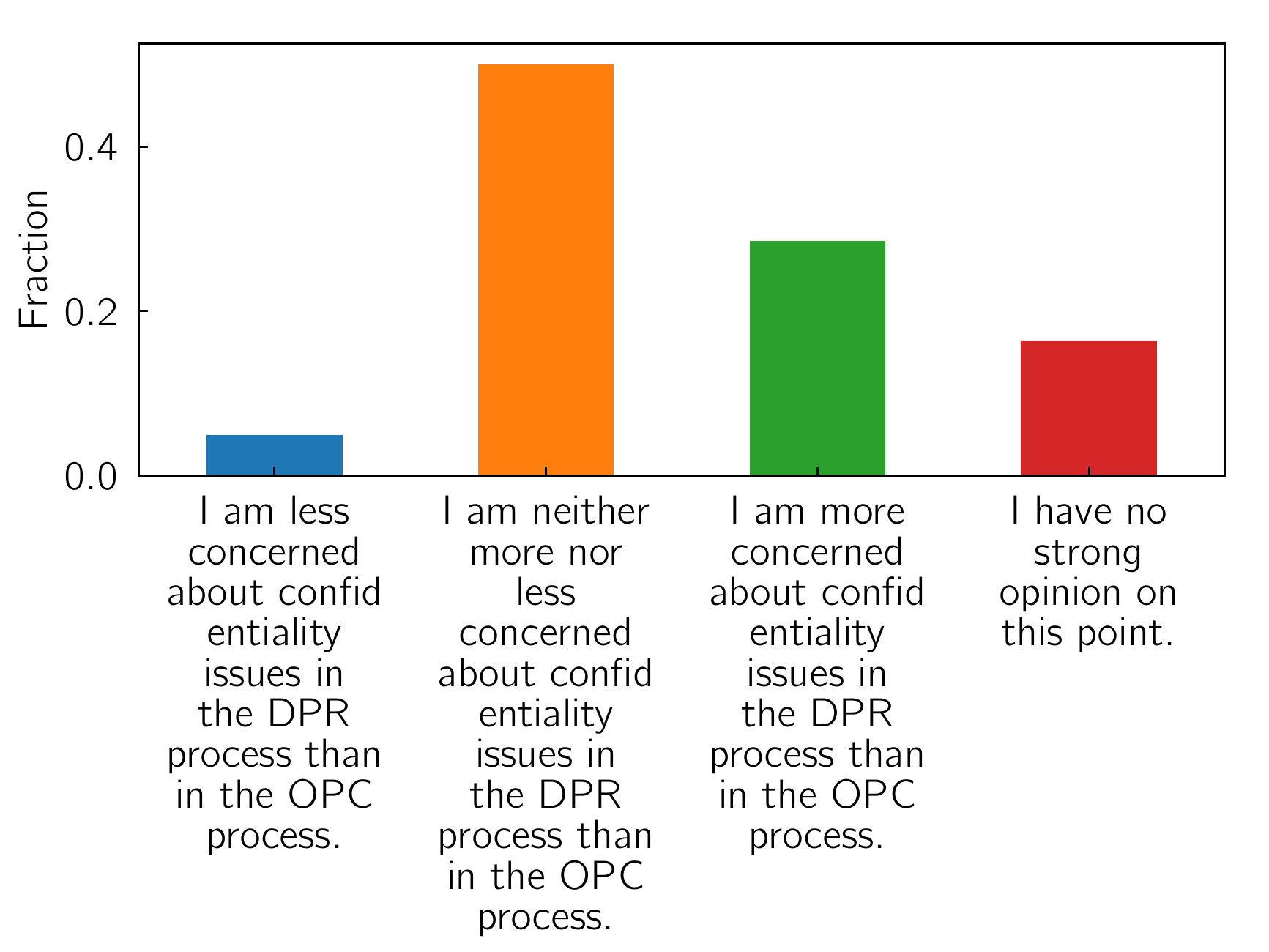}
    \caption{The referee was asked the question \textit{Securing confidentiality in the process:}}
    \label{fig:confidentiality_concern}
\end{figure}


\begin{figure}[p]
    \includegraphics[width=0.5\textwidth]{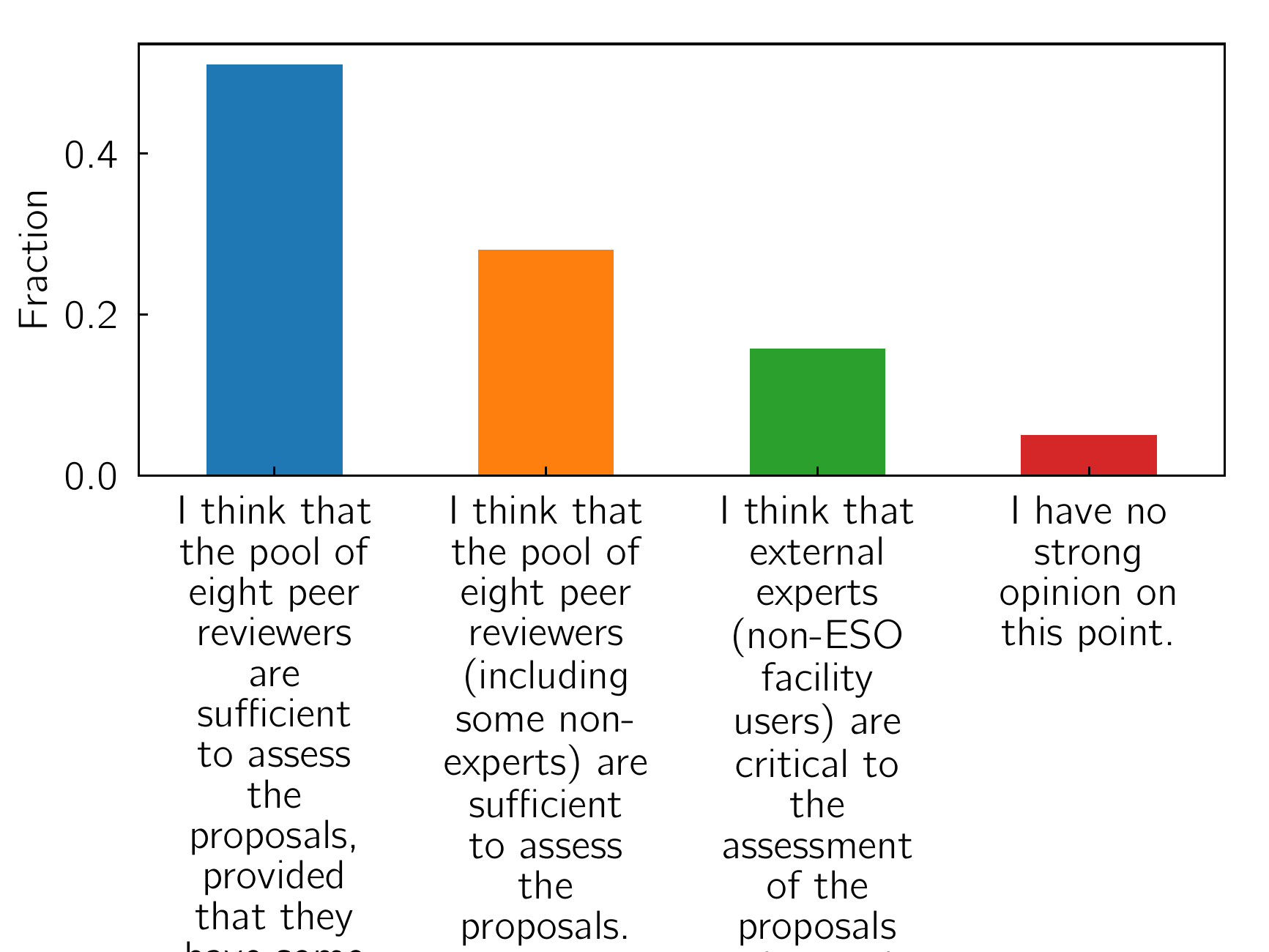}
    \caption{The referee was asked the question \textit{External expertise in the review process:}}
    \label{fig:no_of_reviewers}
\end{figure}


\begin{figure}[p]
    \includegraphics[width=0.5\textwidth]{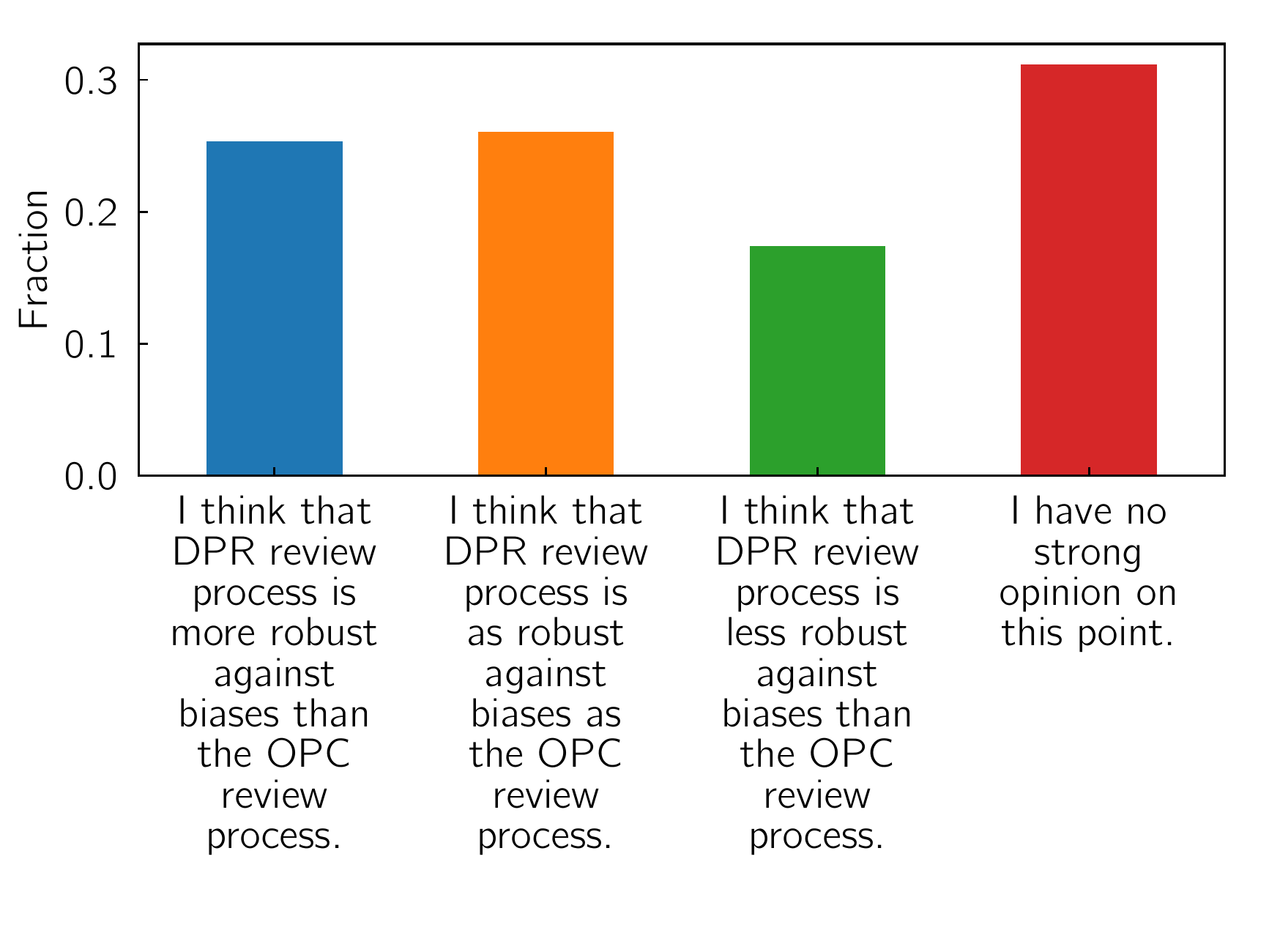}
    \caption{The referee was asked the question \textit{Robustness of the review process against any biases:}}
    \label{fig:bias_robustness}
\end{figure}